\documentclass[a4paper,UKenglish]{dagrep-v2021}

\usepackage{soul}
\usepackage{xcolor}
\newcommand*{\gt}[1]{\parindent0pt \textcolor{gray}{#1}}
\usepackage{tcolorbox}
\usepackage{url}
\usepackage{xurl}
\usepackage{enumitem}
\usepackage{booktabs}
\usepackage{adjustbox}
\usepackage{tabularx}
\newcommand{\footnoteurl}[1]{\footnote{\url{#1}}}
\subject{Report from Dagstuhl Seminar 23031}
\title{Frontiers of Information Access Experimentation for Research and Education}
\titlerunning{23031 -- Frontiers of Information Access Experimentation for Research and Education}

\author[1]{Christine Bauer}
\author[2]{Ben Carterette}
\author[3]{Nicola Ferro}
\author[4]{Norbert Fuhr}
\affil[1]{Utrecht University, NL, \url{c.bauer@uu.nl}}
\affil[2]{University of Delaware and Spotify, US, \url{carteret@acm.org}}
\affil[3]{University of Padua, IT, \url{nicola.ferro@unipd.it}}
\affil[4]{University of Duisburg-Essen, DE, \url{norbert.fuhr@uni-due.de}}

\authorrunning{Christine Bauer, Ben Carterette, Nicola Ferro, Norbert Fuhr}

\seminarnumber{23031}
\semdata{January 15--20, 2023 -- \href{http://www.dagstuhl.de/23031}{http://www.dagstuhl.de/23031}}


\additionaleditors{Guglielmo Faggioli, University of Padua, IT}

\volumeinfo
  {Christine Bauer, Ben Carterette, Nicola Ferro, and Norbert Fuhr}
  {4}
  {Frontiers of Information Access Experimentation for Research and Education}
  {13}
  {1}
  {1}
\DOI{10.4230/DagRep.13.1.1}

\usepackage{acronym}

\acrodef{ACL}[ACL]{Association for Computational Linguistics}
\acrodef{ACM DL}[ACM DL]{ACM Digital Library}
\acrodef{ADM+S}[ADM+S]{Automated Decision-Making and Society}
\acrodef{AI}[AI]{Artificial Intelligence}
\acrodef{ASSIA}[ASSIA]{Asian Summer School in Information Access}
\acrodef{CfP}[CfP]{Call for Papers}
\acrodef{CHI}[CHI]{Computer-Human-Interaction}
\acrodef{CLEF}[CLEF]{Conference and Labs of the Evaluation Forum}
\acrodef{CyCAT}[CyCAT]{Cyprus Center for Algorithmic Transparency}
\acrodef{ECIR}[ECIR]{European Conference on Information Retrieval}
\acrodef{ESS}[ESS]{Experiment Support System}
\acrodef{ESSIR}[ESSIR]{European Summer School on Information Retrieval}
\acrodef{FAccT}[FAccT]{Fairness, Accountability, and Transparency}
\acrodef{FDIA}[FDIA]{Future Directions in Information Access}
\acrodef{FIRE}[FIRE]{Forum for Information Retrieval Evaluation}
\acrodef{GDPR}[GDPR]{General Data Protection Regulation}
\acrodef{IR}[IR]{Information Retrieval}
\acrodef{KPI}[KPI]{Key Performance Indicator}
\acrodef{LLM}[LLM]{Large Language Model}
\acrodef{ML}[ML]{Machine Learning}
\acrodef{MRR}[MRR]{Mean Reciprocal Rank}
\acrodef{NLP}[NLP]{Natural Language Processing}
\acrodef{nDCG}[nDCG]{normalized Discounted Cumulative Gain}
\acrodef{NTCIR}[NTCIR]{NII Testbeds and Community for Information access Research}
\acrodef{PC}[PC]{Program Committee}
\acrodef{PyIRE}[PyIRE]{Python Interactive Information Retrieval Evaluation}
\acrodef{QA}[QA]{Question Answering}
\acrodef{RS}[RS]{Recommender Systems}
\acrodef{RecSys}[RecSys]{ACM Recommender Systems Conference}
\acrodef{SSH}[SSH]{Social Sciences and Humanities}
\acrodef{SPC}[SPC]{Senior Program Committee}
\acrodef{SEME}[SEME]{Search Engine Manipulation Effect}
\acrodef{SIGIR}[SIGIR]{ACM SIGIR Conference on Research and Development in Information Retrieval}
\acrodef{SIGIR-AP}[SIGIR-AP]{ACM SIGIR Conference on Information Retrieval in the Asia Pacific}
\acrodef{SIGPLAN}[SIGPLAN]{ACM Special Interest Group on Programming Languages}
\acrodef{SOTA}[SOTA]{State-of-the-Art}
\acrodef{TREC}[TREC]{Text REtrieval Conference}
\acrodef{UX}[UX]{User Experience}
\acrodef{WiMIR}[WiMIR]{Women in Music Information Retrieval}

\begin{document}

\maketitle

\begin{abstract}
This report documents the program and the outcomes of Dagstuhl Seminar 23031 ``Frontiers of Information Access Experimentation for Research and Education'', which brought together 37 participants from 12 countries.

The seminar addressed technology-enhanced information access (information retrieval, recommender systems, natural language processing) and specifically focused on developing more responsible experimental practices leading to more valid results, both for research as well as for scientific education.

The seminar brought together experts from various sub-fields of information access, namely \ac{IR}, \ac{RS}, \ac{NLP}, information science, and human-computer interaction to create a joint understanding of the problems and challenges presented by next generation information access systems, from both the research and the experimentation point of views, to discuss existing solutions and impediments, and to propose next steps to be pursued in the area in order to improve not also our research methods and findings but also the education of the new generation of researchers and developers.

The seminar featured a series of long and short talks delivered by participants, who helped in setting a common
ground and in letting emerge topics of interest to be explored as the main output of the seminar. This led to the definition of five groups which investigated challenges, opportunities, and
next steps in the following areas: \emph{reality check, i.e. conducting real-world studies}, \emph{human–machine-collaborative relevance
judgment frameworks}, \emph{overcoming methodological challenges in information retrieval and recommender systems through awareness and education}, \emph{results-blind reviewing}, and \emph{guidance for authors}.
\end{abstract}

\newpage

\section{Executive Summary}
\summaryauthor[Christine Bauer, Ben Carterette, Nicola Ferro, Norbert Fuhr]{%
Christine Bauer (Utrecht University, NL, c.bauer@uu.nl)\\
Ben Carterette (University of Delaware and Spotify, US, carteret@acm.org) \\
Nicola Ferro (University of Padua, IT, nicola.ferro@unipd.it) \\
Norbert Fuhr (University of Duisburg-Essen, DE, norbert.fuhr@uni-due.de)
}

\license

Information access---which includes \acf{IR}, \acf{RS}, and \acf{NLP}---has a long tradition of relying heavily on experimental evaluation, dating back to the mid-1950s, a tradition that has driven the research and evolution of the field. However, nowadays, research and development of information access systems are confronted with new challenges: information access systems are called to support a much wider set of user tasks (informational, educational, and entertainment, just to name a few) which are increasingly challenging, and as a result, research settings and available opportunities have evolved substantially (e.g., better platforms, richer data, but also developments within the scientific culture) and shape the way in which we do research and experimentation. Consequently, it is critical that the next generation of scientists is equipped with a portfolio of evaluation methods that reflect the field’s challenges and opportunities, and help ensure internal validity (e.g.,  measures, statistical analyses, effect sizes, etc., to support establishing a trustworthy cause-effect relationship between treatments and outcomes), construct validity (e.g., measuring the right thing rather than a partial proxy), and external validity (e.g., critically assessing to which extent findings hold in other situations, domains, and user groups). A robust portfolio of such methods will contribute to developing more \emph{responsible experimental practices}.

Therefore, we face two problems: Can we re-innovate how we do research and experimentation in the field by addressing emerging challenges in experimental processes to develop the next generation of information access systems? How can a new paradigm of experimentation be leveraged to improve education to give an adequate basis to the new generation of researchers and developers?

This seminar brought together experts from various sub-fields of information access, namely \ac{IR}, \ac{RS}, \ac{NLP}, information science, and human-computer interaction to create a joint understanding of the problems and challenges presented above, to discuss existing solutions and impediments, and to propose next steps to be pursued in the area.

To stimulate thinking around these themes, prior to the seminar, we challenged participants with the following questions:
\begin{itemize}
    \item Which experimentation methodologies are most promising to further develop and create a culture around?
    \item In which ways can we consider the concerns related to \ac{FAccT} in the experimentation practices? How can we establish FaccT-E, i.e. FaccT in Experimentation?
    \item How can industry and academia better work together on experimentation?
    \item How can critical experimentation methods and skills be taught and developed in academic teaching?
    \item How can we foster collaboration and run shared infrastructures enabling collaborative and joint experimentation? How to organize shared evaluation activities taking the opportunity of new hybrid forms of participation? 
\end{itemize}

We started the seminar week with a series of long and short talks delivered by participants, also in response to the above questions. This helped in setting a common ground and understanding and in letting emerge the topics and themes that participants wished to explore as the main output of the seminar.

This led to the definition of five groups which explored challenges, opportunities, and next steps in the following areas
\begin{itemize}
    \item \textbf{Reality check}: The working group identified the main challenges in doing real-world studies in \ac{RS} and \ac{IR} research -- and points to best practices and remaining challenges in both how to do domain-specific or longitudinal studies, how to recruit the right participants, using existing or creating new infrastructure including appropriate data representation, as well as how, why and what to measure.
    \item \textbf{Human-machine-collaborative relevance judgment frameworks}: The working group studied the motivation for using \acp{LLM} to automatically generate relevance assessments in information retrieval evaluation, and raises research questions about how \acp{LLM} can help human assessors with the assessment task, whether machines can replace humans in assessing and annotating, and what are the conditions under which human assessors cannot be replaced by machines.
    \item \textbf{Overcoming methodological challenges in \ac{IR} and \ac{RS} through awareness and education}: Given the potential limitations of today‘s predominant experimentation practices, we find that we need to better equip the various actors in the scientific ecosystem in terms of scientific methods, and we identify a corresponding set of helpful resources and initiatives, which will allow them to adopt a more holistic perspective when evaluating such systems.
    \item \textbf{Results-blind reviewing}: The current review processes lead to undue emphasis on performance, rejecting papers focusing on insights in case they show no performance improvements. We propose to introduce a results-blind reviewing process forcing reviewers to put more emphasis on the theoretical background, the hypotheses, the methodological plan and the analysis plan of an experiment, thus improving the overall quality of the papers being accepted.
    \item \textbf{Guidance for authors}:
    The Information Retrieval community has over time developed expectations regarding papers,
    but these expectations are
    largely implicit.
    In contrast to adjacent disciplines, efforts in the \ac{SIGIR} community have been rather sparse
    and are mostly due to individuals expressing their own views.
    Drawing on materials from other disciplines,
    we have built a draft set of guidelines with the aim of them being understandable, broad, and highly
    concise.
    We believe that our proposal is general and uncontroversial,
    can be used by the main venues, and can be maintained with an open and continuous effort driven by, and for, the community.
\end{itemize}

\tableofcontents


\section{Overview of Talks}

\abstracttitle{Kickoff on Frontiers of Information Access Experimentation
for Research and Education}
\abstractauthor[Ian Soboroff]{%
Ian Soboroff (National Institute of Standards and Technology, US, ian.soboroff@nist.gov)
}
\license
\label{subsec:talk:soboroff}

The goal of this talk is to set out a common starting point for the seminar, and I approach this from the perspective of test collections and information retrieval. I start from the structure of a test collection and describe the pooling and relevance assessment process, highlighting known issues in those processes, including incompleteness, assessor disagreement, shallow pooling, and integrating results from multiple test collections. I close the talk with a list of hard problems in evaluation such as handling low run coverage and the absence of external ground truth.


\abstracttitle{Goodhart’s Law and the Lucas Critique}
\abstractauthor[Justin Zobel]{%
Justin Zobel (University of Melbourne, AU, jzobel@unimelb.edu.au)}
\license
\label{subsec:talk:zobel}

The discipline of \ac{IR} has a deep literature examining how best to measure performance, in particular the practice of assessing retrieval systems using batch experiments based on collections and relevance judgements. However, this literature has only rarely considered an underlying principle: that measured scores are inherently incomplete as a representation of human behaviour. In other disciplines, the significance of the principle has been examined through the perspectives of Goodhart's law and the Lucas critique. Here I argue that these apply to \ac{IR} and show that neglect of this principle has consequences in practice, separate from issues that can arise from poor experimental designs or the use of effectiveness measures in ways that are known to be questionable. Specifically, blind pursuit of performance gains based on the optimisation of scores, and analysis based solely on aggregated measurements, can lead to misleading or meaningless outcomes.

This talk was based on SIGIR Forum paper
``When Measures Mislead: The Limits of Batch Assessment of Retrieval Systems''~\cite{talks-Zobel22Mislead}, available
at \url{https://www.sigir.org/wp-content/uploads/2022/07/p12.pdf}.


\abstracttitle{User-centric Evaluation}
\abstractauthor[Bart P. Knijnenburg]{%
Bart P. Knijnenburg (Clemson University, US, bartk@clemson.edu)
}
\license
\label{subsec:talk:knijnenburg}

I presented an evaluation framework to study the user experience of interactive systems. It involves measuring users’ perception and experiences with questionnaires and then triangulating these with behaviour. The subjective constructs explain why users’ behaviour is different for different systems---this explanation is the main value of our framework.

I also addressed the filter bubble, and proposed to evaluate and build information systems in a way that supports rather than replaces decision-making; covers users’ tastes, plural; and focuses on exploration and preference development rather than consumption.

Finally, I addressed the challenge of designing human subjects studies that preserve research participants' privacy and security while still generating robust results.


\abstracttitle{Offline Evaluation Based on Preferences}
\abstractauthor[Charles L. A. Clarke]{%
Charles L. A. Clarke (University of Waterloo, CA, claclark@uwaterloo.ca)
}
\license
\label{subsec:talk:clarke}

Traditional offline evaluation of search, recommender, and other systems involves gathering item relevance labels from human editors. These labels can then be used to assess system performance using offline evaluation metrics. Unfortunately, this approach does not work when evaluating highly-effective ranking systems, such as those emerging from the advances in machine learning. Recent work demonstrates that moving away from pointwise item and metric evaluation can be a more effective approach to the offline evaluation of systems.


\abstracttitle{The Impact of Human Assessors on Judgements, Labels, Supervised Models, and Evaluation Results}
\abstractauthor[Gianluca Demartini]{%
Gianluca Demartini (The University of Queensland, AU, demartini@acm.org)
}
\license
\label{subsec:talk:demartini}

When we evaluate systems or train supervised models we make use of human annotations (e.g., judgements or labels). In this talk, I have presented examples of how different people may provide different annotations for the same data items. First, I have shown how misinformation judgements are prone to political background bias \cite{Roitero2020,Labarbera2020}. Then, I have shown how human annotators discriminate based on the socio-economic status of the persons depicted in the annotated content \cite{Fan2022}.
The way human annotators are biased also depends on how the annotation task is framed and on what extra information we provide them with \cite{xu2023role}.
Finally, I have shown what it means to train supervised models with such biased labels and how these models behave very differently when they are trained with labels provided by different human annotators \cite{Perikleous2022}. It is thus important for us to start considering tracking information about who the human assessors and annotators are and to include this as meta-data of our test collections \cite{Demartini2021}.


\abstracttitle{A Plea for Result-Less Reviewing}
\abstractauthor[Norbert Fuhr]{%
Norbert Fuhr (University of Duisburg-Essen, DE, norbert.fuhr@uni-due.de)
}
\license
\label{subsec:talk:fuhr}

Scientific experiments aim at testing hypotheses and gaining insights into cause-and-effect for the setting studied.
Unfortunately, most \ac{IR} publications focus on the first aspect, while papers addressing the second aspect get rejected if they fail to show improvements in terms of performance. However, many published papers suffer from severe flaws in their experimental analysis part, which makes their results almost useless.
Focusing on performance numbers, top \ac{IR} conferences and journals accept only papers showing improvements, which also leads to publication bias. As PhD students must publish to get a degree, they might be tempted to cheat if their proposed method does not yield the desired results.

As a way out, we propose to switch to result-less reviewing, which is standard e.g.\ in some psychological journals. Here reviewers cannot see the actual experimental results and have to base their decision on the theoretical background, the hypotheses, the methodological plan and the analysis plan. In case of acceptance, the experimental results are included in the paper published. 

This approach could help to achieve higher scientific quality and better reproducibility of experimental studies in \ac{IR}.


\abstracttitle{Understanding your User, Process Tracing as a User-centric Method}
\abstractauthor[Martijn C. Willemsen]{%
Martijn C. Willemsen (Eindhoven University of Technology \& JADS - ‘s-Hertogenbosch, NL, m.c.willemsen@tue.nl)
}
\license
\label{subsec:talk:willemsen}
In evaluating our information access systems, we get more insights if we combine subjective measures (e.g. satisfaction) with interaction data~\cite{talks-Knijnenburg2012}. However, most interaction data used nowadays, like simple clickstreams, do not provide sufficient insights into the underlying cognitive processes of the user. In this talk, I show how richer process measures (like hovers and eye-tracking) can provide deeper insights into the underlying decision processes of a user. For example, they help to understand when and why users search more superficially or more deeply into a list of results from the algorithm.

\subsubsection{Process tracing in decision making}
In decision-making, process tracing methods are commonly used to better understand human decision processes~\cite{talks-Schulte2017}. In the talk, I demonstrated one technique that I developed myself, called mouselabWEB\footnote{https://github.com/MCWillemsen/mouselabWEB20}. This information board tool allows users to acquire information by hovering over boxes. It can be regarded as a cheap and simple eye-tracker-like tool that can be used in online studies. The tool allows users to easily design a mouselabWEB table and page and takes care of data storage and handling~\cite{talks-Mouselab2019}.
\subsubsection{Process tracing used in Recommender Systems}

We already used process tracing-like measures in earlier \ac{RS} work to better understand the decision processes. In our work on latent feature diversification~\cite{Willemsen2016}, we presented diversified lists of movie recommendations by their titles. Only when hovering the titles, additional movie information and poster were shown. This measured how much effort people spend and how many recommendations were inspected. We found that a top-20 list of recommendations resulted in more effort than a top-5 list, which subsequently increased choice difficulty and reduced satisfaction. In work on user inaction~\cite{Zhao2018}, we investigated why users do not interact with some recommended items, questioning if we should keep showing these recommendations. We found diverse reasons for inaction and showed that some reasons provide good reasons for not recommending the item again, whereas others indicate that it would actually be very beneficial to show the item again in the next round of recommendations. 


\abstracttitle{From Living Lab Studies to Continuous Evaluation}
\abstractauthor[Philipp Schaer]{%
Philipp Schaer (Technische Hochschule Köln, DE, philipp.schaer@th-koeln.de)
}
\license
\label{subsec:talk:schaer}

In this short talk, I briefly introduced the basic idea behind using living labs for information retrieval or recommender system evaluation. I also outlined a framework to extend living labs to enable a continuous evaluation environment. 

\subsubsection{Living labs}
Livings labs were introduced in CLEF and TREC by initiatives like NewsREEL~\cite{DBLP:books/sp/19/HopfgartnerBLKKSL19}, Open\-Search~\cite{DBLP:conf/trec/JagermanBSSTR17} or, more recently, LiLAS~\cite{DBLP:conf/clef/SchaerBCWST21a}, with a particular focus on academic search evaluation. The general motivation behind living labs is to enable in-vivo evaluation in real-world settings and to extend the Cranfield-style in-vitro evaluations. Limitations of Cranfield studies like being static and not incorporating real-world users should be avoided. Instead of using (domain-specific) experts to evaluate retrieval results, the behaviour of real-world users is logged to measure their usage of different system implementations. Approaches like A/B testing or interleaving allow comparing the amount and type of interactions with these different systems to infer the underlying system performance.
By integrating real-world systems and users into the evaluation process, organizers of living lab evaluations can hope to bring more diversity and heterogeneity in the set of evaluators and, therefore, a higher level of realism. In industry, these types of online evaluations in real-world applications are common but not in academia, as access to these systems is usually not possible for external researchers and their systems. Although in principle, systems like STELLA \cite{Breuer2021} would make this possible, it is rarely used. 

Most living lab CLEF and TREC initiatives suffered from a common set of issues, like, the small number of click events gathered in the experiments, therefore long-running experiments, missing user information or anonymous profiles, no differentiating in click events and no possibility to include expert feedback and generally the problem of being confronted with constant change in the systems and their data sets.

\subsubsection{Continuous evaluation}

A framework for continuous evaluation was outlined to overcome some of the previously outlined issues. The framework is based on a living lab installation within a real-world system but extends it with the following components:
\begin{itemize}
    \item Different user profiles -(regular) platform users whose user interaction data is logged and expert users that can directly annotate relevance labels on results in the systems. 
    \item Relevance assessments - The expert assessments will be added to a constantly growing test collection that has to support versioning.
    \item Simulation module - As both expert and regular user feedback is expected to be insufficiently small at the beginning, different user types or interaction patterns can be simulated based on the interaction and relevance data gathered so far.
\end{itemize}

These components within the framework can run over a long time and create a constantly growing set useful for evaluating systems - running in the living lab as an online study or using the distilled/simulated evaluation data available for offline evaluation. 

A first version of this framework will be implemented in the DFG-funded STELLA II project\footnote{https://stella-project.org/}.


\abstracttitle{An Idea for Evaluating Retrieve \& Generate Systems}
\abstractauthor[Laura Dietz]{%
Laura Dietz (University of New Hampshire, US, dietz@cs.unh.edu)
}
\license
\label{subsec:talk:dietz}

Natural language generation models (like GPT*) are here to stay, and they are a huge opportunity to build systems that combine retrieval and language generation in a combined system.

But: how can we evaluate the quality of such systems?

We discuss an idea for a new paradigm, the EXAM Answerability Metric \cite{talks-Sander2021}, which uses a \ac{QA} system along with some human-written exam questions to evaluate whether the systems retrieve good \emph{information} (instead of the right terms).

The paradigm has other advantages such as no need for highly trained assessors, no fixed corpus for retrieval (open web is possible), and comparison of retrieval-only systems and fully-generated systems on equal footing. Moreover, additional systems can be added for evaluation later without bias against non-participating systems. There is the possibility to add additional exam questions at a later point, to increase resolution between systems.

We compare the EXAM evaluation metric to the official TREC quality metrics on the TREC Complex Answer Retrieval Y3 track. We observe a Spearman Rank Correlation coefficient of 0.73. In contrast, ROUGE yields a correlation of 0.01.

There are also many open questions about the evaluation paradigm, I would like to discuss with participants in this Dagstuhl Seminar.


\abstracttitle{Metadata Annotations of Experimental Data with the \texttt{ir\_metadata} Schema}
\abstractauthor[Timo Breuer]{%
Timo Breuer (Technische Hochschule Köln, DE, timo.breuer@th-koeln.de)
}
\license
\label{subsec:talk:breuer}

In this talk, we present the current status of \texttt{ir\_metadata} \cite{DBLP:conf/sigir/BreuerKS22} - a metadata schema for annotating run files of information retrieval experiments. We briefly outline the logical plan of the schema that is based on the PRIMAD model (first introduced as part of the Dagstuhl seminar 16041 \cite{DBLP:journals/dagstuhl-reports/FreireFR16}). The acronym stems from the six components that can possibly affect the reproducibility of an experiment including the Platform, Research Goal, Implementation, Method, Actor, and Data. In addition, we extended the taxonomy with related subcomponents, for which details can be found on the project's website\footnote{https://www.ir-metadata.org/}.

Furthermore, we demonstrate how run files can be annotated in practice, describe the current software support and include example experiments in the form of reproducibility studies. Open points of discussion include what kinds of additional software features could be implemented to reduce the annotation effort or how the schema can be made a community standard in general. By introducing this resource to the community, we hope to stimulate a more reproducible, transparent, and sustainable use of experimental artefacts. 


\abstracttitle{Measuring Fairness}
\abstractauthor[Maria Maistro]{%
Maria Maistro (University of Copenhagen, DK, mm@di.ku.dk)
}
\license
\label{subsec:talk:maistro}

In recent years, the discussion on the fairness of \ac{ML} models has gained increasing attention and involved different research communities, including Information Retrieval (IR) and Recommender Systems (RS).
In the ML community, well-defined fairness criteria have been proposed and applied to the risk assignment score returned by classifiers.
Assume that there are two (or more) groups, denoted by $\mathcal{A}$ and $\mathcal{B}$, defined on attributes that should not be used to discriminate people, e.g., gender, ethnicity, or age.
Kleinberg et al.~\cite{DBLP:conf/innovations/KleinbergMR17} propose $3$ fairness criteria: (1) calibration within groups; (2) balance for the positive class; and (3) balance for the negative class.
Calibration within groups means that the probability score estimated by a classifier is well-calibrated, i.e., if a classifier returns a probability $x$ for people in group $\mathcal{A}$ to belong to the positive class, then an $x$ percentage of people in $\mathcal{A}$ should truly belong to the positive class.
Balance for the positive class states that the average estimated probability for people truly belonging to the positive class should be the same in groups $\mathcal{A}$ and $\mathcal{B}$.
Balance for the negative class is the counterpart defined for the negative class.
Kleinberg et al.~\cite{DBLP:conf/innovations/KleinbergMR17} proves that these criteria are incompatible, except for two non-realistic cases.

The above criteria are not directly applicable when the output of a system is a ranking.
Ekstrand et al.~\cite{DBLP:journals/ftir/Ekstrand0B022} identify several reasons, some of which are mentioned in the following.
First, items are organized in a ranking, where they receive different levels of attention due to the position bias~\cite{DBLP:conf/wsdm/CraswellZTR08}.
Therefore decisions based on model scores, i.e., how to generate the ranking, are not independent and can not be evaluated independently.
Second, users can access \ac{IR} and recommendation systems multiple times over a period of time and decisions based on model predictions are repeated over time. 
Thus, fairness should be evaluated for the whole process, not at a single point in time.
Third, multiple stakeholders are involved with \ac{IR} and \ac{RS} systems and they have different fairness constraints.
For example, users of the system might be concerned about receiving results that are not biased towards some of their attributes, e.g., gender, while providers might be concerned about their items not being underrepresented in the ranking.

Due to the above reasons, there has been a proliferation of fairness definitions and measures, targeting different nuances of the same problem and trying to adapt more general fairness definitions to the ranking problem.
Recent surveys identify more than 6060 different variants of fairness definitions resulting in more than 4040 different fairness measures~\cite{DBLP:journals/corr/abs-2206-03761,DBLP:journals/ipm/AmigoDMB23}. 

In this talk, I argue that there is a need for a better understanding of different fairness definitions and measures. 
I present some open questions and future research directions which include: an exploration of the relationship between bias, data distribution, and fairness~\cite{DBLP:journals/ipm/DeldjooBN21}; an analysis of formal properties and pitfalls of fairness measures as done for \ac{IR} measures~\cite{DBLP:conf/ictir/FerranteFM15}; evaluation approaches able to accommodate multiple aspects, e.g., relevance, fairness and credibility~\cite{DBLP:conf/cikm/MaistroLSL21}; guidelines, benchmarks, and tools to advise researchers and practitioners in designing the most appropriate evaluation protocol for fairness.


\abstracttitle{(Aspects of) Enterprise Search}
\abstractauthor[Udo Kruschwitz]{%
Udo Kruschwitz (University of Regensburg, DE, udo.kruschwitz@ur.de)
}
\license
\label{subsec:talk:kruschwitz}

Search and \ac{IR} is commonly associated with Web search but there are plenty of other areas that fall outside the scope of Web search and which are nevertheless interesting and challenging. One example is enterprise search which describes search within companies or other organisations \cite{Kruschwitz2017}. This is an area that has attracted little attention in academia (as well as in shared tasks and competitions) yet it affects millions of users who try to locate relevant information as part of their everyday work. Key challenges include the silo structure of data sources, privacy issues, the lack of link structure and the fact that there may only be a single relevant document (or none at all) for a given information need. All this has implications, and in the context of this seminar, some of the main challenges include the absence of test collections, problems with data sharing and reproducibility as well as the domain-specific nature of each use case. 


\abstracttitle{Identification of Stereotypes: Retrieval and Monitoring}
\abstractauthor[Paolo Rosso]{%
Paolo Rosso (Technical University of Valencia, ES, prosso@dsic.upv.es)\\
}
\license
\label{subsec:talk:rosso}

In the short talk, I addressed the problem of the retrieval of text fragments containing implicit and subjective information such as stereotypes, framing them, and annotating them. Part of the work was done in collaboration with OBERAXE, the Spanish observatory of racism and xenophobia. Transcribed speeches of the Spanish Congress of Deputies with immigrants as the target were framed as a threat or victims using a taxonomy where the negative/neutral/positive attitudes of the speaker were taken into account. Moreover, social media memes with women as a target were retrieved and annotated. The low inter-annotator agreement shows the necessity to go beyond the aggregated ground truth and consider the pre-aggregated information of each individual annotator in order to give voice also to minorities in disagreement with the opinion of the majority. Using, for instance, the learning with disagreements paradigm should allow the development of more equitable systems in the name of fairness. 


\abstracttitle{Coordinate Research, Evaluation, and Education in Information Access: Towards a More Sustainable Environment for the Community}
\abstractauthor[Nicola Ferro]{%
Nicola Ferro (University of Padua, IT, nicola.ferro@unipd.it)
}
\license
\label{subsec:talk:ferro}

The information access research field is characterized by several areas, such as \ac{IR}, \ac{RS}, and \ac{NLP}. These areas, in turn, offer various venues where the community can meet, discuss, and grow; typically, a mix of \emph{scientific conferences}, \emph{evaluation fora}, and \emph{summer/winter schools}.
For example, in the \ac{IR} area, there are several such venues around the world. In Europe, there is \ac{ECIR}\footnote{\url{https://www.bcs.org/membership-and-registrations/member-communities/information-retrieval-specialist-group/conferences-and-events/european-conference-on-information-retrieval/}} as scientific conference; \ac{CLEF}\footnote{\url{https://www.clef-initiative.eu/}}~\cite{Ferro2019} as evaluation forum; and, \ac{ESSIR}\footnote{\url{https://www.essir.eu/}} as summer school. In America, there is \ac{SIGIR}\footnote{\url{https://sigir.org/general-information/history/}} as scientific conference, which is also the premier international venue for the area; \ac{TREC}\footnote{\url{https://trec.nist.gov/}}~\cite{zz-HarmanVoorhees2005-editor} as evaluation forum; however, they lack a summer/winter school. In Asia, there is the newly born \ac{SIGIR-AP}\footnote{\url{http://www.sigir-ap.org/}} as scientific conference; \ac{NTCIR}\footnote{\url{https://research.nii.ac.jp/ntcir/index-en.html}}~\cite{zz-SakaiEtAl2020-editor} and \ac{FIRE}\footnote{\url{http://fire.irsi.res.in/}} as evaluation fora; and, \ac{ASSIA}\footnote{\url{https://goassia.github.io/}}.

All these venues are independent events, coordinated by their own steering committees (or equivalent bodies), with their own vision and strategic goals. Obviously, being the members of the community shared across the different committees and part of most of them, there is some informal level of coordination among these venues, which are cooperating for the overall growth of the community rather than competing for acquiring ``shares'' of it.

However, the main question of this talk is whether we can make better use of the venues we have in the field in order to fully unveil the potential of (research, evaluation, and education) in a more coordinated way and deliver further benefits to our community in terms of quality and volume of the research produced, robustness of the experimental results achieved, effective and smooth training and education to make our junior members the new leaders.

And, if this were possible in an area, such as \ac{IR}, what would it mean for the information access field at large? How would we cross the boundaries of the different areas?

\subsubsection{Examples of Coordination between Research, Evaluation, and Education}

In the following, we provide some possible examples of coordination between research, evaluation, and education, considering the case of \ac{ECIR}, \ac{CLEF}, and \ac{ESSIR}.

As a preliminary note, all of them happen in Europe, all of them follow an annual cycle, and their schedules match well enough\footnote{The alignment of the schedule is a partially intentional decision by the committees behind these venues.}:
\begin{itemize}
    \item \ac{ECIR}: submission deadline in October, conference in March/April;
    \item \ac{CLEF}: evaluation activities in January-May/June, submission deadline in June/July, conference in September;
    \item \ac{ESSIR}: school in July-August.
\end{itemize}

\paragraph*{ECIR $\leftrightarrow$ CLEF: Research $\leftrightarrow$ Evaluation}

There are already some coordination activities in place between \ac{ECIR} and \ac{CLEF}:
\begin{itemize}
    \item \ac{ECIR} hosts a section dedicated to CLEF labs, in order to stimulate participation in the \ac{CLEF} evaluation activities;
    \item \ac{CLEF} solicits its participants to follow-up their work in the labs with a submission to \ac{ECIR}.
\end{itemize}

This link is possible because the new labs for \ac{CLEF} are selected around July and this matches with the submission deadline to \ac{ECIR} the next October; moreover, the \ac{ECIR} session happens in March/April, which is still in due time for allowing participation in a \ac{CLEF} lab up to May/July. On the other side, \ac{CLEF} activities end in July (labs, papers), even if the actual event is later on in September; therefore, \ac{CLEF} participants have time for planning a follow-up submission to \ac{ECIR} in October.

Why is this link needed? Even if both \ac{ECIR} and \ac{CLEF} are part of the same \ac{IR} area, being it a large community, the audience of \ac{ECIR} and \ac{CLEF} is only partially overlapping. On the other hand, this audience may benefit from participation in both venues, not only because of more opportunities of conducting research but also because of the organized progress of such activities throughout the year, with intermediate delivery points, which help in making it smoother and break-down the overall work.

In his talk, Fuhr, see Section~\ref{subsec:talk:fuhr}, argued for the need for a \emph{result-less reviewing} approach, where papers are assessed on the basis of their methodology, innovation, research questions, soundness of the planned experiments and, if accepted, the actual experiments will be conducted later on, possibly in a follow-up publication.

This could represent another area of coordination between \ac{ECIR} and \ac{CLEF}: result-less papers are submitted at \ac{ECIR} and, if accepted, their experimental part is then submitted to \ac{CLEF} as a follow-up publication. Also in this case the schedule of the two venues aligns well enough to make this possible. And, again, this would allow the community to have more regular and intermediate steps at which to deliver their research, with the additional benefit of focusing each step on a specific aspect of the research and, possibly, improving the overall quality of the output, both the methodology and the experiments.

\paragraph*{ECIR $\leftrightarrow$ ESSIR: Research $\leftrightarrow$ Education}

There is currently no specific joint activity between \ac{ECIR} and \ac{ESSIR}.

A first example of activity could be for \ac{ESSIR} to offer a mentorship program for the students attending it, in order to help them in preparing their submission to \ac{ECIR} and getting feedback about it. Conferences sometimes offer mentorship programs to students but these are often asynchronous exchanges of emails or, at best, remote calls. In this case, students and senior researchers would be back-to-back in the same place for a week and this would allow for a much more smother and productive interaction. This link between \ac{ECIR} and \ac{ESSIR} would be possible because \ac{ESSIR} happens in July/August and the submission deadline for \ac{ECIR} is in October.

During the discussion that followed-up after the presentation, it was correctly asked how this link would compare/relate to a Doctoral Consortium activity. It is true that the two activities would share some commonalities, both being a form of mentorship to students. However, in the case of a Doctoral Consortium, the purpose is to provide students with overall feedback about the PhD theme or thesis; in this case, we would focus on a much more specific goal, which is the submission of a paper to a conference. As a side note, \ac{ESSIR} already hosts a form of Doctoral Consortium which is \ac{FDIA}.

Another form of activities could be to present at \ac{ESSIR} ``digested'' research breakthrough highlights from the latest \ac{ECIR} edition. In organizing a summer/winter school there is always a trade-off between offering foundational and advanced lectures; in both cases, the lectures are expected to cover in a reasonably complete way the topic they are about. This forces school organizers to select some topics and makes it impossible to cover all the frontier of the research in the field. These ``digested highlights'' could be a partial solution: they could provide a taste of other areas of the research frontier, still not being fully-fledged lectures.

\paragraph*{ESSIR $\leftrightarrow$ CLEF: Education $\leftrightarrow$ Evaluation}

There is currently no specific joint activity between \ac{ESSIR} and \ac{CLEF}.

A possible activity could be to organize a permanent educational lab at \ac{CLEF}, focusing on some basic tasks such as ad-hoc retrieval. This would allow us to address another trade-off typical of summer/winter schools: lectures versus hands-on sessions. Indeed, it is often difficult to find the right balance between the two and, due to limited time available or even hardware/software setup, the hands-on sessions are often at risk to be an oversimplification. On the other hand, a permanent lab at \ac{CLEF} could be seen as a very extensive hands-on session of \ac{ESSIR}, giving the possibility of exploring further details, also of practical nature. Moreover, this would allow for addressing some foundational concepts (and ensuring they are well understood) before the school, giving them additional freedom when school organizers have to balance between foundational and advanced topics.

\subsubsection{Towards a More Sustainable Environment for Our Community}

The examples discussed in the previous section provide a very basic idea of what better coordination among our venues could be. At the same time, they should help in making clear that a change in our perspective is required. 

Indeed, we currently adopt a sort of \emph{point-wise} vision, where we target and optimize for each venue separately, and the venues themselves are somewhat organized and managed in isolation. In a sense, this incurs in a \emph{waste of resources}, since we (both organizers and participants) may need to redo some part of the same work when passing from one venue to another and, definitely, we do not exploit any synergy and interaction among venues.

On the other hand, the approach presented in the previous section would require us to adopt a more \emph{flow-wise} vision, consisting of progressive stamps of quality, where the different steps of our research and education activities are part of an organized process, whose ultimate goal is to make them proceed in a smoother way along the pipeline, possibly also improving the quality of the outputs. Moreover, this could be also of further help for junior researchers who often are under the ``publish or perish'' pressure, forcing them to spread submissions to whatever venue, often repeating or slicing their work. In this case, for example, submitting a result-less paper to \ac{ECIR} and the follow-up experiments to \ac{CLEF} would preserve the publication volume but in a more controlled way, aimed at ensuring a better quality of each output, methodology first, and experiment after.

Obviously, this new vision will require training of both authors and reviewers, who should understand the model and how to properly apply it. For example, if a result-less paper is accepted at \ac{ECIR}, when reviewing the experimental part at \ac{CLEF}, its methodology should not be questioned again, especially if the reviewers happen to be different, but the review should focus just on the experimentation and the insights gathered from it.

Overall, this new coordinated vision aims at creating a \emph{more sustainable environment} for our community, reducing the waste of resources for intermediate steps and optimizing the overall effort for delivering an improved quality.


\abstracttitle{Recommender Systems Evaluation 2017--2022}
\abstractauthor[Alan Said]{%
Alan Said (University of Gothenburg, SE, alansaid@acm.org)
}
\license
\label{subsec:talk:said}
Recommender systems research and practice is a fast-developing topic with growing adoption in a wide variety of information access scenarios.
In this talk, I presented a snapshot of the evaluation landscape in \ac{RS} research between 2017 and 2022. The talk is based on a systematic literature review analyzing 64 papers, focusing particularly on the evaluation methods applied, the datasets utilized, and the metrics used. The study shows that the predominant experiment method is offline experimentation and that online evaluations are primarily used in combination with other experimentation methods, e.g., an offline experiment.
The analysis of the snapshot of the last six years of recommender systems research shows that the research community in recommender systems has consolidated the majority of experiments on a few metrics, datasets, and methods.

\newpage
\section{Working Groups}
\abstracttitle{Reality Check -- Conducting Real World Studies}
\label{subsec:reality}
\abstractauthor[Bruce Ferwerda, Allan Hanbury, Bart P. Knijnenburg, Birger Larsen, Lien Michiels, Andrea Papenmeier, Alan Said, Philipp Schaer, Martijn Willemsen]{%
Bruce Ferwerda (Jönköping University, SE, bruce.ferwerda@ju.se)\\
Allan Hanbury (TU Wien, AT, allan.hanbury@tuwien.ac.at)\\
Bart P. Knijnenburg (Clemson University, US, bartk@clemson.edu)\\
Birger Larsen (Aalborg University Copenhagen, DK, birger@ikp.aau.dk)\\
Lien Michiels (University of Antwerp, BE, lien.michiels@uantwerpen.be)\\
Andrea Papenmeier (Universität Duisburg-Essen, DE, andrea.papenmeier@uni-due.de)\\
Alan Said (University of Gothenburg, SE, alansaid@acm.org)\\
Philipp Schaer (Technische Hochschule Köln, DE, philipp.schaer@th-koeln.de)\\
Martijn Willemsen (Eindhoven University of Technology \& JADS, NL, m.c.willemsen@tue.nl)\\
}
\license

Information retrieval and recommender systems are deployed in real world environments. Therefore, to get a real feeling for the system, we should study their characteristics in ``real world studies''. This raises the question: What does it mean for a study to be realistic? Does it mean the user has to be a real user of the system or can anyone participate in a study of the system? Does it mean the system needs to be perceived as realistic by the user? Does it mean the manipulations need to be perceived as realistic by the user? 

\subsubsection{Background \& Motivation}
\label{sec:reality-motivation}

Arguably, the most realistic users can be found on existing systems, which will typically have a sufficiently large user base. However, this raises some additional questions. Firstly, there is the question of how to sample from this user base to obtain a representative sample. Secondly, these users may have some expectations of the system, which may make them somewhat resistant to (drastic) changes. On the other hand, recruiting new users comes with its own set of challenges, discussed further in Section~\ref{sec:recruiting-participants}.

\begin{figure}[t]
    \centering
    \includegraphics[width=\linewidth]{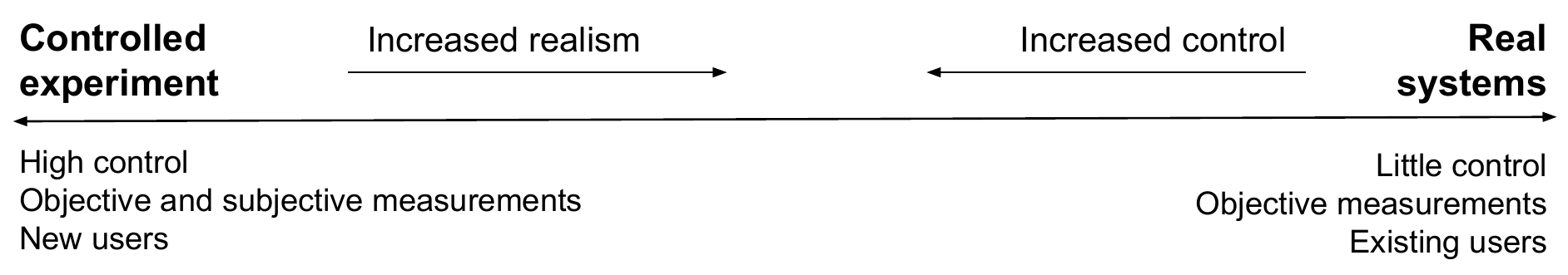}
    \caption{Control versus realism continuum}
    \label{fig:realism-overview}
\end{figure}

In a similar vein, the largest degree of ``system realism'' would be achieved by studying real users of an existing system. For example, log-based studies have been considered the best examples of real world studies~\cite{Kelly2007} since they capture behavior in a real-life setting, with little chance of contamination or bias. However, this limits the amount of control we, as researchers, can exert, and thus the research questions we can pose and answer. On the other hand, highly controlled experiments might lack realism in terms of the system, the user experience (users knowing they are being studied) and the generalizability of the study. Realism in a study is a continuum, as illustrated in Figure~\ref{fig:realism-overview}, ranging from highly controlled experiments towards real systems with real users, and researchers need to identify the appropriate experiment type for their purpose \cite{ZangerleBauer2022}.

One central question in running real world studies is the influence of measurements on the behavior and experience of users. Following the Heisenberg principle~\cite{Heisenberg1927}, it is impossible to measure without influencing. If we study existing users in an existing system, and only use behavioral measures and logs from the system we will not affect users much but it will be hard to answer our question, as the evaluation of our manipulation will be difficult. On the other hand, when we start collecting additional measures, like intermediate surveys, users will know they are part of a study and modify their behavior because of that (Hawthorne effect~\cite{schwartz2013hawthorne}). Also, longer surveys might break the actual flow of system usage and demotivate people. Survey questions might provide the users with insights into the underlying research questions, resulting in unwanted demand characteristics or socially desirable answer patterns. 

However, triangulating objective (behavioral) data with subjective measures will be crucial to understand how users experience the system~\cite{Knijnenburg2012}, so a careful development and usage of a combination of subjective and objective measures is going to be central to balancing realism with adequate measurement. The challenge of `How to measure’ is further discussed in Section~\ref{sec:how-to-measure}. 

Then, we have the realism of the research question and experiment design. In any experiment, we manipulate the system, thus breaking some existing habits or patterns. Especially when studying users of an existing system, the realism of this manipulation is crucial. If users do not experience the manipulation as a realistic feature or implementation, the results may not be representative. Similarly, the degree of information given to the user may also influence the realism of the study. If we provide users with too much information, e.g., a very specific task and scenario to work from, users may perform actions they would not have in a realistic situation. On the other hand, if we provide too little information, e.g., when we introduce a new feature on an existing platform without any instruction, we require users to invest the time and effort to find out how the feature works before they can use it in the way we intended. 

Another important consideration regarding experiment design is the assignment of users to different versions of a system. Should the experience of a single user be kept consistent throughout the entire study? Such between-subjects designs have the advantage of preventing any spill-over effects but users working side by side or communicating about the system might discover there are different versions of the system, accidentally revealing the experimental conditions and goals. Within-subject designs allow users to experience all experimental conditions, which increases statistical power (as we can control for participant variance) but ordering and spill-over effects have to be considered. Moreover, to make a real world study sufficiently realistic and also understand how behavior changes over time and how habits are formed, we will need to consider longitudinal studies which come with their own set of challenges discussed in Section \ref{sec:longitudinal-studies}. 

Even when we carefully design our experiments and research questions and select the appropriate participants, we may arrive at conclusions that do not necessarily generalize beyond the domain. The tension between domain-specific experiments and generalizable findings is further discussed in Section~\ref{sec:realism:domain-specific}.

Finally, the cost of running a real world study is typically many times higher than performing offline evaluation~\cite{ZangerleBauer2022}. Therefore it is important to also consider the available research infrastructure, and promote the development of reusable research infrastructure, as elaborated in Section \ref{sec:research-infrastructure}, and provide datasets in sufficiently general formats to promote reuse, as discussed in Section \ref{sec:data-representation}.

\subsubsection{Recruiting Participants}
\label{sec:recruiting-participants}
Real-world user studies require recruiting efforts to find the ``right’’ participants for the research. As a prerequisite, researchers need to have a clear understanding of the target user group and be able to \textbf{formalize the target user characteristics}. While some research can be conducted on a user sample with few limitations, others pose fine-grained requirements for user characteristics. In both cases, the user group needs to be carefully designed and adapted to the research problem at hand so that the user study is conducted on a sample representative for the user base~\cite{newman2021data}. 

Although some research communities have a broad consensus of what characteristics of participants should be reported, the \ac{RS} and \ac{IR} communities do not yet have a clear checklist of \textbf{reporting sample characteristics} and their information needs. Similarly, very few test collections, like the iSearch collection~\cite{lykke2010collection}, actually report on the context and task users are in. Standardized reporting and metadata would also enable reproducibility~\cite{breuer2022metadata} and data re-use (see Section \ref{sec:data-representation}). Inviting users that fit the recruitment criteria can be challenging. To invite users that fit the user group characteristics, information about the potential participants must be available in a structured format for filtering. Especially in \ac{IR} and \ac{RS}, systems often rely on user profiles~\cite{kobsa1993user}. Such profiles would therefore not only facilitate recruitment but also the usage of the system and avoid the ``cold-start’’ problem~\cite{lika2014facing}. With detailed user profiles, adhering to the GDPR and CCPA and formulating appropriate consent forms become additional points on a researcher’s checklist.

Moreover, participants must be \textbf{recruited at the right moment}: People must be in the right mindset to start with the study. For some user groups, e.g., professionals, finding a good timing to ask for participation is crucial. Participants also must stay motivated throughout the session (or possibly even beyond) to deliver complete data. To gather high-quality data from users in real-life, ensuring that users participate for the right reasons is important too, e.g., participants should have an internal motive (that is, an actual information need) rather than generating data for financial compensation. That said, offering appropriate incentives also works towards data quality and participant motivation~\cite{goritz2006incentives}. For that, a thorough understanding of user needs and motivations is needed. If the task/system provides users with a real benefit and actual value, the payment might not be needed and could even reduce realism and intrinsic motivation. Without such benefits, user behavior might be mostly driven by monetary incentives and divert from user behavior in the wild. However, these aspects are not necessarily in contradiction. Carefully designed, payment combined with benefits might reinforce each other. For example, in a recent longitudinal study on a music genre exploration tool, Liang and Willemsen~\cite{LiangWillemsen2022} recruited new users online and paid them per session, with the system providing the additional benefit of exploring genre exploration and providing them with a personalized playlist. User drop-out was lower than common and engagement remained high across 6 weeks and 4 sessions, despite users having to respond to a medium-sized survey after every session.

\textbf{Recruiting at the right time} can also concern the time of day, week, or season. For example, recruiting during working hours might lead to a lack of users with full-time jobs. Defining filter criteria does not ensure that the diversity of the target user group is covered. Consequently, researchers must monitor the participant group to cover the full bandwidth of the user group under investigation. Neglecting the monitoring of incoming participants could lead to under- or over-representation of certain age, gender, or profession groups~\cite{bergman2016have}.

The \textbf{recruitment channel} is equally important for \ac{IR} and \ac{RS} studies in the wild. Several online recruiting platforms exist and can be used for studies in this field~\cite{alonso2009can}, e.g., MTurk or Prolific, each with their own participant characteristics~\cite{follmer2017role,peer2017beyond}. Other online recruiting channels include social media~\cite{newman2021data}. Offline recruiting for online experiments can pose additional challenges for participants. In some cases, \ac{IR} and \ac{RS} systems are already used in the wild and provide an established user base to invite for studies.

\subsubsection{How to Measure}
\label{sec:how-to-measure}

The abundance of various types of data is both a benefit and a curse of real world studies. Whereas the subsection on data representation (see Section \ref{sec:data-representation} covers the proper management of this data, the current subsection addresses the measurement of data from the perspective of motivation (why do we measure?), best practices (what should we measure, and how can we make measurement easier?), and issues (what makes measurement difficult in realistic studies?). As real world studies often revolve around specific tasks and use contexts (Section~\ref{sec:realism:domain-specific}), we also address the (lack of) generalizability of measurement.

\paragraph{Why to measure}

\textbf{Conduct theory-driven research}
Real-world studies allow us to go beyond optimization of offline algorithmic performance in terms of performance metrics such as \ac{MRR}, \ac{nDCG} and recall, to a fine-grained analysis of how different system parameters can influence the system’s performance at a given task.

Running a real world study requires researchers to think carefully about this ``task'', the right way of measuring how well the system performs at this task, and how the performance is impacted by the different system parameters. Tasks may range from highly domain-specific to more general, as discussed in Section \ref{sec:realism:domain-specific}. This domain-specificity means that if such studies aim to make generalizable contributions to an existing body of scientific knowledge, they should aim to explain why certain system parameters lead to higher performance.

Conducting theory-driven research requires additional measurement of intermediate (or mediating) variables that provide an explanation for the variance in performance indicators caused by system manipulations. Such mediating variables are often inherently user-centred; they can be characterized as subjective system aspects (users’ perceptions of the manipulations) and user experience variables (users’ self-relevant evaluation of the user experience) \cite{Knijnenburg2012}. These can be measured with questionnaires, but there may exist behavioral proxies.

\textbf{Define an evaluation target}
In realistic studies, the evaluation target must shift from system performance to a multi-faceted consideration of stakeholder satisfaction \cite{ZangerleBauer2022}.

As the main goal—and hence the standard metrics—of traditional \ac{IR} and \ac{RS} research is to optimize system performance, it avoids the question of who these metrics are optimized for. In realistic studies, metrics must be optimized to satisfy the stakeholders of the system, and the goals of these stakeholders—and hence the metrics to measure these goals—may not always align. Most prominently, measuring the satisfaction of the end-users of a system has traditionally involved user experience metrics like satisfaction, decision confidence, and self-actualization \cite{Knijnenburg2012}, while system owners tend to be interested in metrics related to conversions, such as click-through rate, session length and basket value \cite{JannachJugovac2019, recommendationspurposejannach2016}.

\paragraph{What to measure}

\textbf{Carefully determine what to measure}
Realistic studies must capture a variety of measures that are closely related to the evaluation target and/or can explain how/why certain system aspects influence the evaluation target.

Realistic studies tend to support a variety of user behaviors, and researchers are encouraged to instrument their research systems to capture these behaviors, such as page visits, ratings, and purchases. At the same time, though, considerations of end-user privacy may prescribe that measurement be limited to the metrics that are essential to answer the research questions. It is important to acknowledge here that a user’s behavior is not always an accurate representation of their own longer-term goals (let alone the goals of the system owner). As the ``true'' evaluation target may be difficult to measure (i.e., ``user satisfaction'' is an inherently latent variable, and ``company profit'' is an aggregate measure that depends on many other variables), researchers must decide which of the measurable behaviors are most closely related to the evaluation target (see also Section \ref{sec:realism-best-practice}). 

An important consideration here is that certain implicit behaviors may also provide valuable insights—especially when taking the importance of explanation into consideration. Users who are ignoring a recommendation, quickly navigating away from a page, or abandoning a shopping cart are providing important insights into their experience.

Users’ subjective evaluations may also be important to measure: such measures may be a more accurate representation of their goals than behaviors, and even in cases where the value of behavioral metrics is clear, subjective evaluations can be used to explain the occurrence of certain behaviors. Subjective evaluations are inherently latent and must be measured using ``indicator variables''~\cite{DeVellisRobertF1991Sd}. The best practice to measure such evaluations is to use multi-item measurement scales, but administering such scales may be considered an intrusive practice (more suggestions on how to best do this are provided below). 

Process data can also be used to explain how an evaluation target is or is not met. Process data consists of particularly granular navigational data—usually at the level of mouse-overs, intermediate clicks, or mouse movements—that can be used as evidence of a user’s decision processes (e.g., which search result to visit, which product to buy, which movie to watch)~\cite{Mouselab2019, Schulte2017}.

\textbf{Make things more measurable}
Realistic studies must trade off depth of measurement with user burden: more insightful measures are often more obtrusive, thereby reducing realism and participation. Below we provide suggestions on how to reduce the obtrusiveness of measurement.

While process measures are very useful to explain users’ decision processes, precise process measures tend to require a certain system structure. For example, users’ attention is easier to measure if certain information is hidden behind a click or a mouse-over if the user must perform a measurable action to acquire said information. More generally, behavioral data tend to be noisy due to the influence of external factors and system factors. The latter can be attenuated by reducing the number of available features and/or the amount of system personalization. Conversely, one can boost the ``signal'' to be measured by making the manipulated system aspect (e.g., a list of recommendations from a variety of different algorithms) more prominent in the system. Importantly, though, all of these practices may reduce the realism of the study.

Moreover, while subjective measures and process measures are invaluable in realistic studies—especially when it comes to explanation—subjective measurement is also more intrusive. Interrupting the user to fill out a questionnaire makes the interaction less realistic, and may cause asymmetric drop-outs from the study. An important consideration in this regard is when to measure users’ subjective experience. The ideal but most intrusive timing is during the interaction; if the measurement occurs after the experience, it will be a retrospective and aggregate account of their experience. Aggregate retrospective evaluations of experiences have been shown to be unduly influenced by strong negative events (peaks), and events that occurred at the end of the experience~\cite{kahneman1993}. Finally, if the measurement occurs too long after the experience, it may no longer accurately reflect the experience, as the user may simply no longer remember the experience. Similarly, in certain contexts users’ subjective evaluations and even their interaction patterns may be inaccurate representations of their true interests—people’s responses may fall prey to desirability bias, framing and default effects, or other heuristic influences that must be accounted for in measurement.

As a final consideration, one could suggest that rather than minimizing (the obtrusiveness of) measurement, one could attempt to promote measurement, e.g., by providing easily accessible and/or gamified feedback elements. Evidently, this may reduce the realism of the study.

\textbf{Provide qualitative insights}
Realistic studies benefit from qualitative evaluations that can be triangulated with quantitative metrics.

The metrics discussed above are well-suited for statistical evaluation—either in a correlational study, an intervention study, or a controlled experiment. When studies are sufficiently large, statistical significance may not be a suitable guideline to decide on the relevance of a finding, as even very small effects become significant when the sample size is large. In this case, researchers should focus on whether the size of the effect constitutes a meaningful contribution. Conversely, some real world studies may not attain the precision or sample size needed for statistical significance. Such studies may still provide valuable insights by treating them as pilot studies for more concerted (but perhaps less realistic) evaluation efforts.

If large sample sizes cannot be attained, a better approach may be to conduct a qualitative study. Regardless, there is immense value in deep, qualitative insights that such studies can provide. For example, one can conduct Grounded Theory studies to establish theories of users’ psychology~\cite{charmaz2006constructing}, or Contextual Design studies to gain a thorough understanding of users’ experiences and their system needs~\cite{holtzblatt2011contextual}. Such studies are particularly useful when investigating evaluation targets that are highly context-dependent and/or not yet very well understood, such as fairness~\cite{JohnMathews2022}, serendipity~\cite{Bjrneborn2017} or surprise~\cite{Kaminskas2014MeasuringSI}. And while statistical methods are often not suitable for qualitative data, established methods exist that allow for systematic comparisons between users and/or systems (cf. ``constant comparison''~\cite{charmaz2006constructing}).

Qualitative studies vary from purely observational studies to in-depth user interviews, and from single sessions to long-running studies where the researcher is ``embedded'' in a team or organization. As realism is often a prime consideration in such studies, other scholars have covered this aspect in much detail~\cite{holtzblatt2011contextual}. Note, though, that the collection and analysis of qualitative data are particularly labour-intensive, especially when they must integrate into a larger real world research infrastructure. It is also important to carefully report on qualitative procedures (e.g., procedures for ``coding'' qualitative data) and findings (e.g., by considering the researchers’ positionality in conducting the study~\cite{charmaz2006constructing} and by providing ample evidence in the form of user quotes).

\paragraph{Towards best practices in measurement}
\label{sec:realism-best-practice}

\textbf{Standardize measurement practices}
To expedite generalizable research with real world systems, the field must adopt a set of theoretically-grounded measurement principles. 

While most system-centric evaluation metrics in \ac{RS} and \ac{IR} have relatively standardized definitions that enjoy mostly universal adoption, this is not true for user behavior and experience metrics. While this is partially due to the highly contextual nature of relevant metrics in such studies, it may still be beneficial to identify a set of standardized metrics—or, at the very least, measurement principles that can improve the robustness of our evaluations and expedite comparisons between studies.

On the subjective side, the field could create a repository of validated measurement scales that have been proven useful in past studies. Care must be taken, though, that such a repository does not become an exclusive source of measurement instruments—there are usually limits to the applicability of existing scales. Researchers could be encouraged to particularly study the measurement principles of existing scales, such as how well they generalize to new tasks, contexts, and user groups (this can be done through the statistical process of ``measurement invariance testing'' \cite{Schoot2012}). Another way to address the context-specificity of measurement is to provide guidelines for researchers to adapt existing scales to their particular context, as well as guidelines for the development of completely new scales~\cite{DeVellisRobertF1991Sd}.

Finally, it is best if the selection, adaptation and development of scales are rooted in a theoretical framework, such as the Knijnenburg et al.~\cite{10.1007/978-3-642-12630-7_47} framework for the user-centric evaluation of recommender systems. This framework should be extended beyond recommender systems and augmented with theoretical considerations regarding users’ long-term behaviors and goals. 

\textbf{Triangulate measures across multiple studies}
To develop a set of robust and relevant metrics, \ac{IR} and \ac{RS} researchers should conduct a variety of studies—offline evaluations, controlled experiments, and A/B tests and observational studies with real world systems—and triangulate the data collected across these evaluation efforts.

Replication is a fundamental principle of robust scientific progress. Researchers who conduct realistic studies have an opportunity to conduct ``conceptual replications''~\cite{Derksen2022}, where they try to replicate the findings from one domain (or one type of study) in their specific real world context. Such conceptual replications can particularly benefit from a theoretical framework like the Knijnenburg et al. framework \cite{10.1007/978-3-642-12630-7_47}, which can provide a high-level understanding of how the user experience of systems comes about (supporting the goal of explanation), provide guidance for the generation of measurement instruments and hypotheses for in-depth empirical research, and serve as a common frame of reference to compare and integrate findings across studies in different real world contexts. Furthermore, the Knijnenburg et al. framework specifically encourages the triangulation of user behaviors with their subjective evaluations—this grounds the subjective evaluations in observable actions, and in turn, explains the observable actions with subjective evaluations.

Relatedly, an important goal of conducting multi-faceted measurements in realistic studies is to test the validity and universality of the system-centric metrics that are commonly used in \ac{IR} and \ac{RS} research. Do these metrics correlate with positive, long-term, real world outcomes? In what contexts do they fail, and are there better system-centric metrics to optimize in these settings? As offline studies are likely not going away anytime soon, realistic studies can provide the all-important ``reality check'' that such studies need to validate their approach. Conversely, real world studies could provide a platform for researchers to test whether the offline performance of their solutions generalizes to a real world context. One could even create leaderboard-style challenges for each real world system to standardize this approach.

\textbf{Measure unobtrusively, where possible}
To maintain realism, researchers should aim to measure things unobtrusively wherever possible.

As mentioned in our introductory subsection (Section~\ref{sec:reality-motivation}), it is impossible to measure users without influencing them. So while subjective evaluations are invaluable to better understand users’ experiences, it would be better for the realism of our studies if such obtrusive measures could eventually be avoided. This could be supported by a concerted effort to establish behavioral proxies for subjective measures: which user behaviors best correlate with, e.g., user satisfaction? For example, Ekstrand et al. \cite{ekstrand2014} showed that objective measures of diversity, novelty and accuracy correlated strongly with subjective measures based on items from a survey. In commercial systems, item ratings may—or may not—be a good proxy for user interests \cite{mcnee02}. In dialogue-based systems, users’ phrasing or tone of voice may be an indicator of their satisfaction or frustration. The answer to this question is likely highly context-dependent, so each real world study should identify its own best behavioral proxy metrics.

Similarly, researchers could benefit from easily measurable proxy metrics for longer-term (behavioral) outcomes. As outlined in Section \ref{sec:longitudinal-studies}, conducting longitudinal studies is a complicated affair, so the establishment of good proxy metrics could help set realistic long-term evaluation goals in studies that run over a shorter time span. Again, the best proxies for longer-term outcomes are likely context-dependent, so each real world study should aim to identify its own best proxies before reverting to shorter studies.

\textbf{Conduct appropriate statistical evaluations}
As real world data is messy and complex, researchers must take care to conduct the appropriate statistical evaluations of their study data.

Using the guidelines for measurement outlined above, researchers conducting realistic studies will likely collect datasets that are complex (i.e., users may have multiple sessions, or may interact in groups) and longitudinal: users are tracked over time, may interact in groups, and can drop out of and into studies at any given moment. Conducting statistical evaluations on such data is not straightforward—aggregating data to a point where simple statistics apply likely wastes much of the benefit of conducting realistic studies, so complex statistical methods are likely required to carefully analyze the data. Calculating the required sample size (both in terms of the number of users and the number of measures per user) is also not straightforward \cite{Bolger12}.

A potential benefit of longitudinal data is that such data can be used to analyze ``cross-lagged panel models'' \cite{Selig12}, where metric A at timestep n is regressed on metric B at timestep n-1 and vice versa. This allows researchers to establish the causal order between metrics.

If studies are conducted on a real world system, then it is important to establish a baseline measurement of user behavior and subjective evaluation. Moreover, if this system is continuously updated, this baseline metric must be continuously updated as well. 

Subsequently, researchers must aim to detect trends in the data that are caused by their interventions. Such trends may be difficult to detect, as external factors (e.g., seasonal patterns) and the effects of multiple overlapping studies influence the study data simultaneously. This means that the data must be ``de-biased'' to isolate the effect of the intended study. Another consideration is that study samples may not be representative (see Section \ref{sec:recruiting-participants}), which may introduce bias in the statistical results. Stratified sampling and weighting may be used to avoid such biases.

A final statistical consideration in real world studies is that most study participants will have an established interaction history with the system before the study starts. Their past experiences may ``spill over'' into subsequent evaluations. It is thus possible that they may be biased against (or in favour of) changes made to the system as part of the experimental study. Ideally, such systems would have a steady stream of new users that can be used to avoid such effects.

\subsubsection{Longitudinal Studies}
\label{sec:longitudinal-studies}
Longitudinal studies conduct continuous measurements on their test subjects over a prolonged period of time. This temporal aspect provides opportunities to increase our understanding of the evolution of user experiences and behaviors over time in a way that does not only capture factors related to users' initial acceptance of a system or technology but also what influences their prolonged usage. Although longitudinal studies provide extended insights on experiences and behaviors and therefore contribute to a more realistic understanding of users, they are often considered too time-consuming and cumbersome to conduct~\cite{kujala2019cross}. We have defined several challenges and opportunities for longitudinal studies.

\textbf{Types of longitudinal studies}~~The strength of longitudinal aspects lies within revealing behavioral and attitudinal changes of users over time. In the most traditional way, longitudinal studies use the same participants over the course of the study (so-called, panel studies). However, the measurement of temporal changes within panel studies comes with its own challenges. For example, researchers must keep participants motivated to continue their participation in the study. These types of longitudinal studies are particularly susceptible to attrition (e.g., missing data due to non-returning dropouts)~\cite{pan2020impact}. Attrition becomes a problem when complete data is systematically different from missing data, as the impact of missing data can accumulate over time.

Time is an important factor when addressing attrition. Dropouts during a longitudinal study typically occur when the study is too long, or the sampling rate is too high (in particular for non-behavioral studies). Hence, careful consideration of temporal aspects within longitudinal studies is crucial to keep participants motivated.
Besides time aspects, there are several alternative types of longitudinal studies~\cite{melo2022longitudinal} that can help to circumvent the negative effects of panel studies:

\begin{enumerate}
\item A cohort study: participants are drawn from a sample consisting of people sharing the same characteristics and events of interest
\item A retrospective study: analyzing historical data (e.g., offline data)
\end{enumerate}

A cohort study allows for flexibility in the participants that one wants to use at a certain point in time as long as the participants show overlap in the characteristics of interest. This would allow for a lightened load on participants that would otherwise continuously be participating in the study. Alternatively, a retrospective study would make inferences based on historical data instead of collecting new data. Existing datasets such as datasets of LastFM\footnote{E.g.,~\url{http://www.cp.jku.at/datasets/LFM-2b/}} and MovieLens\footnote{\url{https://grouplens.org/datasets/movielens/}} could be used to analyze longitudinal behaviors in retrospect.

\textbf{Confounding factors}~~Considering the reliability and the robustness of the collected data, not only the study design but also user and platform aspects play a role. Particularly in paid studies, participants could start multiple sessions to participate by creating multiple accounts or could influence one another when they are acquainted with each other and discuss the study. These activities by participants are difficult to detect and create potential confounds in the collected data. There are also several challenges with platform aspects. For example, adapting and changing the experimental platform based on interactions that were done during the longitudinal study. Adaptation of platform aspects based on participant interactions may contribute to the realism of the study (compared to a static platform) but can also collude how the data should be interpreted.

\textbf{Analysis}~~A challenge with longitudinal studies is how to analyze the data meaningfully. Although behavioral data collection might be continuous (unobtrusive), attitudinal data is collected less frequently as this often involves questionnaires (obtrusive). Hence, the challenge in the analysis is how to distinguish correlation from causation within the collected data. A potential way to address the aforementioned issue is to triangulate the analysis between unobtrusively collected data and obtrusively collected data.

\subsubsection{Domain-specific vs. General}
\label{sec:realism:domain-specific}
In both \ac{RS} and IR, real world experiments are often done in specific domains, for example, \ac{IR} in the patent \cite{Piroi2019} and medical \cite{Mueller2019,SGKA2021} domains and recommender systems in the fashion \cite{recsyschallenge2022} and travel \cite{recsyschallenge2019} domains. The domains are specified by the data used, users, tasks, etc. These domains can be defined at varying levels of granularity, e.g., scientific paper search or recommendation as a domain, vs. a more specific sub-domain such as physics paper search or recommendation. Another example would be medical search as a domain, with medical search for dentists and for radiologists as sub-domains. While classification systems for research areas like DFG Subject Areas\footnote{\url{https://www.dfg.de/en/dfg\_profile/statutory_bodies/review\_boards/subject\_areas/index.jsp}} or the Common European Research Classification Scheme (CERIF)\footnote{\url{https://www.arrs.si/en/gradivo/sifranti/sif-cerif-cercs.asp}} exist and might be a starting point, they do not catch all definitions of domains. 

\begin{figure}[t]
    \centering
    \includegraphics[width=\linewidth]{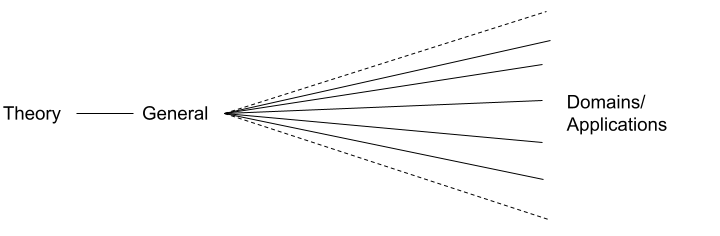}
    \caption{Theory development on a continuum from domain-specific to more general knowledge.}
    \label{fig:theory-domain-specific}
\end{figure}

There is much value in small, in-depth studies, but the results from such studies are hard to generalise. With respect to research infrastructures (see Section \ref{sec:research-infrastructure}) evaluation platforms should be customizable for different applications and domains but are most likely only one-shot implementations that cannot be used in different contexts. The challenge is therefore that domains tend to be treated as silos and there are few attempts to learn general principles that apply across multiple domains. Since the results of domain-specific studies cannot be compared at a numerical level, they must be compared at a conceptual level to allow for generalization. This can be seen as a continuum from general widely-applicable knowledge at one end to domain-specific knowledge at the other end, and the aim would be to shift knowledge from domain-specific to general. The widely applicable knowledge should then also allow theory to be developed---this theory would then allow researchers to make predictions about new domains, which aids the process of building tailored solutions and platforms for specific needs. This is illustrated in Figure~\ref{fig:theory-domain-specific}.

An approach adopted in the DoSSIER project in the area of Professional Search\footnote{\url{https://dossier-project.eu/}} is to classify domains by knowledge task types \cite{DoSSIERKnowledgeTask}, as shown in Figure~\ref{fig_task_template}. This would allow similarities between different domains to be more easily identified, which would assist in the generalization of results. Evaluations of approaches could then be done over similar tasks in different domains, rather than within specific domains, referred to as (semi-)replication\footnote{In the sense of the ACM's definition on reproducibility: ``Different team, different experimental setup'', see \url{https://www.acm.org/publications/policies/artifact-review-and-badging-current}}, conceptual replication, or transitivity. Given the specifications of a new domain, the generalized knowledge and theory could be used to make predictions about how various approaches would work in the domains before any implementation or experiments are done. The ability to make predictions is also important for domains and tasks for which ethics and privacy concerns prevent large-scale experiments from being carried out. 

\begin{figure}
\noindent\fbox{%
	\parbox{\textwidth}{%
\begin{description}
\item[task name:]{\em the unique name assigned to the task, e.g., Prefiling Patentability Search}

\item[definition:]{\em a brief definition of the task}
 
\item[rationale:]{\em why is the task carried out? what should carrying out the task achieve? e.g., the task should lead to the identification of one or more patents that invalidate the query patent.}

\item[initial information available:]{\em what information is available at the beginning to start the search? e.g., a patent application document}

\item[information source:]{\em what information must be searched? e.g., all patent and non-patent information published prior to today.}

\item[searcher:]{\em who usually performs the search? e.g., subject expert or librarian}

\item[query formulation methodology:]{\em how are the queries formulated? e.g., extraction of keywords from the query document and formulation of a Boolean query using synonym expansion lists}

\item[types of tools used:]{\em what tools are commonly used in this type of search? e.g,. clustering results, merging results, Boolean search, ...}

\item[search stopping criteria:]{\em what criteria are used to decide when the search process must be stopped? e.g., a reasonable number of documents returned by a Boolean query}

\item[output of the search:]{\em what does the result list look like? e.g., a list of patents matching the Boolean query in reverse chronological order.}

\item[how/if the search is documented:]{\em is the search documented in some standard way? e.g., queries are placed into a search report along with the number of documents retrieved per query.}

\item[post-processing, interpretation, and analysis of search results:]{\em what is done with the result list once it is obtained? e.g., every patent is checked for relevance by an expert, if relevant it is marked as X or Y...}

\item[any caveats to consider in the analysis or its interpretation:]{\em e.g., the searcher needs to have a good understanding of what the requester is looking for to enable a quick review of the answers for relevance.}

\end{description}
	}
}
\caption{\label{fig_task_template}Task definition template for professional search developed in the DoSSIER Project \cite{DoSSIERKnowledgeTask}.}
\end{figure}

Such a classification would also assist in systematic reviews and meta-analyses across domains. Meta-analysis is a powerful tool to accumulate and summarize the knowledge in a research field~\cite{GZB13}. While meta-analyses are very common in the medical area, they are more challenging in \ac{IR} and \ac{RS} as experiments tend to be less comparable and hence amenable to a statistical meta-analysis. A challenge here would be the different types of studies done, e.g., a controlled randomized trial is likely more easily generalizable than a large search log study. The classification should also facilitate a move toward more task-specific workshops (e.g., ALTARS 2022\footnote{\url{https://altars2022.dei.unipd.it/}}) as a complement to domain-specific workshops (e.g., academic search in medicine or the social sciences~\cite{schaer2021lilas} and legal \ac{IR} workshops). The classification could also assist in identifying domains or task types for which too little experimental work has been done, especially to include domains that are most relevant for communities that are outside the commonly considered WEIRD (white, educated, industrialized, rich, democratic) communities \cite{HHN10}. It could also assist in identifying important theoretical questions and planning experiments that should be conducted to answer them (divide and conquer). 

Challenges foreseen for this approach are:
\begin{itemize}
\item How should domains be differentiated? Medical search for dentists might be different from medical search for radiologists, or they may be considered as part of the broader domain of medical search. Where are the lines between different domains?
\item What are the incentives for researchers to work on generalized insights? Solutions to domain-specific problems are likely more publishable.
\item It is unlikely that we can find generalizable knowledge or theory for every aspect under evaluation. How can such limits be recognized?
\item It makes sense to start this approach at a smaller scale as a proof-of-concept. How do we identify which domains and tasks to start from?
\item Generalizable theory is also about people/users, not only about the systems. What does it mean for users to behave differently in some domains, and how can we generalize knowledge about user behaviors across domains?
\end{itemize}

\subsubsection{Research Infrastructure}
\label{sec:research-infrastructure}

A well-functioning research infrastructure can significantly speed up and improve research in several ways, e.g., by lowering entry requirements, reducing the cost of conducting research, and making it possible to work on common goals from common standards while also increasing comparability between results~\cite{voorhees2005experiment}. Here, we consider challenges when using existing infrastructures and give overall recommendations for creating new research infrastructures that can facilitate real world studies. 

\paragraph{Challenges of using existing infrastructure}

We distinguish three types of research infrastructure used for real world studies. 
First, we have frameworks that can be (re)used to conduct small-scale user studies. Examples are the 3bij3 framework by~\cite{loecherbach20203bij3}, the \ac{ESS} and the \ac{PyIRE}~\cite{oro68811}. We will refer to these as ``frameworks''. Secondly, there is a research infrastructure that is kept continuously running for longer periods of time. Examples are the MovieLens movie recommendation platform~\cite{movielenshistory2015} and the Plista Open Recommendations Platform~\cite{recsyschallenge2013,Kille:ThePlistaDataset:2013}, which has since been discontinued. We will refer to these as “live platforms”. Finally, the \ac{CLEF} includes several labs that address challenges in both the \ac{IR} and \ac{RS} fields with offline datasets collected from real world systems for a specific task~\cite{Barroncedeno2022}, or the \ac{RecSys} challenges, which have run since 2010~\cite{recsyschallenge2022,recsyschallenge2021,said_short_2017}. We will refer to these as ``real-world task datasets''. Below we discuss the key aspects to consider when deciding to reuse existing research infrastructure.

\paragraph{Recruiting participants}
A clear advantage to reusing existing live platforms is that there is often no need to recruit new participants, which comes with its own set of challenges, as discussed in Section~\ref{sec:recruiting-participants}. The platform provides either access to real users on a real product, e.g., Plista, or may have obtained sufficient traction because of its value to the community, e.g., MovieLens. 
Similarly, real world task datasets are usually collected from live platforms, and therefore do not require the recruitment of participants. Frameworks, then, do not share this advantage. 

\paragraph{Customizability/Flexibility}
Frameworks allow for the most flexibility out of all the available options. Provided sufficient knowledge of the tool or some programming experience, frameworks can be customized such that a task of choice can be evaluated, as well as different experimental conditions created at will. At the other end of the spectrum, we find real world task datasets, where the task is set up front and there is no flexibility to change the data collection protocol or decide experimental conditions. In between, we find the live platforms that may have different degrees of flexibility. Flexibility is often at tension with the openness of the platform to the broader research community. On live platforms, users may have some expectations of the system. Therefore, they may be somewhat resistant to change, and therefore offer a limited degree of flexibility. This could be overcome provided a steady stream of new users who do not yet have these expectations of the system, however, on all platforms, only a few users will be converted to loyal users who will use the platform over longer periods of time. 

Examples of this tension between flexibility and openness can be found in the \ac{RS} community. While the NewsReel challenge allowed researchers to directly test algorithms with real users on their platforms, the task was set up front, i.e., obtain the best possible clickthrough rate, and the data collection protocol was fixed. Here, flexibility was limited in favor of broad community access. On the other hand, the MovieLens movie recommendation platform regularly releases new offline datasets but has thus far restricted access to its live platform to researchers within the GroupLens organization. However, research coming out of GroupLens is much more varied: it includes a larger variety of tasks, changes experimental conditions and uses a variety of data collection protocols. Here, flexibility is preferred over broad access. 

\paragraph{Rich data}
When an infrastructure draws on data from running systems with many active users realistic behavioral data can be collected. Collecting additional rich data, which can be of pivotal importance for research, can be a challenge though as system owners may be reluctant to, e.g., allow pop-up questionnaires that might annoy or drive users away. Even when these are allowed, the risk of self-selection bias is high. User behavior in a running system can also appear messy, non-targeted and display many confounding properties not related to the overall research goals. System updates can change the system properties and affect user behavior---especially in longitudinal studies~\cite{schaer2021lilas}.

\paragraph{Recommendations for creating new infrastructure}
When existing research infrastructure is unable to support the researcher’s needs, new research infrastructure has to be built.

Here we put forward some recommendations for building new research infrastructure so that it can benefit the entire research community, as building new infrastructure can be a lengthy and costly process. 

The first challenge lies in obtaining sufficiently large content corpora, e.g., movies, articles or texts. An important consideration here is that after some amount of time, data will inevitably become stale. Therefore, whenever possible, we propose to integrate with APIs that give access to live content corpora that can be kept up-to-date over longer periods of time. The MovieLens platform, for example, integrates with TMDb, and as such has remained relevant for over a decade~\cite{movielenshistory2015}.

Another challenge lies in developing the system, getting the infrastructure up and running, maintaining it and providing support for both users of the system and researchers who wish to use it. Here, we recommend sufficient `realism’: Funding applications should allocate sufficient funds towards software and infrastructure development, as well as the costs of running and supporting research infrastructure over prolonged periods of time. Conversely, funding institutions that wish to support reusable research infrastructure should allow for larger budget applications for the cost of development and running of research infrastructure. An interesting paradox is revealed here: The more successful the platform is with users, the more interesting it becomes for researchers, but also the higher the costs to keep it up and running.

Finally, researchers who wish to create reusable research infrastructure should dedicate significant time and effort towards documenting the system.

\subsubsection{Data Representation}
\label{sec:data-representation}
Information retrieval and recommender systems are critical components of modern information technology, as they allow for the efficient retrieval and recommendation of relevant information. However, for these systems to function effectively, they require underlying data to be present. This is true both in the real world, where these systems are used to process vast amounts of information, as well as in research, where the systems are being developed and tested. Without access to data sets, the research communities would not be able to perform the necessary studies and experiments to further our understanding of these systems.

Given the importance of data in information retrieval and recommender systems research, data representation is one of the cornerstones of this field. In order for datasets to be usable by the research communities, we should strive for a common understanding of what we mean by data, how we represent data, and what we communicate by (and in) data. This includes not only the format of the data but also the semantics and meaning behind the data, as well as the methods used to collect and pre-process the data~\cite{said2014extended}.

Furthermore, data representation also includes the way data is organized, indexed, and stored, as well as how it can be queried and analyzed. By focusing on data representation, we can ensure that the datasets used in information retrieval and recommender systems research are of high quality and that they are accessible and usable by the entire research community. This in turn will facilitate the progress of research in our fields, and ultimately lead to the development of more effective information retrieval and recommender systems.

When sharing data, it is important to communicate the necessary details for understanding the context, use cases, and utility of the data. This includes providing detailed data descriptions, as well as data insights, which can be used by potential data users to understand the utility of the data for the intended research purposes. This information can help users to determine whether the data is appropriate for their research needs, and can also help to facilitate collaboration and sharing of data within the research community. 

To ensure the reproducibility of research and to promote a deeper understanding of the data used, it is essential that researchers provide detailed information about the origin, version, and processing of the data. This includes information about the source of the data, any pre-processing or cleaning that was done, and any specific versions or updates of the data that were used in the research \cite{Bellogin2021}.

One way to achieve this is by adopting the practice of versioning data sets, similar to how software is versioned. This would facilitate easy identification of the specifics of the data set used in a particular study, making it simpler for others to replicate or build upon previous work. Furthermore, it would also allow researchers to clearly communicate which version of the data was used, in turn making it easier for others to access the same data set.

It is also important to remember that data processing is a crucial step in adapting certain datasets to specific use cases. Therefore, introducing the possibility of easily creating and keeping track of unique identifiers for the specific processed data sets used in research studies would facilitate reproducibility of studies. By doing so, researchers can clearly identify the specific processed data set that was used in a particular study, allowing others to easily access and use the same data set for replication or follow-up studies \cite{10.1145/2792838.2792841}.

While keeping track of specific data versions we also need to adopt practices compatible with regulations such as \ac{GDPR}, making sure that users represented in data sets are sufficiently anonymized, and given the opportunity to retrospectively have their data deleted. This may create problematic scenarios if the original data is not sufficiently anonymized. However, this can in turn be used as a motivation for clear and concise privacy policies as to how to generalize, perturb, or as a last resort, censor data in order for it to be released to a wider community~\cite{10.1145/3287168}.

We should remember that data representation within systems may differ immensely between systems. However, when sharing data externally, it is important to ensure that the data representation is realistic in terms of what the data actually express and how. This includes aligning the data types used with the reality, for example, using integers for positive whole numbers and float for non-integer decimal numbers. Additionally, it is important to convey the quality of data realistically and to clearly communicate the purposes for which the shared data is created. This can help potential users to understand the limitations and potential biases of the data and can help to ensure that the data is used appropriately.

We generalize data into two specific data types commonly used in information retrieval and recommender systems, namely, \textbf{living} data, and \textbf{archival} data.

Living data refers to continuously updated data. Living data can be made available in various different formats, including continuous and uniquely identifiable downloadable snapshots, or through a so-called firehose where data is continuously delivered through an API endpoint or similar. While snapshots can provide a unique identifier making it easy to trace back to the exact version of the data, a firehose instead provides an easier way to maintain local data repositories containing up-to-date versions of the source data.

Furthermore, keeping in mind the data representation, it is important to keep the data in a format which is easily understandable, processable and accessible. This includes but is not limited to the type of format (text, image, audio, video etc.), the language of the data, the structure of the data, the size of the data, etc.

Overall, paying attention to data representation and sharing it in a clear and informative manner is crucial for the advancement of research in information retrieval and recommender systems. It can help to ensure that data is used appropriately, and can help to facilitate collaboration and sharing of data among members of the research community.

\subsubsection{Next Steps}
\label{sec:reality-next}

The following steps should be taken to carefully determine the \textbf{goals} of conducting real world studies:

\begin{itemize}
\item Classify domains by knowledge task types
\item Establish context-specific evaluation targets
\item Carefully consider users' information needs when conducting studies
\item Develop a checklist of sample characteristics and user task details that should be collected and reported for each study
\end{itemize}

The following \textbf{resources} would expedite the design, execution and evaluation of real world studies:

\begin{itemize}
\item Provide researchers with access to flexible real world research infrastructure
\item Obtain sufficiently large and rich content corpora that can be used in real world studies
\item Create a repository of validated measurement scales
\item Standardize practices for scale development
\item Establish effective recruitment methods to find the ``right'' participants for a study
\item Develop metrics that are as unobtrusive as possible to measure
\item Design standardized but flexible ways to represent the data and meta-data collected in real world studies
\item Study effective ways to limit attrition in longitudinal studies
\item Produce best-practices guidelines for developing real world systems, getting infrastructures up and running, maintaining them and providing support for both users and researchers
\item Establish guidelines to protect the privacy of research participants
\end{itemize}

The following steps must be taken to allow researchers to \textbf{integrate the findings of real world studies into generalizable knowledge}:

\begin{itemize}
\item Use theory to integrate domain-specific knowledge into a generalized knowledge
\item Define a theoretical framework for measurement
\item Develop an infrastructure for researchers to contribute analyses of and insights about real world datasets in a centralized manner
\item Integrate research within specific domains as well as at the generalized knowledge level using systematic reviews, meta-analyses, task-specific workshops and domain-specific workshops
\item Conduct studies to triangulate qualitative and quantitative insights, behavioral and subjective metrics, and short-term and long-term metrics
\end{itemize}

\newpage

\abstracttitle{HMC: A Spectrum of Human--Machine-Collaborative Relevance Judgment Frameworks}
\label{subsec:evaluation}
\abstractauthor[Charles L. A. Clarke, Gianluca Demartini, Laura Dietz, Guglielmo Faggioli, Matthias Hagen, Claudia Hauff, Noriko Kando, Evangelos Kanoulas, Martin Potthast, Ian Soboroff, Benno Stein, Henning Wachsmuth]{%
Charles L. A. Clarke (University of Waterloo, CA, claclark@plg.uwaterloo.ca)\\
Gianluca Demartini (University of Queensland, AU, demartini@acm.org)\\
Laura Dietz (University of New Hampshire, US, dietz@cs.unh.edu)\\
Guglielmo Faggioli (University of Padua, IT, guglielmo.faggioli@unipd.it)\\
Matthias Hagen (Friedrich-Schiller-Universit\"{a}t Jena, DE, matthias.hagen@uni-jena.de)\\
Claudia Hauff (Spotify, NL, claudia.hauff@gmail.com)\\
Noriko Kando (National Institute of Informatics (NII), JP, Noriko.Kando@nii.ac.jp)\\
Evangelos Kanoulas (University of Amsterdam, NL, e.kanoulas@uva.nl)\\
Martin Potthast (Leipzig University and ScaDS.AI, DE, martin.potthast@uni-leipzig.de)\\
Ian Soboroff (National Institute of Standards and Technology (NIST), US, ian.soboroff@nist.gov)\\
Benno Stein (Bauhaus-Universit\"{a}t Weimar, DE, benno.stein@uni-weimar.de)\\ 
Henning Wachsmuth (Leibniz Universit\"{a}t Hannover, DE, h.wachsmuth@ai.uni-hannover.de)
}
\license

\subsubsection{Motivation}
\label{subsubsec:evaluation-motivation}

\ac{IR} evaluation traditionally needs human assessors to generate relevance judgements. Traditionally, human assessors are asked to judge the relevance of a document with respect to a topic~\cite{harman2011information}. Recently, work looking at preference judgements~\cite{clarke2021assessing,potthast:2019k} has looked at research questions related to how to best evaluate IR systems by asking human assessors which of two results is the better given an information need. 
The recent availability of \acp{LLM} has opened the possibility to use them to automatically generate relevance assessments in the form of preference judgements. While the idea of automatically generated judgements has been looked at before \cite{buttcher2007reliable}, new-generation \acp{LLM} drive us to re-ask the question of whether human assessors are still necessary.

New models tend to fail in a different and more diverse way compared to traditional approaches. Failure points for old models were more uniform and clear, with new systems it is harder to predict in which ways the model will fail.
In most cases, \acp{LLM} (especially for what concerns generative aspects) focus on entertainment tasks. Models tend to report false facts in such a convincing way that they need to be carefully read by some expert to identify lacking factuality (e.g., Michel Foucault simulation\footnote{https://www.youtube.com/watch?v=L6c0xeAqEz4E}).

\begin{figure}[tb]
    \centering
    \includegraphics[width=1\textwidth]{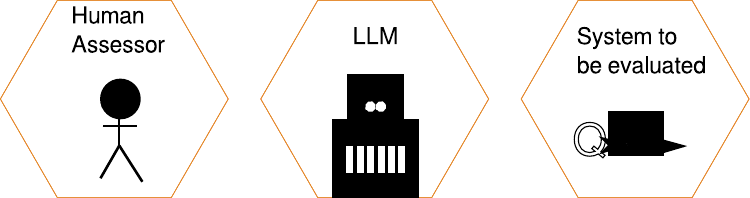}
    \caption{The three most relevant components in our system: the human assessor, the \acf{LLM} that can help humans or replace them in annotating documents for relevance, and the system that we want to evaluate using the newly produced relevance judgements.}
    \label{fig:chm-components}
\end{figure}

Our motivation to investigate the possibility of using \acp{LLM} in order to provide automatic annotations stems from some fundamental research questions that can be summarized as follows. 
\begin{itemize}
    \item \textbf{RQ1}: In which way automatic approaches, and in particular \acp{LLM}, can help assessors with the assessment task to yield the most reliable annotations while improving the efficiency of the annotation process? 
    This question raises other interesting related inquiries. For example, if we were to build such a mixed human-machine annotation paradigm, which held out (not provided to the IR system) supporting information about the topic would yield the best and fastest annotations? What weighting between human and \acp{LLM} and AI-assisted annotations is ideal?
    \item \textbf{RQ2}: Can machines (either in the form of \acp{LLM} or in general as \ac{AI} models) replace humans in assessing and annotating?
    This question raises also concerns about what annotation target (e.g., relevance labeling, summarization, paragraph highlighting, exam questions \cite{Sander2021}) would yield the best and fastest annotations.
    \item \textbf{RQ3}: What are the conditions under which human assessors cannot be replaced by machines? Alternatively, in which role can the Human assessor most productively provide relevance assessments?
\end{itemize}

Answering the questions mentioned would also require finding viable solutions for a set of additional questions and open issues that touch a number of IR evaluation process steps.
\begin{itemize}
    \item Assessors And Collections:
    \begin{itemize}
        \item How to use LLM to help assessors: some examples of possible usages include, summarising text, associating keywords and identifying the content of long podcasts to help assessors annotate the documents, for example by highlighting relevant fragments of text/podcast or segments with correct answers.
        \item What is the effective role of the human assessor in annotating material for generative models? Should the annotator provide input at the beginning of the pipeline, by annotating the original documents, or are they more useful downstream, after the task has been carried out? 
        \item Generative models can be used to create new collections: corpora, conversations, queries, abstracts and so on.
    \end{itemize}
    \item LLM and generative models to retrieve information in a broader sense:
    \begin{itemize}
        \item \ac{IR} tasks that employ \acp{LLM} have the means to provide more details: often a single answer is not satisfactory for the user. How to support the user in exploring the results further (for example via links and connected pages). Generative models can help, but is this helpful when the model simply generates the response without knowing where it comes from?
        In many cases, the user is not interested in receiving only the direct/short answer, but rather in seeing which documents contain it and related pieces of information to expand their knowledge.
    \end{itemize}
    \item \acp{LLM} as an evaluation tool:
    \begin{itemize}
        \item The model is biased: how can we use it to evaluate itself? If a model has been trained on biased data, then also the evaluation is prone to the same biases. How to detect and account for such biases? 
    \end{itemize}
    \item Evaluating \acp{LLM} and their trustworthiness:
    \begin{itemize}
        \item Can we find a way to understand and measure to what level we can trust the results of a generative model?
        \item How to carry out fact-checking, for example by identifying the source of information of a generative model and verifying that it is presented accurately.
        \item Distinguish between human and machine-generated data: Important for many tasks, such as journalism, where it is of uttermost importance to verify the information. Human-generated data is more trusted.
    \end{itemize}
\end{itemize}

We argue that the collaboration between humans and \ac{ML}, especially under the form of \acp{LLM}, could be abstracted in the form of a spectrum. On the two extremes of this spectrum, we have either the human or the machine entirely tasked to annotate documents for relevance with respect to a query. Within the spectrum, humans and \acp{LLM} interact to a different extent. Theoretically, such a spectrum corresponds also to moving from highly expansive annotations in terms of human effort, cost and time, but with high-quality annotations, to a much less expensive annotation procedure with also a decreased annotation quality.
We also argue that something exists beyond the spectrum; it corresponds to the scenario in which the machine overcomes the human, by producing relevance judgments without any form of bias. We observed this phenomenon happening already in several tasks and scenarios, and therefore we can aspect this to happen also with respect to the construction of the relevance judgments.

The remainder of this chapter is organized as follows: Subsection~\ref{subsubsec:evaluation-gaps} reports details on the current state of the art and limitations associated with the current usage of \acp{LLM} ad \ac{AI} in annotating documents. Subsection~\ref{subsubsec:CHM-Relevance-Judgements} illustrates our proposal of a spectrum of possible interactions between the human and the machine, to provide more efficient and effective annotations and relevance judgments. Subsection~\ref{subsubsec:next-steps} outlines a possible experimental protocol that would allow us to verify at what point modern \acp{LLM} and whether they can be used to produce automatically relevance judgements.

\subsubsection{State of the Art, Idea, and Gaps}
\label{subsubsec:evaluation-gaps}

\paragraph{Using LLMs to Generate Annotations and Label Automatically}

Potential uses of \acp{LLM} to annotate documents, extract snippets, summarize and, in the end, annotate documents for relevance. If this can be made to work reliably, it opens up many opportunities for evaluation. For example, the \ac{LLM} can be used directly to evaluate the output of other large language models (for example in summarization). 

Assessments can arise from different sources, with different levels of quality and collection costs as follows. 

\begin{itemize}
    \item Human assessors or, in the enterprise scenario, final users. This scenario, at the current time, is the most expansive, but also likely to produce high-quality annotations.
    \item Human assessors aided by mild automatic support systems (e.g. remove redundancy, encourage consistency)
    \item Half of the judgments are produced by human assessors and half of the judgments are produced automatically.
    \item Automatic annotation of a collection, which is verified and corrected by human intervention.
    \item At some point even a fully automatic assessment.
\end{itemize}

An additional axis describes the type of annotations. Typically an annotation is a graded relevance judgment, but for example in EXAM \cite{Sander2021}, humans are used for generating questions instead. This can be generalized by asking human assessors for something different than traditional annotation while some Machine Learning (ML) converts the human responses into relevance assessments. This follows the paradigm of Competence Partitioning of Human-Machine-Collaboration where humans and machines are performing tasks they are best at (not vice versa).

One concern is that fully automatic assessment with \acp{LLM} can be very expensive, which is also the reason why we consider the application of \acp{LLM} as part of the retrieval process.
In such a case, we could reduce the cost by considering a teacher-student training paradigm (knowledge distillation) in which a large and expensive \ac{LLM} is used to train a smaller model that is less expensive to run.

Not all \ac{IR} tasks focus on topics. For example, one may want to search for podcasts where two or more people interact or with a particular style.
Another issue is regarding truth. For example, finding a podcast for the query “does lemon cure cancer?” that talks about healing cancer with lemon might be on topic. Nevertheless, it is unlikely to be factually correct, and therefore not relevant to correctly answering the information needs. To overcome this issue, assessors have to access external information to determine the trustworthiness of a source, or the truthfulness of a document.
In a similar way, we can assume our LLM is used as an oracle that accesses external facts, verified by humans. To properly support different tasks, human intervention can be plugged into the collection and annotation of additional facts, to define relevance.

There are open questions for the special case of 100\%-machine/0\%human. How is this ranking evaluation different from being an approach that produces a ranking? (circularity problem).
We can use multiple \acp{LLM}, possibly based on different rationales, such that it is possible to define an inter-annotation systems agreement, in which different systems are used to verify if there is an agreement between each other. An alternative approach is to endow the evaluation with additional information about relevant facts/questions/nuggets that the system (under evaluation) does not have access to. 

It is yet to be understood what the risks associated with such technology are: it is likely that in the next few years, we will assist in a substantial increase in the usage of \acp{LLM} to replace human annotators. Nevertheless, a similar change in terms of data collection paradigm was observed with the increased use of crowd assessor. Up to that moment, annotations were typically made by in-house experts. Then, such annotation tasks were delegated to crowd workers, with a substantial decrease in terms of quality of the annotation, compensated by a huge increase in annotated data. 
It is a concern that machine-annotated assessments might degrade the quality, while dramatically increasing the number of annotations available.

The Cranfield paradigm~\cite{Voorhees2001} is based on simplifying assumptions that make manual evaluation feasible: \textit{1)} independence of queries; \textit{2)} independence of relevance of documents; \textit{3)} Relevance is static (and not changing in time). 
Recently, the field is diverging from this paradigm, for example with \ac{TREC} CAR and \ac{TREC} CAsT/iKAT where the information needs are developing as the user learns more about the domain. The TREC Evaluation of CAST describes a tree of connected information needs, where one conversation takes a path through the tree. The Human-Machine evaluation paradigm might make it feasible to assess more connected (and hence, realistic) definitions of relevance.

\subsubsection{Collaborative Human-Machine Relevance Judgments}
\label{subsubsec:CHM-Relevance-Judgements}
We can describe a spectrum of Collaborative-Human-Machine paradigms to create relevance judgments, where the weighting of human contributions vs machine contributions changes along the spectrum.

\begin{figure}[t]
    \centering
    \includegraphics[width=1\textwidth]{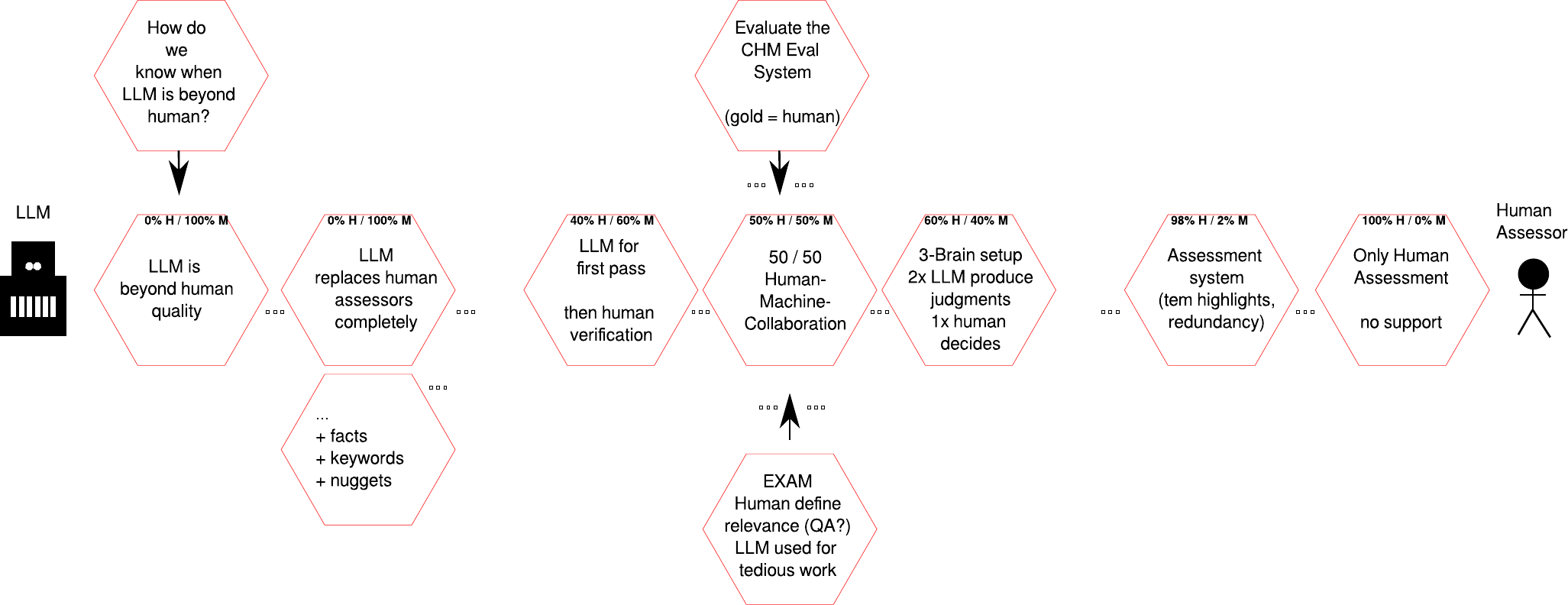}
    \caption{A spectrum of Collaborative-Human-Machine paradigms to create relevance judgments.}
    \label{fig:chm-spectrum}
\end{figure}

\begin{itemize}
    \item \textbf{Only Human (100\%H / 0\%M)}: On one extreme, the human will do all assessments manually without any kind of support. 
    \item \textbf{Human with assessment system (99\%H / 1\%M)}: This is a more realistic case for how TREC assessment is conducted, where humans have full control of what is relevant but are supported in the following ways: Humans can define “scan terms” that will be highlighted in the text, can limit view the pool that is already judged, ordering documents so that similar documents are near one another, produce readable presentations of retrieve content. 
    \item \textbf{Human with document summaries (80\%H/ 20\%M)}: A text summarization model produces a generative summary representation of the document to be judged. The human assessor judges the representation, which is more efficient to do.
    \item \textbf{EXAM (60\%H / 30\% M)}: For each query, the human defines information nuggets that are relevant (e.g. exam questions). The machine is trained to automatically determine how many test nuggets are contained in the retrieved results (e.g. via a Q/A system).
    \item \textbf{Equal contribution (50\%H / 50\%M)}: A theoretic midpoint in the collaborative spectrum. Humans perform tasks that humans are good at. Machines perform the tasks that machines are good at. It is yet to be concretely defined what this might be.
    \item \textbf{3-Brain Setup (32\%H / 58\%M)}: Two machines each generate an assessment, and a human will select the best of the two assessments (+verification). Human decision trumps machines’.
    \item \textbf{LLM for first pass + human verification (30\%H / 60\%M)}: A first-pass assessment of the LLM is automatically produced as a suggestion. This can also be an assessment-supporting surrogate prediction like a rationale. The human assessment is based on this suggestion, but the human will have the final say.
    \item \textbf{LLM replaces humans completely (0\%H / 100\%M)}: We explore the possibility that a fully automatic assessment system might be as good as a human in producing high-quality relevance judgments. 
    \item \textbf{LLM is beyond human (0\%H / 100\%M)}: Given known biases in human assessments, we contemplate the possibility that the automatic assessments might even surpass the human in terms of quality. While not feasible at the current time, this is an important case to consider when we evaluate the HMC evaluation.
\end{itemize}

\paragraph{Use LLMs to Help Humans in Annotating Documents}
LLMs could be successfully applied in helping human assessors with annotating data. For example, \acp{LLM} might be particularly useful in recognizing near duplicates and using them to verify if the two near duplicates share the same relevance annotation -- with the human entering the loop only in those cases where the system has a high degree of uncertainty.

Related to the case of (100\%H / 0\%M), we have the \textit{human-in-the-loop}, helping the system in realizing its annotation goal. Such help might include providing annotated facts or verifying the annotation after a first pass from the system. In the 50\%/50\% case, equal contributions, we have a substantial equilibrium between both the human and the machine. We refer to this scenario as \textit{competence partitioning}: the task is assigned to either the human or the machine, depending on who is currently better at the current moment.
On the other side of the spectrum (\%M > \%H), the scenario is called \textit{model-in-the-loop}: the model offers its contribution in organizing the data, where the human is used as a verification step. The concern is that any bias in the \ac{LLM} might be affecting the relevance assessments, as the human will not be able to correct for information it will not see.

An alternative approach to the collaborative one is a complementary one, where the human and the machine both produce judgments, but different ones. This then becomes a task allocation problem where the aim is to predict who among the human and the machine assessor is best suited for any given judgment.

\paragraph{Beyond Human Performance}
We could expect that, at a certain point in the future, the \acp{LLM} will overcome humans in a number of tasks that can be reconducted to annotate the documents. Humans are likely to make mistakes when annotating documents and are limited in the time dedicated to the annotation. In contrast, \acp{LLM} are likely to be more self-consistent and potentially capable of annotating all the documents perfectly. Machines can also annotate a much larger number of data points.

Furthermore, we have a series of assumptions, such as the fact that relevance does not change through time, that are enforced to make evaluation tractable. These assumptions can be relaxed if the machine annotates automatically.

It is an open issue in recognizing when the human is failing. All the above strategies assume human annotations are the gold standard without errors. This assumption is strong: the \ac{LLM}, having access to more information, might be able to correct human mistakes.

We are likely to reach the limit of measurement: we will not be able to use differences between the current evaluation paradigms to evaluate such models. A problem is that if we surpass the quality of only human-annotated data, we will not be able to detect this if we use only human-annotated data as a gold standard. will not suffice and will fail in providing a gold standard.

Another research question is to identify optimal competence partitioning. One idea is to use the \ac{LLM} to generate rationales for explaining the relevance. While humans are often considered experts for rational generation, recent advancements, including chatGPT, suggest that we are on the verge of a shift of paradigm, with \acp{LLM} constantly improving in identifying why a document is (non)-relevant, either considering information with the document, or other relevant external pieces of information.

\paragraph{Trust, Correctness, and Inter-annotator Agreement}
One important difference between humans and automatic assessors concerns the assessment sample size. While it is possible to hire multiple assessors to annotate the documents and, possibly, resolve disagreements between annotators, this is not that trivial in the automatic assessor case.
We can expect that \acp{LLM} which are trained on similar corpora will likely produce correlated answers — but we don’t know whether these are correct. A possible solution to this would include the usage of different subcorpora based on different sets of documents.
This, in turn, could lead to personalized \acp{LLM}, fine-tuned on data from different types of users, which would allow to auto-annotate documents directly according to the user’s subjective point of view, while also helping with increasing the pool of annotations collected. While this technology is not available yet, mostly due to computational reasons, we expect it to be available in a few years. 

A related idea that can be implemented today is to allow \acp{LLM} to learn by observing human annotators performing the task or following an active learning paradigm. The \ac{LLM} starts with mild suggestions to the user on how to annotate the documents, then it continues to learn by considering actual decisions made by the annotator and finally improving the quality of the suggestions provided.

\subsubsection{Next Steps}
\label{subsubsec:next-steps}
Tables \ref{tbl:evaluation-topic-54_8} and \ref{tbl:evaluation-topic-67_8} report two examples of document annotation done with two well-known LLMs: YouChat\footnote{\url{https://you.com/}} and ChatGPT\footnote{\url{https://chat.openai.com/}}. It is interesting to notice that, in both cases, both models provided the correct answer, correctly identifying the passage which was annotated as more relevant. 
It is possible to observe that, while ChatGPT simply repeats the relevant passage, YouChat is capable of correctly identifying the reason why a passage is more relevant than the other.

\begin{table}[t]
\centering
\caption{Response of two well-known \acp{LLM} in comparing two passages as answer for the query 54\_8 from CAsT '19: ``What are anemia's possible causes?''}
\label{tbl:evaluation-topic-54_8}
\begin{tabular}{|l|p{0.8\textwidth}|}
\hline
Prompt & Select the passage that answers the question better.\newline - If both answers are similar, select the one with the least extraneous information.\newline - If both answers are still similar, select the one with the best formatting.\newline \newline Question:\newline \newline What are anemia's possible causes?\newline \newline Passage 1:\newline \newline If you have anemia, your blood does not carry enough oxygen to the rest of your body. The most common cause of anemia is not having enough iron. Your body needs iron to make hemoglobin. Hemoglobin is an iron-rich protein that gives the red color to blood. It carries oxygen from the lungs to the rest of the body. Anemia has three main causes: blood loss, lack of red blood cell production, and high rates of red blood cell destruction
\newline \newline
passage 2:\newline \newline Broadly, causes of anemia may be classified as impaired red blood cell (RBC) production, increased RBC destruction (hemolytic anemias), blood loss and fluid overload (hypervolemia\\
\hline
ChatGPT & Passage 1:\newline \newline If you have anemia, your blood does not carry enough oxygen to the rest of your body. The most common cause of anemia is not having enough iron. Your body needs iron to make hemoglobin. Hemoglobin is an iron-rich protein that gives the red color to blood. It carries oxygen from the lungs to the rest of the body. Anemia has three main causes: blood loss, lack of red blood cell production, and high rates of red blood cell destruction. \\ 
\hline
YouChat & Passage 1 is the better answer, as it provides an explanation of what anemia is, and then lists the three main causes in detail, which is the most relevant information for the question. Passage 2, while providing some good background information, does not provide the same level of specificity as Passage 1, and includes extraneous information that is not directly relevant to the question.\\
\hline
\end{tabular}
\end{table}

\begin{table}[t]
\centering
\caption{Response of two well-known \acp{LLM} in comparing two passages as answer for the query 67\_8 from CAsT '19: ``What is a DC half smoke?''}
\label{tbl:evaluation-topic-67_8}
\begin{tabular}{|l|p{0.8\textwidth}|}
\hline
Prompt & Select the passage that answers the question better.\newline - If both answers are similar, select the one with the least extraneous information.\newline - If both answers are still similar, select the one with the best formatting.\newline \newline Question:\newline \newline What is a DC half smoke?\newline \newline Passage 1:\newline \newline However, celebrities at Ben's are not limited to American citizens. In 1998, former DC mayor Marion Barry described having traveled to Ghana and meeting the Mayor of Accra, an alumnus of Howard University, whose greeting was, ``Glad to have you in Accra. Is Ben's Chili Bowl still there?'' And when French President Nicolas Sarkozy and his wife Carla Bruni-Sarkozy visited Washington, D.C. in March 2010, they reportedly each had two of Ben's half smokes during their visit to the restaurant.\newline \newline Passage 2:\newline \newline A half-smoke is a ``local sausage delicacy'' found in Washington, D.C. and the surrounding region. Similar to a hot dog, but usually larger, spicier, and with more coarsely-ground meat, the sausage is often half-pork and half-beef, smoked, and served with herbs, onion, and chili sauce.\\
\hline 
ChatGPT & Passage 2:\newline \newline A half-smoke is a local sausage delicacy found in Washington, D.C. and the surrounding region. Similar to a hot dog, but usually larger, spicier, and with more coarsely-ground meat, the sausage is often half-pork and half-beef, smoked, and served with herbs, onion, and chili sauce. \\ 
 \hline
YouChat & Passage 2 is the better answer as it more directly answers the question and does not include any extraneous information.\\
\hline
\end{tabular}
\end{table}

To assess the feasibility of the proposed approaches, next steps would include an experimental comparison of the different Collaborative-Human-Machine paradigms. This should include multiple test collections (e.g., TREC-8 and TREC Deep Learning), multiple types of judgments (e.g., binary, graded, preference), and multiple models (e.g., GPT-2, GPT-3, chatGPT, etc.). Comparison between human-generated judgments and machine-generated judgments may be performed both using inter-assessor agreement metrics as well as IR system ranking correlation methods.

\newpage

\abstracttitle{Overcoming Methodological Challenges in Information Retrieval and Recommender Systems through Awareness and Education}
\label{subsec:education}
\abstractauthor[Christine Bauer, Maik Fr{\"o}be, Dietmar Jannach, Udo Kruschwitz, Paolo Rosso, Damiano Spina, Nava Tintarev]{%
Christine Bauer (Utrecht University, NL, c.bauer@uu.nl)\\
Maik Fr{\"o}be (Friedrich-Schiller-Universit{\"a}t Jena, DE, maik.froebe@uni-jena.de)\\
Dietmar Jannach (University of Klagenfurt, AT, dietmar.jannach@aau.at)\\
Udo Kruschwitz (University of Regensburg, DE, udo.kruschwitz@ur.de)\\
Paolo Rosso (Technical University of Valencia, ES, prosso@dsic.upv.es)\\
Damiano Spina (RMIT University, AU, damiano.spina@rmit.edu.au)\\
Nava Tintarev (Maastricht University, NL, n.tintarev@maastrichtuniversity.nl)
}
\license

\subsubsection{Background \& Motivation}
\label{subsubsec:education-motivation}
In recent years, we have observed a substantial increase in research in \ac{IR} and \ac{RS}. To a large extent, this increase is fueled by progress in \ac{ML} (deep learning) technology. As a result, countless papers are nowadays published each year which report that they improved the state-of-the-art when adopting common experimental procedures to evaluate \ac{ML} based systems. However, a number of issues were identified in the past few years regarding these reported findings and their interpretation. For example, both in \ac{IR} and \ac{RS}, studies point to methodological issues in \emph{offline} experiments, where researchers for example compare their models against weak or non-optimized baselines or where researchers optimize their models on test data rather than on held-out validation data~\cite{armstrong2009improvements,ferraridacremaetal2019,sun2020arewe,wei2019critically}. 

Besides these issues in offline experiments, questions concerning the \emph{ecological validity} of the reported findings are raised increasingly. Ecological validity measures how generalizable experimental findings are to the real world. An example of this problem in information retrieval is 
the known problem of mismatch between offline effectiveness measurement and user satisfaction measured with online experimentation \cite{chen2017meta,hassan2010beyond,mao2016when,sanderson2010do,zhang2020models}
or 
when the definition of relevance does not consider the effect on a searcher and their decision-making. For example, the order of search results, and the viewpoints represented therein, can shift undecided voters toward any particular candidate if high-ranking search results support that candidate \cite{epstein2015SearchEngineManipulation}. This phenomenon---often referred to as the \textit{\ac{SEME}}
---has been demonstrated for both politics~\cite{epstein2015SearchEngineManipulation,epstein2017SuppressingSearchEngine} and health~\cite{allam2014ImpactSearchEngine,pogacarPositiveNegativeInfluence2017}. By being aware of the phenomena, methods have been adapted to measure its presence \cite{TimDraws2023,draws2021assessing}, and studies to evaluate when and how it affects human decision-makers \cite{draws2021not}. 
Similar questions of ecological validity were also raised in the RS field regarding the suitability of commonly used computational accuracy metrics as predictors of the impact and value such systems have on users in the real world. Several studies indeed indicate that the outcomes of offline experiments are often \emph{not} good proxies of real-world performance indicators such as user satisfaction, engagement, or revenue~\cite{DBLP:conf/ercimdl/BeelL15,Gomez-Uribe:2015:NRS:2869770.2843948,jannach2021mcnamara}.

Overall, these observations point to a number of open challenges in how experimentation is predominantly done in the field of information access systems. 
Ultimately, this leads to the questions of \emph{(i)}
how much progress we really make despite the large number of research works that are published every year \cite{armstrong2009improvements,lin2021significant,Zobel22Mislead}
and \emph{(ii)} how effective we are in sharing and translating the knowledge we currently have for doing \ac{IR} and \ac{RS} experimentation \cite{Ferro2022,sakai2018laboratory}. 
One major cause for the mentioned issues, for example, seems to lie in the somewhat narrow way we tend to evaluate information retrieval and recommender systems: primarily based on various computational effectiveness measures. In reality, information access systems are interactive systems used over longer periods of time, i.e., they may only be assessed holistically if the user's perspective (task and context) is taken into account, cf.~\cite{Lykke22Role,white2016interactions,edu-ZangerleBauer2022}. Studies on long-term impact furthermore need to consider the wider scope of stakeholders~\cite{bauer2019_multimethod_multistakeholder,jannach2021mcnamara}. Moreover, for several types of information access systems, the specific and potentially competing interests of multiple stakeholders have to be taken into account~\cite{bauer2019_multimethod_multistakeholder}. Typical stakeholders in a recommendation scenario include not only the consumers who receive recommendations but also recommendation service providers who for example want to maximize their revenue through the recommendations~\cite{JannachAdomaviciusVAMS2017,jannach2021mcnamara}.

Various factors contribute to our somewhat limited view of such systems, e.g., the difficulties of getting access to real systems and real-world data for evaluation purposes. Unfortunately, the IR and RS research communities to a certain extent seem to have accepted to live with the limitations of the predominant evaluation practices of today. Even more worryingly, the described narrow evaluation approach has become more or less a standard in the scientific literature, and there is not much debate and---as we believe---sometimes even limited awareness of the various limitations of our evaluation practices.

There seems to be no easy and quick way out of this situation, even though some of the problems are known for many years now \cite{Ekstrand2011Rethinking,hassan2010beyond,KonstanGedas2013Toward,sanderson2010do}. 
However, we argue that improved \emph{education} of the various actors in the research ecosystem (including students, educators, and scholars) is one key approach to improve our experimentation practices and ensure real-world impact in the future. As will be discussed in the next sections, better training in experimentation practices is not only important for students, but also for academic teachers, research scholars, practitioners and different types of decision-makers in academia, business, and other organizations. This will, in fact, help address the much broader problem of reproducibility\footnote{\url{https://www.wired.com/story/machine-learning-reproducibility-crisis/}} and replicability
\footnote{\url{https://cacm.acm.org/magazines/2020/8/246369-threats-of-a-replication-crisis-in-empirical-computer-science/abstract}} we face in Computer Science \cite{Cockburn20Threats,zz-DagstuhlSeminar16041} in general and in AI in particular~\cite{DBLP:conf/aaai/GundersenK18}.

This chapter is organized as follows: Next, in Section~\ref{subsubsec:education-stakeholders} we briefly review which kinds of actors may benefit from better education in information access system experimentation. Afterwards, in Section~\ref{subsubsec:education-next-steps}, we provide concrete examples of what we can do in terms of concrete resources and initiatives to increase the awareness and knowledge level of the different actors. Finally, in Section~\ref{sec:education-challenges}, we sketch the main challenges that we may need to be aware of when implementing some of the described educational initiatives.

\subsubsection{Actors} 
\label{subsubsec:education-stakeholders}
As in any process related to the advancement, communication, and sharing of knowledge, knowing how to properly design and carry out correct and robust experimentation concerns people with various different roles. 
This covers a broad spectrum including 
academia, industry, and public organizations, e.g., from a lecturer in \ac{IR} and \ac{RS} introducing evaluation paradigms to undergrad students and data scientists---not necessarily experienced in \ac{IR} and \ac{RS}---choosing metrics aligned to business \acp{KPI} by looking at textbooks and Wikipedia pages. We have identified a number of actors that are involved in the education to experimentation in information access, who are listed below. Note that this categorization is not exhaustive nor exclusive, as actors may have multiple roles.

\begin{tcolorbox}[colframe=white]
\paragraph*{Students}
This category embraces the different stages of academic training. Starting from students enrolled in \ac{IR} \& \ac{RS} courses \cite{markov2019what}, including, for instance, undergraduate students in Computer Science degrees and Master's students in Data Science, \ac{AI}, and Human-Computer Interaction. It also includes students enrolled in a doctoral degree, i.e., PhD students, including those jointly co-supervised with industry.

\paragraph*{Educators}
Academic roles related to education, such as course coordinators, lecturers, teaching assistants, as well as research student supervisors.

\paragraph*{Scholars}
Researchers and academics involved in academic services, including reviewers, journal editors, program chairs, grant writers, etc.

\paragraph*{Practitioners}
Data scientists, developers, \ac{UX} designers, and other practitioners outside academia, that may need support in their lifelong learning.

\paragraph*{Decision-makers}
People that make strategic decisions in processes, policies, products \mbox{and}/or human resources (e.g., managers in industry or policy-makers) that may benefit from having a better understanding of \ac{IR} and \ac{RS} core concepts in evaluation and experimentation.
\end{tcolorbox}

\begin{figure}[h!tp]
    \centering
    \includegraphics[width=1\textwidth]{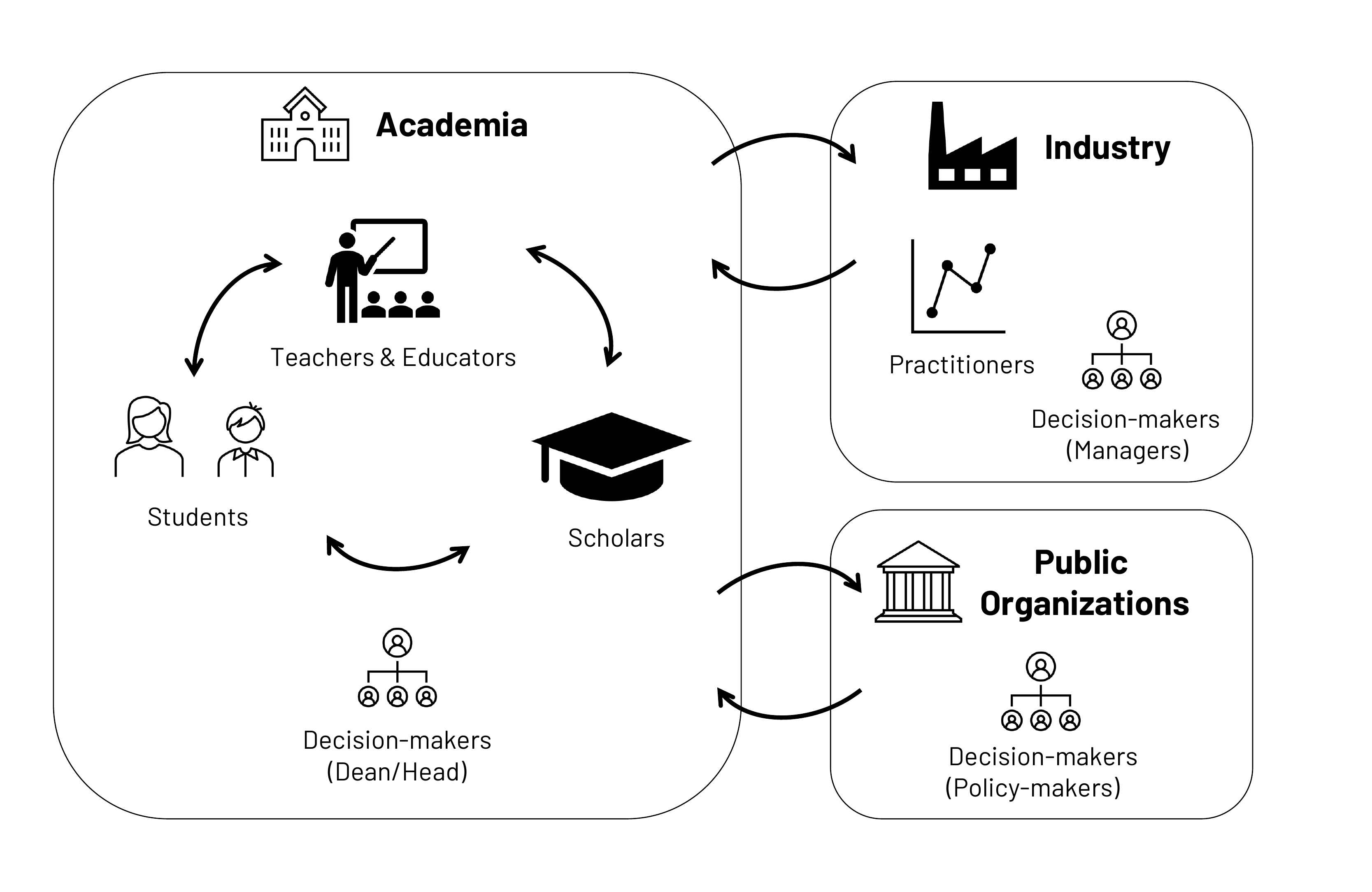}
    \caption{Interaction among actors involved in \ac{IR} and \ac{RS} experimental education.\label{fig:education:stagkeholders}}
    \label{fig:actors}
\end{figure}

Figure~\ref{fig:education:stagkeholders} shows the interaction among the identified actors. In academia, students, educators, and scholars are in continuous interaction through learning, teaching, and supervision processes, which are overseen and/or led by decision-makers such as deans, heads of departments, etc. In industry, decision-makers such as product and team managers, as well as practitioners, make use of training and education resources and initiatives to support experimentation in real-world domains. The cyclic arrows represent the active participation in the creation and development of those resources and initiatives. Decision-makers in public organizations, such as policy-makers, are also key actors in the definition of curricula, which has a direct impact on how and to which extent experimentation in \ac{IR} and \ac{RS} is included in Data Science, Computer Science, \ac{CHI}, and \ac{AI} programs.

\subsubsection{What can we do?}
\label{subsubsec:education-next-steps}

In this section, we first provide examples of helpful \emph{resources} to improve education in IR and RS evaluation. Then, we outline several possible \emph{initiatives} that contribute to increasing awareness about current methodological issues and to disseminate knowledge about experimentation approaches.


\paragraph{Resources}

The resources with which the actors interact are a way to share, maintain, and promote best practices while ensuring a low barrier of entry to the field. Given that those resources might be widely used in education, research (experimentation, etc.), and even production systems, resources have great potential to continuously grow the knowledge of future generations of scholars, practitioners, and decision-makers. 

{\bf General Teaching Material}. Textbooks quickly may become outdated,\footnote{In contrast to that, the main textbook in the area of natural language processing has for years only been available as an online draft and is continuously being updated: \url{https://web.stanford.edu/~jurafsky/slp3/}} but have the advantage that these typically reach a wide audience, whereas slides and tutorials that cover evaluation methodology in more depth might only reach smaller audiences. Often, today's online lectures primarily report on `mainstream' information retrieval (e.g., offline studies, common metrics), but foster reflection and discussion only to a very limited extent. More comprehensive resources should be made publicly available and shared across universities, summer schools, and meetups.\footnote{For instance, Sebastian Hofst{\"a}tter released Open-Source Information Retrieval Courses: \url{https://github.com/sebastian-hofstaetter/teaching}.} Finally, having the \ac{IR} and \ac{RS} community actively contribute to the curation of material in sources that are widely used by the general public---and, thus, also by students---as a starting point to get a basic understanding of a topic (e.g., Wikipedia) is advisable. Further, contributing to the documentation of software such as Apache Solr,\footnote{\url{https://solr.apache.org/}} Elasticsearch,\footnote{\url{https://www.elastic.co/es/elasticsearch/}} Surprise,\footnote{\url{https://surpriselib.com/}} Implicit,\footnote{\url{https://implicit.readthedocs.io}} etc. (see the report by Ferro et al.~\cite{FerroFuhrEtAl2018} for more that are widely used in practice), can help to make non-experts more aware of the best practices in IR and RS experimentation.

Apart from introducing modern information retrieval systems, \textbf{teaching material} should give more attention to a wider set of application fields of \ac{IR}, including recommender systems and topics related to query and interaction mining and understanding, and online learning to rank~\cite{markov2019what}. To date, also online evaluation falls short in such resources although it is essential in the spectrum of evaluation types~\cite{markov2019what}. Students need to be introduced to concepts such as reproducibility and replicability, and it is essential that students understand what makes a research work impactful in practice. To lower the entry barrier to the field, students should be taught how to use available tools and environments that enable quick prototyping, and that have real-world relevance. Teaching fairness, privacy, and ethical aspects, both in designing experiments and also in how to evaluate them, is also important.\footnote{\ac{CyCAT} project: \url{https://sites.google.com/view/biasvisualizationactivity/home}}

Moreover, the participation in \textbf{shared tasks (challenges or competitions)} of evaluation campaigns in \ac{IR} (e.g., TREC,\footnote{\url{https://trec.nist.gov/}} \ac{CLEF},\footnote{\url{https://www.clef-initiative.eu/}} \ac{NTCIR},\footnote{\url{https://research.nii.ac.jp/ntcir/}} or \ac{FIRE}\footnote{\url{https://fire.irsi.res.in/fire/}}) and \ac{RecSys} (e.g., the yearly ACM RecSys challenges\footnote{\url{https://recsys.acm.org/challenges/}}) should be fostered. To facilitate the participation of students, it is worthwhile to make the timelines of such challenges and competitions compatible with the academic (teaching) schedules (e.g., in terms of semesters). Students will be provided with the datasets used in the benchmarks and will be able to learn more on evaluation methodologies (for instance, students from Padua, Leipzig, and Halle participated in Touché~\cite{bondarenko:2022f,bondarenko:2021d} hosted at \ac{CLEF}). At the same time, it is important to critically reflect with students on the limitations and dangers of competitions~\cite{church:2022} and encourage them to go beyond leaderboard \ac{SOTA} chasing culture---e.g., only optimizing on one metric or a limited set of metrics without reflection of the suitability of these metrics in a given application context~\cite{voorhees:2020,jannach2021mcnamara}. Hence, it is important that a student's (or student group's) grade does not depend on their rank in the leaderboard but to a large degree on their approach, reasoning, and reflection to counteract \ac{SOTA} chasing and help students to focus on insights. Inspired by result-blind reviewing in Section~\ref{subsec:reviewing}, we might refer to this as `result-blind grading'.

{\bf Test collections}\footnote{In \ac{IR}, an offline test collection is typically composed of a set of topics, a document collection, and a set of relevance judgments.} and {\bf runs/submissions}---typically combined with novel evaluation methodologies---are the main resources resulting from shared tasks or evaluation campaigns. Integrating the resulting test collections into tools such as \texttt{Hugging Face datasets}~\cite{lhoest:2021}, \texttt{ir\_datasets}~\cite{Macavaney2021} or \texttt{EvALL}~\cite{amigo2017evall} allows for unified access to a wide range of datasets. Furthermore, some {\bf software components} such as \texttt{Anserini}~\cite{Yang2017}, \texttt{Capreolus}~\cite{Capreolus2020}, \texttt{PyTerrier}~\cite{pyterrier2021}, \texttt{OpenNIR}~\cite{macavaney:wsdm2020-onir}, etc., can directly load test collections integrated into \texttt{ir\_datasets} which substantially simplifies data wrangling for scholars of all levels. For instance, PyTerrier allows for defining end-to-end experiments, including significance tests and multiple-test correction, using a declarative pipeline and is already used in research and teaching alike (e.g., in a master course with 240~students~\cite{pyterrier2021}). Other resources for performance modeling and prediction in \ac{RS}, \ac{IR}, and \ac{NLP} can also be found in the manifesto of a previous Dagstuhl Perspectives Workshop~\cite{FerroFuhrEtAl2018}. The broad availability of such resources makes it tremendously easier to replicate and reproduce approaches that were submitted to a shared task (challenge) before. Further, it lowers the entry barrier to experiment with a wider set of datasets and approaches across domains as switching between collections will be easy. New test collections can be added with limited effort. Still, further promoting the practice of sharing code and documentation,\footnote{\url{https://www.go-fair.org/fair-principles/}} or using software submissions with tools such as TIRA~\cite{froebe:2023a,potthast:2019p} in shared tasks is important.

{\bf Combining and integrating the resources} listed above in novel ways has the potential to reduce or even remove barriers between research and education, ultimately enabling Humboldt's ideal to combine teaching and research. Students who participate in shared tasks as part of their curriculum already go in this direction~\cite{elstner:2023}. Continuously maintaining and promoting the integration of test collections and up-to-date best practices for shared tasks 
into a shared resource 
might further foster student participants because it becomes easier to ``stand on the shoulders of giants'' yielding to the cycle of education, research, and evaluation that is streamlined by ECIR, CLEF, and \ac{ESSIR} (see Section~\ref{subsec:talk:ferro}).

%
%
%
%
%
%
%
%
%

\paragraph{Initiatives}

We have identified a range of actors, and we argue that addressing the problems around education requires a number of different initiatives some of which target one particular type of actor but more commonly offer benefits for different groups. These initiatives should not be seen in isolation as our vision is in line with what has been proposed in Section~\ref{subsec:talk:ferro} which calls for coordinated action around education, evaluation, and research. Here we will discuss instruments we consider to be essential on that path. There is no particular order in this discussion other than starting with well-established popular concepts.

{\bf Summer schools} are a key instrument primarily aimed at graduate students. \ac{ESSIR}\footnote{\url{https://www.essir.eu}} is a prime example of a 
summer school focusing on delivering up-to-date educational content in the field of \ac{IR}; the Recommender Systems Summer School is organized in a similar manner focusing on \ac{RS}. Beyond the technical content, summer schools do 
also serve the purpose of community-building involving different actors, namely students and scholars. Annually organized summer schools appear most effective as they make planning easier by integrating them into the annual timeline of \ac{IR}- and \ac{RS}-related events. This is in line with the \emph{flow-wise} vision discussed earlier in Section \ref{subsec:talk:ferro}.

Summer schools also provide a good setting to embed (research-focused) {\bf Mentoring} programs and {\bf Doctoral Consortia}. This allows PhD students as well as early-career researchers to learn from experts in the field outside their own institutions. Both instruments are well-established in the field. However, even though the established summer schools are repeatedly organized, these often happen on an irregular basis (sometimes yearly, sometimes with longer breaks) and using different formats. This irregular setting makes it difficult to integrate it into a PhD student's journey from the outset. 
Currently, Mentoring is often merely a by-product of other initiatives such as Summer Schools and Doctoral Consortia. It may be a fruitful path to see mentoring programs as an independent (yet, not isolated) initiative. For instance, the ``\ac{WiMIR} Mentoring program''\footnote{\url{https://wimir.wordpress.com/mentoring-program/}} sets an example of a sustainable initiative that is organized independently of other initiatives and on yearly basis. A similar format seems a fruitful path to follow in the \ac{IR} and \ac{RS} communities, where it is advisable to facilitate exchange across (sub-)disciplines and open up the initiative to the entire community. We note that---similar to the \ac{WiMIR}---mentoring may not only address PhD students but is well suited also for later-career stages. 

While the \ac{IR} and \ac{RS} communities have a tradition of research-topic-driven {\bf Tutorials} as part of the main conferences, {\bf Courses} that address skills and practices beyond research topics (similar to courses hosted by the \ac{CHI} conference\footnote{\url{https://chi2023.acm.org/for-authors/courses/accepted-courses/}}) would be an additional fruitful path to follow. Such courses may, for instance, address specific research and evaluation methods on an operational level\footnote{See, e.g., \ac{CHI} 2023's C12: Empirical Research Methods for Human-Computer Interaction \url{https://chi2023.acm.org/for-authors/courses/accepted-courses/\#C12}, C18: Statistics for \ac{CHI} \url{https://chi2023.acm.org/for-authors/courses/accepted-courses/\#C18}} or how to write better research papers for a specific outlet or community\footnote{See, e.g., \ac{CHI} 2021's C02: How to Write \ac{CHI} Papers~\cite{HowtowriteCHI2012}}. With regard to support in writing better papers, see also Section~\ref{subsec:guidance}. 
In Bachelor and Master education, more resources in the form of Formal Educational Materials could be developed. For example, students could benefit from The Black Mirror Writers' Room exercise\footnote{\url{https://discourse.mozilla.org/t/the-black-mirror-writers-room/46666}} which helps convey ethical thinking around the use of technology. Participants choose current technologies that they find ethically troubling and speculate about what the next stage of that technology might be. They work collaboratively as if they were science fiction writers, and use a combination of creative writing and ethical speculation to consider what protagonist and plot would be best suited to showcase the potential negative consequences of this technology. They plot episodes, but then also consider what steps they might take now (in regulation, technology design, social change) that might result in \emph{not} getting to this negative future. More experienced Bachelor students and Master students could have assessments similar to paper reviews as part of their curriculum to practice critical thinking.


Topically relevant {\bf Meetups} ranging from informal one-off meetings to more regular thematically structured events offer a much more flexible and informal way to learn about the field. Unlike summer schools they bring together the community for an evening and cater for a much more diverse audience involving \emph{all} actors with speakers as well as attendees from industry, academia and beyond. Talks range from specific use cases of \ac{IR} in the industry (e.g., search at Bloomberg), to the latest developments in well-established tools (such as Elasticsearch) to user studies in realistic settings. There is a growing number of information-retrieval-related and recommender-systems-related Meetups\footnote{See, e.g., \url{https://opensourceconnections.com/search-meetups-map/}, \url{https://recommender-systems.com/community/meetups/}} and many of which have become more accessible recently as they offer virtual or hybrid events. Meetups offer a low entry barrier in particular for students at all levels of education and they help participants obtain a more holistic view of the challenges of building and evaluating \ac{IR} and \ac{RS} applications. Loosely incorporating Meetups in the curriculum, in particular when there is alignment with teaching content (e.g., \textbf{joint seminars}), has been demonstrated to be effective in our own experience. These joint initiatives may go beyond the dissemination of content, but also involve practitioners as well as decision-makers in terms of facilitating (or hindering) strategic alliances or setting strategic themes.

Knowledge Transfer through \textbf{collaboration between industry and academia} is another instrument offering a mutually beneficial collaboration between three key actors: PhD students, academic scholars, and practitioners in the industry. By tackling real-world problems (as defined by the industrial partner) using state-of-the-art research approaches in the fields of \ac{IR} and \ac{RS} (as provided by the academic partner) knowledge does not just flow in one direction but both ways. In the context of our discussion, this is an opportunity to gain insights into evaluation methods and concerns in the industry. There are well-established frameworks to foster knowledge transfer such as Knowledge Transfer Partnerships\footnote{\url{http://ktp.innovateuk.org}} in the UK with demonstrated impact in IR\footnote{\url{https://www.gov.uk/government/news/media-tracking-firm-wins-knowledge-transfer-partnership-2015}} and beyond.


Knowledge transfer should also be facilitated and supported at a higher level at conferences and workshops. This is where the \ac{RS} community is particularly successful in attracting 
industry contributions to the \ac{RecSys} conference series. In \ac{IR}, there is still an observable 
gap between key academic conferences such as \ac{SIGIR} and practitioners' events like Haystack (\emph{``the conference for improving search relevance''}\footnote{\url{https://haystackconf.com}}). The annual Search Solutions conference is an example of a successful forum to exchange ideas between all different actors.\footnote{\url{https://www.bcs.org/membership-and-registrations/member-communities/information-retrieval-specialist-group/conferences-and-events/search-solutions/}}

With a view to improving evaluation practices in the long-term, the reviewing process and practices play an important role. Hence, {\bf addressing reviewers and editors} is essential. Reviewers are important actors in shaping what papers will be published and which not. And it is essential that good evaluation is acknowledged and understood while poorly evaluated papers are not let through. Similarly, it is crucial to have reviewers who acknowledge and understand information retrieval and recommendation problems in their broader context (e.g., tasks, users, organizational value, user interface, societal impact) and review papers accordingly.
Hence, it is essential to develop educational initiatives concerning evaluation that address current and future reviewers (and editors) accordingly. Promising initiatives include the following:
\begin{itemize}
    \item Clear reviewer guidelines acknowledging the wide spectrum of evaluation methodology and the holistic view on information retrieval and recommendation problems. For example, CHI\footnote{ACM CHI Conference on Human Factors in Computing Systems} and \ac{ACL}\footnote{Association for Computational Linguistics} provide detailed descriptions of what needs to be addressed and considered in a review and what steps to take.\footnote{CHI 2023 Guide to reviewing papers \url{https://chi2023.acm.org/submission-guides/guide-to-reviewing-papers/}; \ac{ACL}'s How to Review for \ac{ACL} Rolling Review \url{https://aclrollingreview.org/reviewertutorial}; Ken Hinckley's comment on what excellent reviewing is~\cite{Hinckley2016}.} Care has to be taken, though, that such guidelines are kept concise to not overwhelm people before even starting to read. Further suggestions on results-blind reviewing and guidance for authors can be found in Sections~\ref{subsec:reviewing} and Section~\ref{subsec:guidance} respectively.
    \item Next to reviewers, meta-reviewers and editors is another entity to address, which can be done in a similar manner as addressing reviewers. These senior roles can have strong momentum in inducing change---but have a strong power position in preventing it. Stronger resistance might be expected on that (hierarchical) level. Seemingly, only a few conferences and journals---for instance, \ac{ACL}\footnote{\ac{ACL}'s Action Editor Guide to Meta-Reviewing \url{https://aclrollingreview.org/aetutorial}}---seem to offer clear guidelines for the meta-reviewing activity.
    \item Similar to courses on research methods or addressing paper-writing skills, it is advisable to provide courses that specifically address how to peer review.\footnote{\url{https://chi2023.acm.org/for-authors/courses/accepted-courses/\#C16}}
    \item Mentored reviewing is another promising initiative to have better reviews that, on the one hand, better assess submitted papers and, on the other hand, are more constructive to induce better evaluation practices for future research. Mentored reviewing programs are, for instance, established in Psychology\footnote{\url{https://www.apa.org/pubs/journals/cpp/reviewer-mentoring-program}}. The MIR community\footnote{\url{https://www.ismir.net}} has a New-to-ISMIR mentoring program\footnote{\url{https://ismir2022.ismir.net/diversity/mentoring}} that mainly addresses paper-writing for people who are new to the community but will likely also have an impact on reviewing practices. Similar programs could be established in the \ac{IR} and \ac{RS} communities with a particular focus on evaluation aspects. It is worthwhile to note that a recent study (in \ac{ML} and \ac{AI}) indicates that novice reviewers provide valuable contributions in the reviewing process~\cite{Stelmakh2020_novicereviewers}.
    \item Summer schools mainly address (advanced) students and are also a good opportunity to include initiatives addressing reviewing.
\end{itemize}

{\bf General Public Dissemination} is another important aspect that needs to be addressed. Communication in the lay language of our field is very important. Editing and curating better relevant Wikipedia pages on evaluation measures for information retrieval\footnote{\url{https://en.wikipedia.org/wiki/Evaluation_measures_(information_retrieval)} [Accessed: 20-Jan-2023]} and recommender systems\footnote{\url{https://en.wikipedia.org/wiki/Recommender_system\#Evaluation} [Accessed: 20-Jan-2023]} will increase the potential of reaching a wider audience, including potential future students. Other actions can concern publishing papers in magazines with a wider and differentiated audience, such as \emph{Communications of the ACM}\footnote{\url{https://cacm.acm.org/}}, \emph{ACM Inroads}\footnote{\url{https://inroads.acm.org/}}, \emph{ACM XRDS: Crossroads}\footnote{\url{https://xrds.acm.org/}}, \emph{IEEE Spectrum}\footnote{\url{https://spectrum.ieee.org/}}. One of the final goals is to make IR and RS more popular to both attract students to the field and grow a healthy ecosystem of professionals at various levels.

We have described actors, resources, and initiatives that we think are worth considering in moving forward as a community towards creating more awareness, as well as sharing and transferring knowledge on experimental evaluation for \ac{IR} and \ac{RS}. We summarize the participation (either primary or secondary actors) in generating and consuming these resources and initiatives in Table~\ref{table:mapping}. This is not intended as a definitive list but aimed to represent the primary and secondary actors which are involved.

\begin{table}
\caption{Actors generating or consuming resources and initiatives related to education in evaluation for IR and RS. \checkmark and (\checkmark) indicate primary and secondary actors, respectively.}\label{table:mapping}
\begin{center}
\adjustbox{max width=\textwidth}{%
\begin{tabular}{ p{5cm}ccccc}
 \toprule
\textit{Actors:} & Students & Educators & Scholars & Practitioners & Decision-makers\\
 \midrule
 \multicolumn{6}{c}{\textit{Resources}}\\
\midrule
Teaching Materials & \checkmark & \checkmark & && (\checkmark) \\

Shared tasks/challenges/competitions & \checkmark& \checkmark & \checkmark & \checkmark &\\

Test collections \& runs/submissions & \checkmark & \checkmark & \checkmark & \checkmark & \\

Software (components) & \checkmark & \checkmark &\checkmark &\checkmark &\\
\midrule
  \multicolumn{6}{c}{\textit{Initiatives}}\\
   \midrule
   Mentoring: Summer schools and Doctoral Consortia & \checkmark  & & \checkmark  & (\checkmark ) &\\
   
Tutorials and courses & \checkmark & & \checkmark & \checkmark &\\
 
 Meetups & (\checkmark ) & (\checkmark ) & \checkmark  & \checkmark  & \checkmark \\
 
  Joint seminars & \checkmark & \checkmark &  & \checkmark & (\checkmark )\\
 
    Collaboration between industry and academia & \checkmark  & & \checkmark  & \checkmark & \\
 
   Reviewing & (\checkmark ) & & \checkmark  & &\\
 
   General public dissemination & (\checkmark )& (\checkmark ) & \checkmark  & \checkmark  & \checkmark \\
 \bottomrule
   \end{tabular}
   }
\end{center}
\end{table}

\subsubsection{Challenges \& Outlook}
\label{sec:education-challenges}

Given the importance of reliable and ecologically valid results, one may ask oneself which obstacles occur in the path of developing better education for experimentation and evaluation of information access systems. 
We see different potential barriers (and possibilities) for the different actors: students, educators, scholars, practitioners, and decision-makers. We will investigate each actor in turn.


{\bf Scholars.} As has also been identified in a previous Dagstuhl seminar~\cite{FerroFuhrEtAl2018}, it is significantly harder to test the importance of assumptions in user-facing aspects of the system, such as the presentation of results or the task model, as it is prohibitively expensive to simulate arbitrarily many versions of a system and put them before users.
User studies are therefore also at higher risk of resulting in hypotheses that cannot be clearly rejected (non-significant results), leading to fear of criticism and rejection from paper reviewers. There are some proponents of Equivalence Testing \cite{lakens2017equivalence}\footnote{See also \url{https://cran.r-project.org/web/packages/TOSTER/TOSTER.pdf}} and Bayesian Analysis \cite{van2021jasp} in Psychology which may also be useful in Computer Science.

As \acp{LLM} are becoming a commodity, policies to educate and guide authors and reviewers in how different \ac{AI} tools can (or cannot) be used for writing assistance should be discussed and defined.\footnote{For instance, see the \ac{ACL} 2023 Policy on \ac{AI} Writing Assistance: \url{https://2023.aclweb.org/blog/ACL-2023-policy/}.} These guidelines may inspire educators on how to characterize the role of these tools in learning \& teaching environments, including assessment design and plagiarism policies\footnote{\url{https://www.theatlantic.com/technology/archive/2022/12/chatgpt-ai-writing-college-student-essays/672371/}}.

In addition, a current culture of `publish or perish' incentivizes short-term and incremental findings\footnote{\url{https://harzing.com/resources/publish-or-perish}}, over more holistic thinking and thoughtful comparative analysis. The problem of `\ac{SOTA}-chasing' has also been discussed in other research areas, e.g., in \ac{NLP} \cite{church:2022}. Change in academic incentive systems both within institutions and for conferences and journals change slowly but they do evolve.


{\bf Students and Educators.} Thankfully, institutions are increasingly recognizing the need for reviewing studies before they are performed, such as Ethics and Data Management plan\footnote{Further proposals for methodological review are also under discussion in Psychology, but will likely take longer to reach Computer Science: \url{https://www.nature.com/articles/d41586-022-04504-8}}.
In Bachelor and Master education, in particular, this means that instructors may require training in writing such documents, and institutions appreciate and are equipped for timely review. Therefore, planning of education would benefit from allowing sufficient time for submission, review, and revision. 

In that context, teaching evaluation methodologies may require some colleagues to retrain, in which case some resistance can be expected. Improving access to training initiatives and materials at post-graduate level can support colleagues who are willing but need additional support. Various forms of informal or even organized exchange between teachers may be a helpful instrument to grow the competency of educators.

Furthermore, certain evaluation concepts and methodologies cannot be taught before certain topics are covered in the curriculum. A student in recommender systems may need to understand the difference between a classification and regression problem; or the difference between precision and recall (for a given task and user it may be more important to retrieve accurate results, or to retrieve a wider range of results) before they can start thinking about the social implications.

Moreover, some students are prone to satisfice, thinking that ``good enough is good enough'': there are many methodologies available for evaluation, and the options are difficult to digest in a cost-effective way at entry-level---highlighting the need for availability of tutorials and low-entry level materials as indicated earlier in Section~\ref{subsubsec:education-next-steps}. Embedding participation to shared tasks and competitions (e.g., \ac{CLEF} labs or \ac{TREC} tracks) which provide a common framework for robust experimentation may help overcome this challenge---although the synchronization between the semester and participation timelines may not be straightforward.

Finally, there is a growing number of experiments in developing multi-disciplinary curricula -- with the appreciation that different disciplines bring to such a program. Successful initiatives include group projects consisting of students in both \ac{SSH} and Computer Science. In fact, one of the underlying principles of the continuously growing \emph{iSchools consortium}\footnote{\url{https://www.ischools.org}} is to foster such interdisciplinarity. 
The challenge here is not only the design of the content but also accreditation and support from the strategic level of institutions.

{\bf Practitioners.} Maintenance of resources used to translate knowledge about models and methodologies for evaluation is challenging given the fast pace of the field. This can make it hard to compare results across studies and to keep up with the \ac{SOTA} of best practices in experimentation. In this regard lowering the entry barrier to participating in initiatives such as shared tasks/challenges \cite{ferro2019happened,edu-zz-HarmanVoorhees2005-editor} and maintaining documentation of resources commonly used by non-experts are increasingly helpful.

Another issue is the homogeneity of actors. Often there is no active involvement of actors outside a narrow academic Computer Science sphere, who otherwise might have indicated assumptions or limitations early on. It can be challenging to set up productive collaborations between industry and academia, as well as across disciplines. Typical issues include, for instance, common terminology used in a different way, or different levels of knowledge of key performance indicators. Co-design in labs has set a good precedent in this regard. Examples are ICAI in the Netherlands\footnote{\url{https://icai.ai/}}, its extension in the new 10-year ROBUST initiative\footnote{\url{https://icai.ai/ltp-robust/}}, and the Australian Centre of Excellence for \ac{ADM+S}\footnote{\url{https://www.admscentre.org.au/}}, where PhDs in multiple disciplines (Social Sciences \& Humanities, Computer Science, Law, etc.) are jointly being trained in shared projects. 

Research Advisory Boards are another effective instrument to draw in practitioners but here the challenge is to make the most of the little time that is usually available for the exchange of ideas between practitioners and academics.

{\bf Decision-makers.} The output of evaluation and experimentation in \ac{IR} and \ac{RS} may be used to inform decision-making on the societal level. Consequently, if the evaluation is poorly done, or the results incorrectly generalized, the implications may also be poor decision-making with far-reaching impacts on society, e.g. \cite[Ch. 10]{Kahneman11Thinking}. 

The ability of the other actors to support education on evaluation is constrained and shaped by decision-makers.
Policy-makers in public organizations and program managers or deans in academia play a crucial role in curriculum design. Scholars and educators will have to communicate effectively the importance of experimental evaluation in information access in order to inform the decision-making process. The challenge here is to initiate change in the first place and to drive such changes. Any new initiative will necessarily involve not just a single decision-maker but more stakeholders and committees making this a more effortful but possibly also more impactful process than many of the other initiatives we have identified. 

Additionally, decision-makers within academic institutions, namely libraries and career development centres, can play an important role towards developing the competency of students and educators. Making best practices in evaluation available as a commodity through these channels will require making resources more accessible for non-experts in \ac{IR} and \ac{RS}.

\subsubsection{Concluding Remarks}

Education and dissemination represent key pillars to overcoming methodological challenges in Information Retrieval and Recommender
Systems. What we have sketched here can be interpreted as a general roadmap to create more awareness among and beyond the \ac{IR} and \ac{RS} communities. We hope the recommendations---and the identified challenges to consider---on what we can do will help to support education for better evaluation in the different stages of the lifelong learning journey.
We acknowledge that facets such as incentive mechanisms and processes in institutions are often slow-moving. The vision proposed in this section is therefore also aimed at a longer-term (5--10 years) perspective.

\abstracttitle{Results-blind Reviewing}
\label{subsec:reviewing}
\abstractauthor[Joeran Beel, Timo Breuer, Anita Crescenzi, Norbert Fuhr, Meijie Li]{%
Joeran Beel (University of Siegen, DE, joeran.beel@uni-siegen.de)\\
Timo Breuer (Technische Hochschule Köln, DE, timo.breuer@th-koeln.de) \\
Anita Crescenzi (University of North Carolina at Chapel Hill, US, amcc@unc.edu)\\
Norbert Fuhr (University of Duisburg-Essen, DE, norbert.fuhr@uni-due.de) \\
Meijie Li (University of Duisburg-Essen, DE, meijie.li@uk-essen.de)
}
\license

\subsubsection{Motivation}
\label{subsubsec:reviewing-motivation}
Campbell and Stanley defined experiments as ``that portion of research in which variables are manipulated and their effects upon other variables observed” (p. 1 in \cite{Campbell1963}).'' Scientific experiments are used in confirmatory research to test a priori hypotheses as well as in exploratory research to gain new insights and help to generate hypotheses for future research \cite{Shadish2002}.
In information access research, the ultimate goal is to gain insights into cause and effect. Unfortunately, many reviewers of information access experiments place undue emphasis on performance, rejecting papers that contain insights if they fail to show improvements in performance.
The focus on performance numbers not only leads to publication bias. It also puts additional pressure on early-career researchers who must publish or perish, thus being tempted to cheat if their proposed method does not yield the desired results. Moreover, reviewers pay little attention to the experimental methodology and analysis \cite{Fuhr17} in case the results are impressive. Focusing primarily on performance (and in particular aggregated performance) can lead to a neglect of insights; gaining insights is critical to move the information access field forward and 
essential to be able to make performance predictions \cite{rev-FerroFuhrEtAl2018}.

We think that one important step to change the situation is if we alter the review process such that there is more emphasis on the theoretical background, the hypotheses, the methodological plan and the analysis plan of an experiment, while improvement or decline of performance should play less of a role when deciding about the quality of a paper. It is hoped that this will lead to a higher scientific quality of publications, more insights, and improved reproducibility (as there is less incentive for beautifying results). 
As Woznyj et al. \cite{Woznyj2018} note in their survey of editorial board members, overall there are  positive attitudes towards results-blind reviewing and advantages for the scientific community outweigh concerns.


In order to move the review focus away from performance improvement, appealing to reviewers alone will not be sufficient. A more drastic measure is the change of the review process such that reviewers decide about acceptance vs.\ rejection of a paper without knowing the outcome of the experiments described.

\subsubsection{Current Situation and Gaps}
\label{subsubsec:reviewing-gaps}

As part of \ac{IR} or \ac{RS} conferences, the peer-reviewing process usually involves the review of the full paper using double-blinded reviewing, i.e., both authors and reviewers remain anonymous to each other. Before submission, authors are informed about possible reviewing criteria and areas of interest in the \ac{CfP} that can be found on the conference website. Upon submission, the paper should contain all of the relevant information regarding the motivation, the research methodology or study design, the experimental results, and finally, a discussion that puts the results into context.

For each submission, usually, a group of three reviewers is assigned. All of them should align their reviews to those criteria mentioned in the \ac{CfP} and, depending on the submission system, express their opinion in written text or by pre-defined answers regarding particular aspects. In addition, they can assign (overall) scores. The final decision is based on a discussion among reviewers, which is governed by an additional meta-reviewer, and consolidation with the program chairs.

Even though this traditional review model has been established for several years, it can imply negative impacts on the stakeholders or the scientific community as a whole. Under the assumption that reviewers overemphasize positive outcomes, the authors might be inclined to ``search for'' performance gains in system-oriented experiments at the cost of scientific rigor and reasoning. Even more, there is the danger of fraud or selecting positive outcomes, considering the need to publish in order to proceed in an academic career.

\begin{table}[t] 
\caption{Comparison of traditional and emerging approaches to peer review: results-blind, preregistered reports, and registered reports.}
\label{tab:review-comparison}
\begin{tabular}{p{12em} p{5.5em} p{5.5em} p{5.5em} p{5.5em}}
\hline
 & Traditional & Results-Blind & Preregistered & Registered Report \\ 
 \hline
protocol preregistration & optional & optional & yes (in journal repository) & no \\ 
protocol publication (separate from research article) & no & no & no & yes \\ 
peer review of research protocol before data collection & no & no & yes & yes \\ 
peer review of paper with blinded results & no & yes & no & no \\ 
peer review of full paper & yes & yes (if in-principle acceptance) & yes with focus on results (if in-principle acceptance) & yes (if in-principle acceptance) \\ 
Example publication(s) & ACM SIGIR, ACM CHIIR & BMC Psychology & PLOS Biology & PLOS ONE \\ 
\hline
\end{tabular}
\end{table}

Alternatives to the traditional review process have emerged with an initial round of peer review of a manuscript with the results blinded or a study protocol and a subsequent round of peer review of the full paper including results. Table \ref{tab:review-comparison} shows the traditional peer review model with our recommended results-blind reviewing and two other variants, each of which we describe below.
The Center for Open Science notes that, as of January 2023, over 300 journals have adopted one or more variants of this approach.\footnote{https://www.cos.io/initiatives/registered-reports} In addition, several preliminary analyses of their implementation have been conducted and published (e.g., ~\cite{Findley2016,Maggin2020,Woznyj2018}).


A results-blind review involves an in-principle acceptance or rejection decision based on peer review of the paper \emph{with the results blinded} from the reviewers (see the third column of Table~\ref{tab:review-comparison}). 
The reviewers can put more emphasis on judging the merits of the general motivation, the study design, and what kinds of scientific insights could be gained from the experiments. 
If the paper is accepted in-principle, it proceeds to a second stage of peer review of the \emph{paper with the results} included for reviewers. 
The final decision about the acceptance is based on the second stage of the review in which the reviewers have access to the experimental outcomes. 
 


Other peer-reviewing models have emerged in recent years as part of the growing awareness of preregistration\footnote{https://www.cos.io/initiatives/prereg}$^{,}$\footnote{https://plos.org/open-science/preregistration/} and its adoption \cite{Nosek2018}.
One such approach to peer review involves the review and in-principle acceptance of the study protocol including the methods and analysis plan before data is collected or analysis begins. 
Variants of this approach include preregistered research articles 
and registered reports 
for confirmatory research~\footnote{For examples of how preregistered research articles and research reports have been implemented, see the summary provided by PLOS. https://plos.org/open-science/preregistration/}. Although preregistered reports and registered reports are typically used for confirmatory research, there are variants for exploratory research and some journals also use a separate approach for exploratory research projects which do not have a confirmatory component (e.g., an Exploratory Report article type in journal \emph{Cortex}).


Preregistered research articles involve researchers submitting a research study protocol including the rationale and hypotheses, methodology including analysis plan, and materials to a journal for review and simultaneous depositing into a repository often associated with the journal (see the fourth column of Table~\ref{tab:review-comparison}). The preregistered protocol is peer-reviewed with a focus on methods and the analytic approach, and a provisional in-principle acceptance conditional upon the execution of the study as designed. The researchers execute the study, analyze the results, and submit a full manuscript. After peer review of the new sections, the completed manuscript is published. 

Registered Reports also involve submission and peer review of a study protocol (see the third column of Table~\ref{tab:review-comparison}). A key difference from preregistered articles is that accepted protocols are published immediately and a future article with the results of the study is given an in-principle acceptance. After the study execution, the full manuscript is submitted and reviewed.

\subsubsection{Next Steps}

We propose several changes to the reviewing processes for information access papers to reduce publication biases. Our recommendations are that information access scholarly community: 
\begin{enumerate}
    \item adopts a pilot test of results-blind reviewing for a conference or journal, 
    \item considers starting from our initial process recommendation for results-blind reviewing,
    \item ask authors, conference organizers, and reviewers to place more emphasis within papers on the insights that can be gained from their research, 
    \item considers allowing additional space for additional details about study methodology, and 
    \item considers whether to implement a two-stage review process in which research proposals and/or preregistered research reports are reviewed with a tentative acceptance decision before data collection and analysis are conducted. 
\end{enumerate}
Each of these is described in more detail below.



\paragraph*{Recommendation 1: Pilot test of results-blind reviewing in conference(s) or journal(s)} 

\begin{figure}[t]
    \centering
    \includegraphics[width=0.8\textwidth]{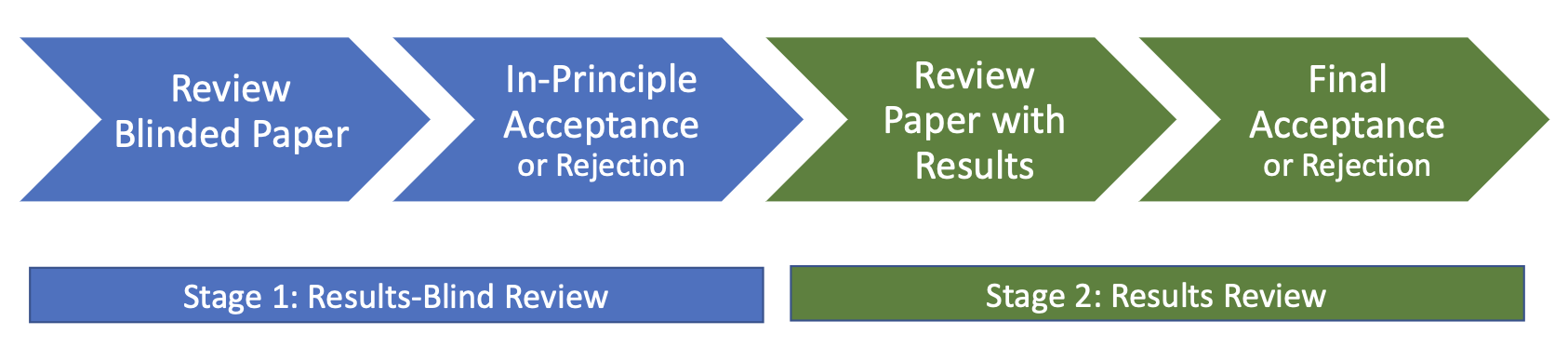}
    \caption{Proposed two-stage process for results-blind reviewing (figure adapted from BMC\footnote{https://www.biomedcentral.com/collections/RFPR})}
    \label{fig:two_stages}
\end{figure}

Our first and most important recommendation is that the information access research communities (i.e., \ac{IR} and \ac{RS} communities) adopt a results-blind approach to peer reviewing for conference(s) and/or journal(s). We recommend that the community start with a pilot test of results-blind reviewing in an established conference track, perhaps with a new paper track with an earlier deadline to allow for a two-stage review process. In results-blind reviewing, the authors submit two versions of their manuscript:  one version of the paper with the full results, and one version with the results blinded.
The two submitted  versions are the basis of a results-blind reviewing process with two major stages (see Figure~\ref{fig:two_stages}). 

Stage 1 consists of the Results-Blind Review. The results-blind version of the manuscript is reviewed and an in-principle acceptance (or rejection) is made. During Stage 1, as in the traditional reviewing process, the paper is reviewed by multiple reviewers who also make acceptance recommendations. In the case of conferences, the in-principle acceptance (or rejection) decision is made after discussion with the \ac{SPC}/meta reviewer and in the \ac{PC} meeting. 
Papers that receive an in-principle acceptance proceed to Stage 2.

Stage 2 consists of the Results Review. The paper containing the results is reviewed by the same set of reviewers with a focus on the results. In the case of a conference, the final acceptance (or rejection) decision is made after a discussion period with the \ac{SPC} and in the \ac{PC} meeting.


\paragraph*{Recommendation 2: Initial process recommendation for a results-blind reviewing pilot}

Below, we recommend a high-level process for how a results-blind reviewing process pilot might be implemented and important considerations for conference organizers and reviewers as well as authors. 
\subparagraph{Conference organizers}
Once the decision for results-blind reviewing has been made, conference organizers would have to take the following steps:
\begin{itemize} 
\item First, the \ac{CfP} for the new track should be written. As the proposed results-blind reviewing process with two stages of review will take longer to complete, an earlier deadline for this track should be set.
\item Criteria for both stages of the review (blinded and with results) should be defined. Special attention should be given to the criteria for changing an initial acceptance recommendation into a rejection.
\item Author instructions for the results-blind reviewing track have to be formulated, describing not only the new reviewing criteria and process but also specific instructions on how to prepare the blinded version of an article. For the results-blind version of the paper, the authors will need to blind all mentions of the results (e.g., in the abstract, introduction, discussion, and conclusion in addition to in a results section) in a way that it is not technically possible to recover the blinded text. There should be a way for reviewers to easily determine the differences between the results-blind version of the paper and the one with the results.
\item Reviewers for the results-blind reviewing track have to be recruited. In the beginning, additional or different expertise will be required for this track. A special introduction of training for the reviewers might be necessary in order to make them familiar with the new process and criteria.
\item The reviewing software will need to be configured for multiple stages of review for the results-blind reviewing. In the first stage of reviewing, only the blinded version of the papers should be distributed to reviewers (see below for the process for reviewers).
\item After the final decision by the PC, the authors will be provided with the review and informed about the final accept or reject decision. In the case of a rejection decision, authors should also be notified at which stage the paper was rejected. 
\item The organizers should give special recognition to the \ac{PC} member of the track (on the conference Web site and in the proceedings)
\item The success of the new track and the process should be evaluated.
\end{itemize}

\subparagraph{Reviewers} Once the reviewers are provided with instructions about the general process and received additional training, we recommend the following process:
\begin{itemize}
    \item In the first stage, the reviewers are provided with the results-blind version of the submission and complete their review including a recommendation about the in-principle acceptance.
    \item Once the reviews are complete, a discussion phase with the \ac{SPC} follows, leading to a recommendation for each paper.
    \item The \ac{PC} for the track meets and makes an initial decision (in-principle acceptance or rejection) for each paper. 
    \item For the second reviewing stage, only in-principle accepted papers are considered. Reviewers get the full versions of the papers they reviewed before. They add an additional part to their review focusing on the results which were previously blinded. Also, they make a second recommendation about acceptance.
    \item As for the first phase, a discussion phase with the \ac{SPC} follows leading to a recommendation for each paper.
    \item The track PC meets for the second time and makes the final decision for each paper.
\end{itemize}

\subparagraph{Authors}
Authors will have to understand the new reviewing scheme, and possibly be trained/educated for preparing manuscripts that satisfy the new reviewing criteria. They will have to prepare and submit two versions of a paper, a version with the results as in the traditional model as well as one in which the results are blinded. 

\paragraph*{Recommendation 3: Emphasize insights in papers} 
We recommend that authors, conference organizers, and reviewers place additional emphasis on communicating expected insights to be gained from experiments.
Guidelines (and review forms) should ask the reviewers to comment on the theoretical background, the hypotheses, the methodological plan and the analysis plan of the experiment(s) described. Special attention should be given to the expected insights to be gained from experiments, i.e.\ regarding cause and effect.

\paragraph*{Recommendation 4: Extra space for methods information} 
Another recommendation is for the community to consider explicitly allowing methodological appendices for authors to provide additional methodological details outside of page and/or word limits and to include these appendices with the text of the paper and not as supplementary materials. 
While not needed for all publications, this would be very beneficial for some types of studies so that the authors can include all study materials. For example, in user studies, researchers may administer multiple questionnaires, conduct a semi-structured interview, and read from a script. It is not uncommon for researchers to administer multiple questionnaires and conduct a semi-structured interview. 

This would be especially important if adopting a results-blind reviewing process as careful scrutiny of the study design and all study materials is needed to ascertain whether the authors will be able to answer the research questions. For example, due to page limits, it is common for authors to describe the topics of an interview but uncommon to include the full text of an interview guide due to page limits.

In addition, this would have an additional benefit for other researchers who wish to replicate the study. 
While, for example, authors can currently make supplementary materials available in \ac{ACM DL}, these materials are not included in the downloadable version of the article or when reading online in the ACM DL in the eReader or HTML formats. 

\paragraph*{Recommendation 5: Consider a two-stage review process adapted from preregistered or registered reports} 
Although our primary recommendation is for conference organizers or journal editors to embrace a results-blind reviewing approach, we also recommend that they consider piloting a conference track or article type in which the study protocol undergoes peer review and is accepted in-principle before data collection or analysis begins. This may be more appropriate for certain types of research (e.g., user studies).

\subsubsection{Conclusion}
At first glance, the new result-blind reviewing scheme might seem to be only attractive for papers describing failed experiments, while authors with successful results would go to the established tracks. In order to avoid this impression, it is essential that the new scheme is piloted as a highly visible and prestigious track in an established conference.
Furthermore, it should be clearly communicated that the results-blind reviewing scheme aims at establishing high standards for the design, execution and analysis of experiments while shielding the reviewers from being blinded by shiny experimental results. 
Thus, it is our hope that papers published in this track will be regarded as high-quality publications which thoroughly address research questions and clearly demonstrate the insights that may be gained from the research.

\newpage

\abstracttitle{Guidance for Authors}
\label{subsec:guidance}
\abstractauthor[Giorgio Maria Di~Nunzio, Maria Maistro, Christin Seifert, Juli\'an Urbano, Justin Zobel]{%
Giorgio Maria Di~Nunzio (University of Padova, IT, giorgiomaria.dinunzio@unipd.it)\\
Maria Maistro (University of Copenhagen, DK, mm@di.ku.dk)\\
Christin Seifert (University of Duisburg-Essen, DE, christin.seifert@uni-due.de)\\
Juli\'an Urbano (Delft University of Technology,  NL, j.urbano@tudelft.nl)\\
Justin Zobel (University of Melbourne, AU, jzobel@unimelb.edu.au)
}
\license

\subsubsection{Motivation}
\label{subsubsec:ag-reviewing-motivation}

The IR community has over time developed a strong shared culture of
expectations of published papers, particularly in our leading venues.
However, these expectations are not explicit and
the evidence of submitted papers is that many authors are not aware of
what elements, or omissions, are likely to be of concern to reviewers.
While accepted papers do provide an indication of what an author should
do, they are, of course, uneven, and the small set of papers that an author 
is consulting in their new work could easily be unrepresentative of the best 
IR work as a whole.

In this section, our aim is to provide a basis for general guidance for
authors and reviewers, with a focus on people who are new to the community.
It should communicate to authors and reviewers a range of
factors that the community regards as significant.
Such guidance, if well designed, should help authors to lift the standard
of their work and provide context should it not be accepted; for reviewers,
especially those new to the task, it can provide checklists and (at a high
level) advice about the field from beyond their immediate research environment.

Some elements in papers have attracted specific criticism in publications; this is particularly true of effectiveness measurement, where a long history of research on method has argued for and against a range of
measures, forms of evidence for statistical validity, treatment of test
collections, and so on.
Such literature is critical to improving the quality of our research but does not necessarily represent a settled, shared view of best practice.

In our view, it is essential that general advice be constructive, readily understandable by new \ac{IR} authors and reviewers, and---to the extent that is possible---not the subject of active debate.
In the following, we have sought to follow this principle.
We first explain the basis of the draft guidance for authors that we have developed and then present that guidance.
How this work might develop over time is considered under ``next steps''.

\subsubsection{Flaws in Submitted IR Papers}
\label{subsubsec:flaws}


For our goal of developing draft guidelines for authors for the community,
we have multiple sources of inspiration.
As a first step, it is valuable to understand and list the kinds of issues that lead experienced reviewers to criticize papers,
that is, to collect the opinions from the community based on their experience from different roles as scientists: authors, readers, reviewers and meta-reviewers. 
Another valuable source of information consists in existing guidelines in adjacent research fields, as they reflect a common agreement of what constitutes a good scientific paper in that community and point out commonly agreed issues that may lead to rejection. 

By collecting, consolidating, and harmonising the collected information, we aim to establish a strong foundation for the synthesis of a new set of draft guidelines that comprehensively capture the community-agreed strengths aspects of good scientific papers as well as issues that commonly lead to rejection;
and separately to identify significant emerging aspects that are not yet captured by existing guidelines.\footnote{As an example, \ac{ACL} 2023 includes a ``Policy on AI Writing Assistance'' in their call for papers\linebreak \url{https://2023.aclweb.org/blog/ACL-2023-policy/}.}
To obtain concise, comprehensive, understandable, and actionable guidelines for early-career researchers, we translated the identified issues, points of criticism, and guideline items, which have been described at varying levels of detail, into observations on elements that papers should include and on elements that can lead to rejection. 

We designed the following approach to create our guidelines:
(1)~search of existing guidelines;
(2)~brainstorming to identify common pitfalls;
(3)~categorization of the outcomes from the brainstorming exercise and comparison
of these with existing guidelines; and
(4)~consolidation and integration with existing SIGIR guidelines.\footnote{\url{https://sigir.org/sigir2023/submit/call-for-full-papers/checklist-to-strengthen-an-ir-paper/}}
Throughout each step of the process, we adhere to the principle of keeping only issues that we believe to be widely agreed upon within the community.

We now describe our approach.

\paragraph*{Identifying existing guidelines}
We started by searching for existing guidelines for authors and reviewers that have been proposed in adjacent research communities.
In our search for existing guidelines, we considered the following sources.
\begin{itemize}
\item The ACM Special Interest Group on Information Retrieval (SIGIR) developed recommendations to strengthen IR papers. These are rather general suggestions concerning presentation and experimentation. 
We used them as the initial stage and extend them to design our list of recommendations for authors (see Section~\ref{subsubsec:guidance}).
\item Empirical Evaluation Guidelines from the \ac{SIGPLAN}.\footnote{\url{https://www.sigplan.org/Resources/EmpiricalEvaluation/}}
This is a checklist that presents best practices meant to support both authors and reviewers within the community.
The checklist includes some broad categories (e.g., appropriate presentation of results) and examples of violations for each subcategory (e.g., a misleading summary of results).
These are reported in Appendix~\ref{appdx:authorguide:sigplan}.
\item The Special Interest Group on \ac{CHI} SIGCHI\footnote{\url{https://chi2022.acm.org/for-authors/presenting/papers/guide-to-reviewing-papers/}} published a guide for reviewing papers submitted to the CHI conference.
This is a general overview of both quality considerations (e.g., whether the paper contribution is
sufficiently original), and more practical considerations related to the paper length and the review process.
SIGCHI also suggested the Equitable Reviewing Guide,\footnote{\url{https://chi2022.acm.org/for-authors/presenting/papers/equitable-reviewing-guide/}} which is a list of recommendations to help reviewers write fair reviews.
Some of their points include reflecting on personal bias or considering that many authors are not native English speakers, thus being lenient on writing style and typos.
\item The \ac{ACL} presented an online tutorial to instruct reviewers on the \ac{ACL} Rolling Review process.\footnote{\url{https://aclrollingreview.org/reviewertutorial}}
This tutorial presents some practical suggestions (e.g., planning the reading and reviewing time to avoid rushed reviews), as well as suggestions to evaluate the quality of the paper and a list of common reasons for rejection, which often lead to author complaints because such reasons are not actual weaknesses but rather easy, unreasonable
grounds for rejection.
\item Ulmer et al.~\cite{Ulmer2022_experimental-standard-dl-and-nlp} present a list of best practices and guidelines for experimental standards within \ac{NLP}.
These guidelines contain some broad categories, (e.g., data), and minimal requirements and recommendations for each category (e.g., publish the dataset accessibly and indicate changes).
These are reported in Appendix~\ref{appdx:authorguide:exp-dl}.
\end{itemize}

\paragraph*{Brainstorming to identify common issues}
After our search for guidelines, we ran a brainstorming exercise among contributors of the working group.
The goal of this exercise was to identify concerns and flaws that we, as reviewers, would not want to find in IR papers and can very likely lead to rejection. 
This list of reflections is included in Appendix~\ref{appdx:authorguide:reflections}. 

We extended the brainstorming exercise to all participants in the Dagstuhl seminar through an online survey.
We asked participants to list ``things we don't like to see in papers'', and
provided some examples for guidance and the full list of SIGPLAN categories for inspiration.
We received $35$ items.
Comments concerning strategic issues, such as ``I prefer to have a new paper category'' were omitted from further analysis; others were integrated into our findings.
As mentioned above, we adhere to the principle of keeping only issues that we did not
regard as controversial issues or the subject of debate, with the aim of omitting points that might lead to
disagreement in the community.

\paragraph*{Integration and categorization}

Inspired by the SIGPLAN and \ac{NLP} guidelines, we developed an initial set of broad categories to organize the
issues we identified above.
We then mapped each item in our list of reflections to the corresponding category.
We did the same for the suggestions collected from the participant survey, as well as for the pertinent points identified in the SIGPLAN and \ac{NLP} guidelines and the SIGIR guidance.
In this process, we focused on issues that specifically relate to IR papers and set aside more general issues
such as ``captions of tables should be clear''.

There were several rounds of review to clarify and consolidate similar items, with minor re-categorizations when needed. 
The final result of this process is a  list of what we believe are recognised as common flaws in IR papers. The final list consists of $57$ items organized in the following $9$ categories (see Appendix~\ref{appdx:authorguide:flaws}):
(1)~Design, motivation and hypothesis;
(2)~Literature;
(3)~Model and method;
(4)~Data, data gathering and datasets;
(5)~Metrics;
(6)~Experiments;
(7)~Analysis of results and presentation;
(8)~Repeatability, reproducibility, and replicability; and
(9)~Conclusions and claims.

Finally, we used this list of concerns to propose an update to the existing SIGIR guidelines.
This is described in the next section.




\subsubsection{Draft Guidance for Authors}
\label{subsubsec:guidance}

Some years ago, \ac{SIGIR} introduced brief guidance for authors as ``Things that strengthen an IR paper''.\footnote{The earliest mention we are aware of is from \ac{SIGIR} 2021, \url{https://sigir.org/sigir2021/checklist-to-strengthen-an-ir-paper/}.}
One of us (Zobel) recently updated this guidance for \ac{SIGIR-AP}'23, in consultation with
the other Program Chairs, but we note that it represented the views of just a couple
of individuals.
The \ac{SIGIR} guidance proposed, at a high level, aspects to consider in presentation and experiments.
The \ac{SIGIR-AP} revision primarily addressed some aspects---omissions, oversights, and shortcomings---that are offered as grounds for rejection.

Here, we took the SIGIR-AP draft guidance as a starting point and reviewed it against the list of concerns that we set out in Section~\ref{subsubsec:ag-reviewing-motivation}.
We also took note of generic writing advice that is widely available and decided to omit elements that we regarded as pertinent to computer science research in general.
This led to the following, which we propose as a basis for the advice provided by venues that publish \ac{IR} work.

We have sought to make the advice broad, understandable, and constructive; but it is of necessity brief and some readers may seek more detail.
For that reason, when the advice (or a revision of~it) is used, it might also be helpful to link to a version of the lists of concerns in Appendix~\ref{appdx:authorguide:flaws}.

Our proposed draft guidance is as follows.

\begin{tcolorbox}[colframe=white]
\paragraph*{Motivation and claims}
\begin{itemize}
\item The problem is well characterised and motivated, and the potential impact is discussed.
\item The proposed application of the work is contextualised by pertinent knowledge from that
domain, including potential ethical, social, or environmental impacts.
\item The research goals and original contributions (that is, the elements that are a contrast to
the prior art) are stated and are clearly distinguished from prior work.
\item The claims are properly scoped and supported.
\item There are explicit statements of what was done and what was not.
\end{itemize}
\end{tcolorbox}

\begin{tcolorbox}[colframe=white]
\paragraph*{Presentation}
\begin{itemize}
\item The literature review considers competitive previous solutions for the problem, that is, it
is not limited to consideration of other work on the same technology as that explored in the
submission.
\item There is a reasoned justification for each of the choices made in each step of the research
and each element of the method.
\item Results are presented in keeping with the norms in the field as exemplified in strong
prior work.
\item A substantive, focused, and insightful discussion accompanies the results taking into account
limitations and scope of the work.
\end{itemize}
\end{tcolorbox}

\begin{tcolorbox}[colframe=white]
\paragraph*{Experiments}
\begin{itemize}
\item The experimental design and its scale are appropriate to the problem.
\item In comparative studies, appropriate baselines are used; they are deployed and optimized in
ways comparable to those used for the proposed method.
\item The experimental results are reliable and generalizable, and preferably show illustrative individual
cases as well as aggregated results.
\item Where appropriate, a diversity of data sets are used, including public-domain data sets
used in prior work.
\item Sufficient details (with data and code where appropriate) are provided to enable other
researchers to assess and reproduce the experiments; this includes the nature, source, and
collection process for the data, and the data preparation steps. 
\end{itemize}
\end{tcolorbox}

\begin{tcolorbox}[colframe=white]
\paragraph*{Results and analysis}
\begin{itemize}
\item The evaluation methods and measures address the research questions; the use of redundant or
highly correlated measures should be avoided.
\item Statistical analysis is used and reported appropriately.
\item Development data, training data, and test data are distinguished from each other.
\item User studies are based on adequately sized, representative cohorts; data is gathered in
ways that meet ethical norms, or where appropriate in keeping with prescribed ethics practices.
\item Final results were obtained after all development was complete, that is, not selected
because they are the best outcomes amongst a larger set of experiments or hand-fitted to the
data.
\end{itemize}
\end{tcolorbox}

 \begin{tcolorbox}[colframe=white]
\paragraph*{Common problems that lead to rejection}

Issues with papers in relation to the recommendations above can lead to rejection.
Other problems that can lead to rejection are as follows.
 
\begin{itemize}
\item Literature reviews that lack critical analysis of prior work or that largely consist of
lists of papers, that is, do not have an insightful discussion.
\item Contributions that consist of small modifications to established techniques, particularly
where the contribution is a straightforward variation of the established technique or where there
are numerous prior papers exploring similar variations.
\item Methods that appear to be developed and hand-tuned on a specific data set without
discussion or demonstration of their lessons for future work or of how the methods would be more
generally applicable.
\item Justification of a method solely by its score in experiments, lacking an a~priori rationale
for why the method is worth exploring.
\item Experiments where the data volumes are too small to support the conclusions.
\item Any form of academic fraud, misrepresentation, or dishonesty.
\end{itemize}
\end{tcolorbox}

\subsubsection{Next Steps}

Guidance and lists of issues should be living documents that reflect a current and uncontroversial agreement in the community.
Therefore, they should be open to change because there can always be some disagreements and expectations
of authors can change over time, in some cases quite quickly, especially as the subjects of research shift to focus on new topics.
For that reason, no set of advice should be regarded as fixed, but revision should be undertaken consultatively and with a spectrum of colleagues.

We suggest that the detailed list of issues of concern in Appendix~\ref{appdx:authorguide:flaws} be made available in some form as educative for reviewers. 
We stress here that it is not our intention that reviewers simply reject papers because of these issues. It could also provide a resource at forums such as doctoral consortia.

We thus believe that it would be valuable for the community to:
\begin{itemize}
\item Ensure that the guidelines are prominent in the calls-for-papers at our major conferences and journals, or otherwise disseminated.
\item Encourage the \ac{SIGIR} executive committee to take ownership of the guidelines and to occasionally
convene a panel to produce an update.
\item Use these resources educatively for new members of the community and for new reviewers.
\end{itemize}

In this exercise, we have not produced guidance for reviewers, which in other disciplines tends to consist of two parts: general advice on how to approach the task and specifics for the field.
An example that we found was produced by the \ac{ACL}, as discussed above; a particular strength of these guidelines in our view is the enumeration of unfair grounds for rejection.
We believe that such guidance would be of value to our community, and could make use of the materials we have presented here.

\newpage

\begin{participants}
\participant Christine Bauer\\ Utrecht University, NL
\participant Joeran Beel\\ University of Siegen, DE
\participant Timo Breuer\\ Technische Hochschule Köln, DE \\
\participant Charles L. A. Clarke\\ University of Waterloo, CA \\
\participant Anita Crescenzi\\ University of North Carolina at Chapel Hill, US
\participant Gianluca Demartini\\ The University of Queensland, AU
\participant Giorgio Maria Di~Nunzio\\ University of Padua, IT
\participant Laura Dietz\\ University of New Hampshire, US
\participant Guglielmo Faggioli\\ University of Padua, IT
\participant Nicola Ferro\\ University of Padua, IT
\participant Bruce Ferwerda\\ Jönköping University, SE
\participant Maik Fr{\"o}be \\ Friedrich-Schiller-Universit{\"a}t Jena, DE
\participant Norbert Fuhr\\ University of Duisburg-Essen, DE
\participant Matthias Hagen\\ Friedrich-Schiller-Universit{\"a}t Jena, DE
\participant Allan Hanbury\\ TU Wien, AT
\participant Claudia Hauff\\ Spotify, NL
\participant Dietmar Jannach \\ University of Klagenfurt, AT
\participant Noriko Kando\\National Institute of Informatics, JP
\participant Evangelos Kanoulas\\ University of Amsterdam, NL
\participant Bart P. Knijnenburg\\ Clemson University, US
\participant Udo Kruschwitz\\ University of Regensburg, DE
\participant Meijie Li \\ University of Duisburg-Essen, DE
\participant Maria Maistro \\ University of Copenhagen, DK
\participant Lien Michiels \\ University of Antwerp, BE
\participant Andrea Papenmeier \\ University of Duisburg-Essen, DE
\participant Martin Potthast\\ Leipzig University, DE
\participant Paolo Rosso \\ Technical University of Valencia, ES
\participant Alan Said \\ University of Gothenburg, SE
\participant Philipp Schaer \\ Technische Hochschule Köln, DE \\
\participant Christin Seifert \\ University of Duisburg-Essen, DE
\participant Ian Soboroff \\ National Institute of Standards and Technology, US
\participant Damiano Spina \\ RMIT University, AU
\participant Benno Stein\\ Bauhaus-Universit{\"a}t Weimar, DE
\participant Nava Tintarev \\ Maastricht University, NL
\participant Juli{\'a}n Urbano\\ Delft University of Technology, NL
\participant Henning Wachsmuth\\ Leibniz Universit{\"a}t Hannover, DE
\participant Martijn Willemsen \\ Eindhoven University of Technology \& JADS, NL 
\participant Justin Zobel\\ University of Melbourne, AU
\end{participants}

\begin{center}
    \vspace*{1em}
    \includegraphics[width=0.8\textwidth]{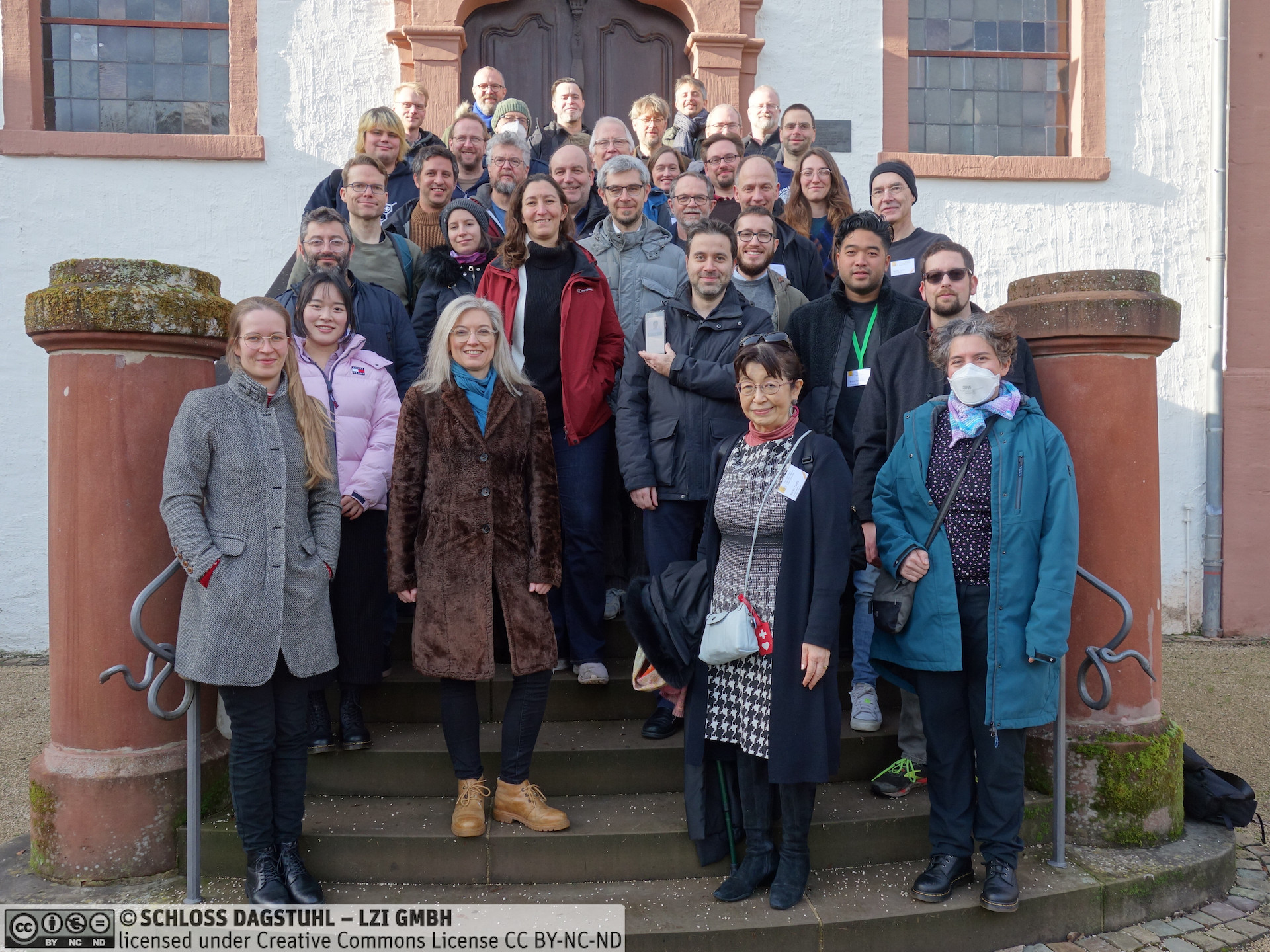}
\end{center}
\clearpage

\section{List of Acronyms}
\begin{acronym}
\acro{ACL}{Association for Computational Linguistics}
\acro{ACM DL}[ACM DL]{ACM Digital Library}
\acro{ADM+S}[ADM+S]{Automated Decision-Making and Society}
\acro{AI}{Artificial Intelligence}
\acro{ASSIA}{Asian Summer School in Information Access}
\acro{CfP}{Call for Papers}
\acro{CHI}{Computer-Human-Interaction}
\acro{CLEF}{Conference and Labs of the Evaluation Forum}
\acro{CyCAT}{Cyprus Center for Algorithmic Transparency}
\acro{ECIR}{European Conference on Information Retrieval}
\acro{ESS}{Experiment Support System}
\acro{ESSIR}{European Summer School on Information Retrieval}
\acro{FAccT}{Fairness, Accountability, and Transparency}
\acro{FDIA}{Future Directions in Information Access}
\acro{FIRE}{Forum for Information Retrieval Evaluation}
\acro{GDPR}{General Data Protection Regulation}
\acro{IR}{Information Retrieval}
\acro{KPI}{Key Performance Indicator}
\acro{LLM}{Large Language Model}
\acro{ML}{Machine Learning}
\acro{MRR}{Mean Reciprocal Rank}
\acro{NLP}{Natural Language Processing}
\acro{nDCG}{normalized Discounted Cumulative Gain}
\acro{NTCIR}{NII Testbeds and Community for Information access Research}
\acro{PC}{Program Committee}
\acro{PyIRE}{Python Interactive Information Retrieval Evaluation}
\acro{QA}{Question Answering}
\acro{RS}{Recommender Systems}
\acro{RecSys}{ACM Conference on Recommender Systems}
\acro{SSH}{Social Sciences and Humanities}
\acro{SPC}{Senior Program Committee}
\acro{SEME}{Search Engine Manipulation Effect}
\acro{SIGIR}{ACM SIGIR Conference on Research and Development in Information Retrieval}
\acro{SIGIR-AP}[SIGIR-AP]{Information Retrieval in the Asia Pacific}
\acro{SIGPLAN}{ACM Special Interest Group on Programming Languages}
\acro{SOTA}{State Of The Art}
\acro{TREC}{Text REtrieval Conference}
\acro{UX}{User Experience}
\acro{WiMIR}{Women in Music Information Retrieval}
\end{acronym}
\newpage


\section{Author Guidance Appendix}
\label{appdx:authorguide}

This Appendix consists of annotated materials that helped to inform the list of concerns described in
Section~\ref{subsec:guidance}, and the list of concerns itself.
As explained in the main text, our aim was to gather suggestions of guidance and issues from a range of
sources and consolidate them into a resource for~IR.

\subsection{SIGPLAN Empirical Evaluation Guidelines with Annotations}
\label{appdx:authorguide:sigplan}

The following is our annotation of the \ac{SIGPLAN} Guidelines.\footnote{https://www.sigplan.org/Resources/EmpiricalEvaluation/}
In this annotation,
we \ul{underlined} aspects deemed particularly worth reflecting in guidelines for~\ac{IR}.
\st{Strikethrough} was used for aspects that we felt did not translate to our community well, and \gt{greyed text}
for aspects, we felt to be valuable but needing adaptation for an IR context.

\subsubsection*{Clearly stated claims}

\noindent
S1: Claims not explicit

\begin{itemize}
\item
Claims must be explicit in order for the reader to assess whether the empirical evaluation supports them.
\st{Missing claims cannot possibly be assessed.
Claims should also aim to state not just what is achieved but how.}
\end{itemize}

\noindent
S2: Claims not appropriately scoped

\begin{itemize}
\item
\ul{The truth of a claim should clearly follow from the evidence provided.}
\st{Claims that are not fully supported mislead readers.}
\gt{ ’Works for all Java’ is over-broad when based on a subset of Java. Other examples are ’works on real hardware’ when evaluating only with (unrealistic) simulation, and ’automatic process’ when requiring human intervention. }
\end{itemize}

\noindent
S3: Fails to acknowledge limitations

\begin{itemize}
\item
A paper should acknowledge its limitations to place the scope of its results in context.
\st{Stating no limitations at all, or only tangential ones, while omitting the more relevant ones may mislead the reader into drawing overly-strong conclusions. This could hold back efforts to publish future improvements and may lead researchers down to wrong paths.}
\end{itemize}

\noindent
S4: Suitable comparison
\begin{itemize}
\item 
\ul{Fails to compare against the appropriate baseline.}
\st{Empirical evidence for a claim that a technique/system improves upon the state-of-the-art should include a comparison against an appropriate baseline. The lack of a baseline means empirical evidence lacks context. A 'straw man' baseline that is misrepresented as state-of-the-art is also problematic, as it would inflate apparent benefit.}
\end{itemize}

\noindent
S5: Comparison is unfair
\begin{itemize}
\item
Comparisons to a competing system should not unfairly disadvantage that system. \gt{Doing so would inflate the apparent advantage of the proposed system. For example, it would be unfair to compile the state-of-the-art baseline at -O0 optimization level, while using -O3 for the proposed system.}
\end{itemize}

\subsubsection*{Principled Benchmark Choice}

\noindent
S6: Inappropriate suite
\begin{itemize}
\item\ul{Evaluations should be conducted using appropriate established benchmarks where they exist} \st{so that claimed results are more likely to generalize. Not doing so may yield results that are not sufficiently general.}
\gt{Established suites should be used in context; e.g., it would be wrong to use a single-threaded suite for studying parallel performance.}
\end{itemize}

\noindent
\gt{S7: Unjustified use of non-standard suite(s)}
\begin{itemize}
\item \gt{
The use of standard benchmark suites improves the comparability of results. However, sometimes a non-standard suite, such as one that is subsetted or homegrown, is the better choice. In that case, a rationale, and possible limitations, must be provided to demonstrate why using a standard suite would have been worse.}
\end{itemize}

\noindent
\st{S8: Kernels instead of full applications}
\begin{itemize}
\item
\st{Kernels can be useful and appropriate in a broader evaluation. However, a claim that a system benefits applications should be tested on such applications directly, and not only on micro-kernels, which may lack important characteristics of full applications.}
\end{itemize}

\subsubsection*{Adequate Data Analysis}

\noindent
S9: Insufficient number of trials
\begin{itemize}
\item
\gt{Modern systems with non-deterministic performance properties may require many trials (e.g., of a single time measurement) to characterize their behavior adequately. Failure to do so risks treating noise as a signal. Similarly, more trials may be needed to get the system into an intended state (e.g., into a steady state that avoids warm-up effects).}
\end{itemize}

\noindent
\ul{S10: Inappropriate summary statistics}
\begin{itemize}
\item
Summary statistics such as mean and median can usefully characterize many results. But they should be selected carefully because each statistic presents an accurate view only under appropriate circumstances. An inappropriate summary may amplify noise or hide an important trend.
\end{itemize}

\noindent
\ul{S11: No data distribution reported}
\begin{itemize}
\item
A measure of variability (e.g., variance, std. Deviation, quantiles) and/or confidence intervals, is needed to understand the distribution of the data. Reporting just a measure of central tendency (e.g., a mean or median) can mislead the reader, especially when the distribution is bimodal or has a fsignificant variance.
\end{itemize}

\subsubsection*{Relevant metrics}

\noindent
\ul{S12: Indirect or inappropriate proxy metric}
\begin{itemize}
\item
\ul{Proxy metrics can substitute for direct ones only when the substitution is clearly, explicitly justified.} \gt{For example, it would be misleading and incorrect to report a reduction in cache misses to claim actual end-to-end performance or energy consumption improvement.}
\end{itemize}

\noindent
S13: Fails to measure all important effects
\begin{itemize}
\item
All important effects should be measured to show the true cost of a system. \gt{For example, compiler optimizations may speed up programs at the cost of drastically increasing compile times of large systems, so the compile time should be measured as well as the program speedup. Failure to do so distorts the cost/benefit of the system.}
\end{itemize}

\subsubsection*{Appropriate and Clear Experimental Design}

\noindent
\ul{S14: Insufficient information to repeat}
\begin{itemize}
\item
\ul{Experiments evaluating an idea need to be described in sufficient detail to be repeatable.} All parameters (including default values) should be included, as well as all version numbers of software, and full details of \st{hardware} platforms. \st{Insufficient information impedes repeatability and comparison of future ideas and can hinder scientific progress.}
\end{itemize}

\noindent
\st{S15: Unreasonable platform}
\begin{itemize}
\item
\st{The evaluation should be on a platform that can reasonably be said to match the claims; otherwise, the results of the evaluation will not fully support the claims. For example, a claim that relates to performance on mobile platforms should not have an evaluation performed exclusively on servers.}
\end{itemize}

\noindent
\ul{S16: Ignores key design parameters}
\begin{itemize}
\item
Key parameters should be explored over a range to evaluate sensitivity to their settings. \gt{Examples include the size of the heap when evaluating garbage collection and the size of caches when evaluating a locality optimization.} All expected system configurations \st{(e.g., from warmup to steady state)} should be considered.
\end{itemize}

\noindent
\st{S17: Gated workload generator}
\begin{itemize}
\item
\st{Load generators for typical transaction-oriented systems should be 'open loop', to generate work independent of the performance of the system under test. Otherwise, results are likely to mislead because real-world transaction servers are usually open-loop.}
\end{itemize}

\noindent
\ul{S18: Tested on training set}
\begin{itemize}
\item
\gt{When a system aims to be general but was developed with close consideration of specific examples, it is essential that the evaluation explicitly perform cross-validation, so that the system is evaluated on data distinct from the training set. For example, static analysis should not be exclusively evaluated on programs used to inform its development.}
\end{itemize}

\subsection{Experimental Standards for Deep Learning Guidelines with Annotations}
\label{appdx:authorguide:exp-dl}

The following is our annotation of the 
highlighted material from the
Experimental Standard for Deep Learning\footnote{\url{https://arxiv.org/pdf/2204.06251.pdf}}~\cite{Ulmer2022_experimental-standard-dl-and-nlp}.
A question mark~(!) indicates the ``must'' category from the original paper and
a plus~(+) indicates recommendations from the original paper.
We use \st{striketrough} for items we deemed not specifically relevant for the IR community,
and \gt{gray text} for relevant items with a valuable issue that needs to be adapted to be
made pertinent to the IR community.

\noindent\subsubsection*{Data}
\begin{description}
\item{\bf D01 !} Consider dataset and experiment limitations when drawing conclusions (Schlangen, 2021);
\item{\bf D02 !} Document task adequacy, representativeness and pre-processing (Bender and Friedman, 2018);
\item{\bf D03 !} \gt{Split the data such as to avoid spurious correlations;}, 
\item{\bf D04 +} Publish the dataset accessibly \& indicate changes;
\item{\bf D05 +}  Perform exploratory data analyses to ensure task adequacy (Caswell et al., 2021);
\item{\bf D06 +} Publish the data set with individual-coder annotations, besides aggregation;
\item{\bf D07 +} \gt{ Claim significance considering the dataset’s statistical power (Card et al., 2020).}
\end{description}

\noindent\subsubsection*{Codebase \& Models}
\begin{description}
\item{\bf D08 \st{!}} Publish a code repository with documentation and licensing to distribute for replicability;
\item{\bf D09 !} Report all details about hyperparameter search and model training;
\item{\bf D10 !} Specify the hyperparameters for replicability
\item{\bf D11 +} \gt{Publish model predictions and evaluation scripts.};
\item{\bf D12 +} \st{Use model cards};
\item{\bf D13 +} \gt{Publish models;} 
\end{description}

\noindent\subsubsection*{Experiments and Analysis}
\begin{description}
\item{\bf D14 !} \gt{Report mean \& standard deviation over multiple runs;}
\item{\bf D15 !} Perform significance testing \st{or Bayesian analysis} and motivate your choice of method;
\item{\bf D16 !} \gt{Carefully reflect on the amount of evidence regarding your initial hypotheses.}
\end{description}

\noindent\subsubsection*{Publications}
\begin{description}
\item{\bf D17 !} \gt{Avoid citing pre-prints (if applicable);}
\item{\bf D18 \st{!}} Describe the computational requirements;
\item{\bf D19 \st{!}} Consider the potential ethical \& social impact;
\item{\bf D20 +} \st{Consider the environmental impact and prioritize computational efficiency;}
\item{\bf D21 +} Include an Ethics and/or Bias Statement.
\end{description}

\subsection{Quick reflections}
\label{appdx:authorguide:reflections}

Our next resource was an unstructured collection of material we gathered by discussing of our individual experience as reviewers.
We also gathered similar kinds of comments from other attendees, which we omit here (they are
much less structured) but incorporated into the list of concerns below.

\begin{itemize}
    \item No analysis of outliers or inspection of spread and diversity of results (aka just report the mean score).
    \item Lit reviews that are lists of papers without reflection, analysis or connections to the current work (gaps, bridges, etc); addition of max number of citations to each statement.
    \item Unreflective use of ‘the rubric’ as a way of writing the paper; no insights, no meaningful analysis, no meaningful identification of contribution.
    \item Justification by score.
    \item Don’t show examples of the method or only show the positive ones.
    \item Unjustified experimental settings such as hyperparameter choices, or a long sequence of unjustified design choices/decisions., and how they may be perpetuated thanks to citing work.
    \item Graph overload -- thousands of results without explanation, choice of illustrative cases -- lost in visualisation. Also, graphs that make no sense.
    \item Confident, bold statements of goals that are impossible to interpret in concrete terms.
    \item Model and problem are not related to each other.
    \item Problem and measures are not related.
    \item Scale of data absurdly out of keeping with the problem that the paper sets out to solve.
    \item Claims are overstated by comparison to the data.
    \item Na\"ive, outdated baselines -- a single strong competitive baseline is better than a family of simplistic baselines.
    \item No consideration of the possibility or scale or presence of random error.
    \item Assumption that training data is perfect; use of cross-fold validation (the dataset defines the task) to draw general conclusions.
    \item Doing of user studies just to get a check-mark for making it real.
    \item Failure to get ethics clearances when required.
    \item Use of crowd-sourcing for experiments that require a laboratory setting.
    \item Use of students enrolled in a subject as experimental subjects when representativeness is required.
    \item Inadequate description of the data, lack of clarity on source and availability, and likewise for the code.
    \item Basic issues with clarity and obscurity; obfuscation.
    \item Badly implemented baseline or implementation is not comparable.
    \item Failure to consider Goodhart’s law.
    \item Inference from aggregate data.
    \item Comparison between systems with different scales of hyperparameters (time-constrained tuning vs. grid search)
    \item Papers that just show summary statistics and don’t show any examples. 
    \item Lack of understanding of what is needed for repeatability, reproducibility, and replicability.
    \item Lack of distinction between development data and test data; selective presentation of results that are favourable.
\end{itemize}

\subsection{Common Flaws in Submitted IR Papers}
\label{appdx:authorguide:flaws}

Our analysis of the materials above, and reading of other resources for authors in cognate fields, provided the basis for the categorisation of areas of concern.
These areas of concern were subsequently analysed to inform the Guidance for Authors included in the main
text. 
In this analysis, we identified bullet points that we regarded as essential; these are marked with
a star~($*$).

\subsubsection*{Design, motivation, and hypothesis}

\begin{itemize}
\item Basic issues with clarity and obscurity; make the design, motivation, and hypotheses difficult to understand and unintentionally obfuscate the main content of the research. 
\item Confident, bold statements of goals that are impossible to interpret in concrete terms.
\item Unreflective use of ‘the rubric’ as a way of writing the paper (i.e., a specific set of details about what is needed to structure a paper) with no insights, no meaningful analysis, and without any meaningful identification of contribution.
\item Inclusion of elements just to follow a template, such as unhelpful user studies, use of ablation when it doesn’t relate to the conclusions, and graphs showing irrelevant data.
\item Lack of a clearly stated problem or research goal.
\item Lack of appreciation that method design relies on domain knowledge; lack of inclusion of extra-disciplinary knowledge where relevant. $*$
\item No acknowledgement of the social or ethical impact of the work. $*$
\end{itemize}

\subsubsection*{Literature}
\begin{itemize}
\item Literature reviews that are mere lists of papers without reflection, analysis or connections to the current work; unreasonably large numbers of citations to each statement. $*$
\item Citing of papers which would clearly fail the above guidelines. 
\item Obvious gaps in the bibliography due to poor literature search, such as missing foundational or key papers that are relevant to the work, recent citations or older citations that are still current.
\end{itemize}

\subsubsection*{Model and method}
\begin{itemize}
\item Model (or the method or solution) and the problem are not related to each other.
\item A long sequence of unjustified design choices and decisions, or justification from prior work that does not apply. $*$
\item Lack of examples of how the model is going to work.
\item Not clear how the method is distinct from and connected to, prior work. $*$
\end{itemize}

\subsubsection*{Data, data gathering, and datasets}
\begin{itemize}
\item Inadequate description of the data, lack of clarity on creation, source or availability. $*$
\item Inappropriate choice of human subjects (e.g., the researchers themselves, or students in cases where they do not represent the target populations).
\item Use of crowd-sourcing for experiments that require a laboratory or controlled settings.
\item Use of survey instruments that are not a good match to the problem, or that haven’t been validated for it.
\item Failure to get ethics clearances when required, lack of consideration of ethics, bias, confidentiality or privacy. $*$
\item Scale of data clearly out of keeping with the problem.
\item Lack of multiple datasets when readily available and appropriate to the problem. $*$
\item Use of the wrong dataset, or no exploration of its suitability for the problem.
\end{itemize}

\subsubsection*{Metrics}
\begin{itemize}
\item Problems and chosen measures are not related (for example, a classification problem and the use of inappropriate measures for this kind of problem). $*$
\item Selective, post hoc use of metrics to find positive results.
\item Reporting of multiple, correlated metrics as if they represented independent sources of evidence.
\item Invented metrics, especially when they are not explained or difficult to interpret.
\end{itemize}

\subsubsection*{Experiments}
\begin{itemize}
\item Lack of distinction between data partitions, such as training, validation, and test set. $*$
\item Results that come from overfitting to the wrong data partition, especially hand-tuned models for that data, or results that are hand-picked from a large volume of trials.
\item No exploration of the sensitivity of the method to the values of key (hyper-)parameters.
\item Unjustified decisions in the experimental setting, such as hyperparameter settings. 
\item Use of default parameters for baselines while tuning the same parameters for the proposed model. $*$
\item Lack of consideration about testing systems with very different numbers of hyperparameters (e.g., time-constrained tuning vs. grid search).
\item Poor or naive choice of baselines (e.g., a single strong competitive baseline is better than a family of simplistic baselines).
\item Badly implemented baselines, or implementation is not comparable.
\end{itemize}

\subsubsection*{Analysis of results and presentation}
\begin{itemize}
\item Reporting only summary statistics without specific examples, positive examples and negative examples.  $*$
\item No consideration for variability and diversity of results and outliers (i.e., reporting only mean scores). $*$
\item Only quantitative results, without studying whether modeling assumptions are reasonably held up and a qualitative discussion of error sources.
\item Selective presentation of results that are favorable. $*$
\item Selective, post hoc use of statistical analysis to find positive results; reporting of results as ``nearly significant''.
\item No consideration of the presence and scale of random error.
\item Overstatement of the statistical precision of results.
\item Data overload: unnecessarily large numbers of graphs and tables, or insufficient explanation as to how to interpret them.
\item Poor statistical analysis, such as wrong choice of significance test, lack of consideration of power or effect size, statistical testing when sample size is unsuitable, or missing to mention what hypotheses are being tested and how.
\item Superficial analysis or without interpretation.
\end{itemize}

\subsubsection*{Repeatability, reproducibility and replicability}
\begin{itemize}
\item Lack of communication of what is needed for repeatability, reproducibility, and replicability; e.g., missing parameter settings, missing explanation of data preparation and pre-processing. $*$
\item Failure to use an appropriate standard dataset.
\item Failure to use a standard implementation (e.g., baselines, evaluation software).
\item Lack of recognition of the value of publishing data and code. 
\item Inadequate description of code, lack of clarity on source and availability, documentation, licensing, key metadata, or not versioned.
\end{itemize}

\subsubsection*{Conclusions and claims}
\begin{itemize}
\item Inference of general conclusions from aggregated data without individual analysis.
\item Assumption that training data are perfect (e.g., that they are an ideal setting and representative of all possible data). $*$
\item Claims of performance on unseen data based on cross-validation results.
\item Claims that do not follow from the results. $*$
\item Justification of innovation entirely by numerical results. $*$
\item Use the current results to reformulate the initial hypotheses.
\item No consideration of limitations of the proposed solution or experimentation.
\item No noting of excessive or large-scale computational requirements.
\end{itemize}

\end{document}